\documentclass[]{aa} 
\usepackage{natbib} 
\usepackage{graphicx}
\usepackage{amssymb}
\usepackage{amsmath}
\usepackage{txfonts}
\usepackage{epsfig}
\usepackage{enumitem}

\def\inte{{\em INTEGRAL}}
\def\xmm{{\em XMM-Newton}}

\def\beppo{{\em BeppoSAX}}

\def\swift{{\em Swift}}

\def\suzaku{{\em Suzaku}}

\def\erg{erg~s$^{-1}$}

\begin{document}

   \title{The accretion environment of Supergiant Fast X-ray Transients probed with \xmm}

   \author{E. Bozzo 
          \inst{1}
        \and F. Bernardini
          \inst{2}
        \and  C. Ferrigno
           \inst{1} 
         \and M. Falanga 
           \inst{3}  
          \and P. Romano
          \inst{4}  
          \and L. Oskinova
          \inst{5,6}              
          }
   \institute{ISDC Data Centre for Astrophysics, department of Astronomy, University of Geneva, Chemin d’Ecogia 16,
             CH-1290 Versoix, Switzerland; \email{enrico.bozzo@unige.ch}
         \and
         New York University Abu Dhabi, PO Box 129188, Abu Dhabi, UAE
         \and
        International Space Science Institute (ISSI) Hallerstrasse 6, CH-3012 Bern, Switzerland
          \and   
        INAF - Osservatorio Astronomico di Brera, via Emilio Bianchi 46, I-23807 Merate (LC), Italy    
          \and 
         Institut f\"ur Physik und Astronomie, Universit\"at Potsdam, Karl-Liebknecht-Strasse 24/25, 14476 Potsdam, Germany
          \and 
          Kazan Federal University, Kremlevskaya Str., 18, Kazan, Russia
             }
   
   \date{Submitted: 2017 Jan 5; Accepted 2017 Aug 31}

  \abstract{Supergiant fast X-ray transients are a peculiar class of supergiant X-ray binaries characterized by a remarkable 
  variability in the X-ray domain, widely ascribed to the accretion from a clumpy stellar wind.} 
  {In this paper we performed a systematic and homogeneous analysis of the sufficiently bright X-ray flares observed 
  with \xmm\ from the supergiant fast X-ray transients to probe spectral variations on timescales as short as a few hundred of seconds. 
  Our ultimate goal is to investigate if SFXT flares and outbursts are triggered by the presence of clumps and eventually 
  reveal whether strongly or mildly dense clumps are required.}  
  {For all sources, we employ a technique developed by our group already exploited in a number of our previous papers, making use of an adaptive 
  rebinned hardness ratio to optimally select the time intervals for the spectral extraction. A total of twelve observations 
  performed in the direction of five SFXTs are reported, providing the largest sample of events available so far.} 
  {Using the original results reported here and those obtained with our technique from the analysis of two previously published 
  \xmm\ observations of IGR\,J17544-2619 and IGR\,J18410-0535, we show that both strongly and mildly dense clumps 
  can trigger these events. In the former case, the local absorption column density may increase by a factor of $\gg$3, while in the latter case, 
  the increase is only by a factor $\sim$2-3 (or lower). An increase in the absorption column density is generally recorded during 
  the rise of the flares/outbursts, while a drop follows when the source achieves the peak flux. In a few cases, a re-increase of the absorption column 
  density after the flare is also detected, and we discovered one absorption event related to the passage of an unaccreted clump in front of the compact object.  
  Overall, there seems to be no obvious correlation  between the dynamic ranges in the X-ray fluxes and absorption column densities in supergiant fast X-ray transients, 
  with an indication that lower densities are recorded at the highest fluxes.} {The spectral variations measured in all sources are in agreement with the idea that 
  the flares/outbursts are triggered by the presence of dense structures in the wind interacting with the X-rays from the compact object (leading to photoionization).  
  The lack of correlation  between the dynamic ranges in the X-ray fluxes and absorption column densities can be explained by the presence of accretion inhibition 
  mechanism(s). Based on the knowledge acquired so far on the SFXTs, 
  we propose a classification of the flares/outbursts  from these sources to drive future observational investigations. 
  We suggest that the difference between the classes of flares/outbursts is related to the fact that the mechanism(s) inhibiting accretion 
  can be overcome more easily in some sources compared to others. We also investigate the possibility that different stellar wind structures, 
  rather than clumps, could provide the means to temporarily overcome the inhibition of accretion in  
  supergiant fast X-ray transients.}


   \maketitle

\section{Introduction}
\label{sec:intro}

Supergiant Fast X-ray Transients (SFXTs) are a sub-class of the so-called ``classical'' Supergiant High Mass X-ray Binaries 
(SgXBs) hosting a neutron star (NS) accreting from the wind of an O-B supergiant companion 
\citep[see, e.g.,][for a recent review]{nunez17}. At odds with classical systems, the SFXTs are known to be largely sub-luminous 
(a factor of $\gtrsim$10-100) and to display a much more prominent variability in X-rays. In particular, SFXTs 
spend most of their lifetime in a low quiescent states (with X-ray luminosities as low as 10$^{32}$~\erg) and 
undergo sporadic few hours-long outbursts reaching luminosities of $L_{\rm X}$$\simeq$10$^{36}$-10$^{38}$~\erg.  
Flares with intermediate luminosities between quiescence and outburst 
($L_{\rm X}$$\simeq$10$^{34}$-10$^{35}$~\erg) are often observed and last a few thousands seconds at the most \citep[see, e.g.,][]{romano14}.  
The largest dynamic range in luminosity recorded so far is that of the SFXT prototype IGR\,J17544-2619, reaching 
a factor of $\sim$10$^6$ \citep{romano15}. For comparison, the classical SgXBs display a typical dynamic range of 
$\lesssim$100 \citep{walter15}. 
 \begin{table*} 
 \scriptsize	
 \begin{center} 	
 \caption{Log of all \xmm\ observations considered in this paper for the different SFXT sources. We provide the 
 relevant reference for the observations already published by other authors (without performing an adaptively rebinned  
 HR-resolved spectral analysis) and marked with ``this work'' 
 data that are publicly available but were not published yet. In the column ``Comments'' we indicated if the source 
 was showing bright or moderately bright flares, providing also some information about the level of background in the most 
 critical cases.}  	
 \label{tab:log} 	
 \begin{tabular}{lllllll} 
 \hline 
 \hline 
 \noalign{\smallskip}  
 Source & Obs. ID. & Start & Stop & Exp. & Reference & Comments \\
 \noalign{\smallskip} 
        &          & (UTC) & (UTC) & (ks) &  & \\
 \hline 
 \noalign{\smallskip} 
IGR\,J18450-0435 & 0306170401 & 2006-04-03 14:44 & 2006-04-03 20:22 & 19.0 & \citet{zurita09} & bright flares \\
                 & 0728370801 & 2014-10-13 23:29 & 2014-10-14 05:50 & 21.6 & this work & bright flares \\
                 & 0728371001 & 2014-10-15 16:20 & 2014-10-15 21:50 & 18.5 & this work & faint flares \\
 \noalign{\smallskip} 
 \hline 
  \noalign{\smallskip}
IGR\,J17544-2619 & 0148090501 & 2003-09-11 18:44 & 2003-09-11 22:09 & 11.0 & \citet{riestra04} & bright flares \\
 \noalign{\smallskip}
                 & 0154750601 & 2003-09-17 17:13 & 2003-09-11 19:59 & 8.0 & \citet{riestra04} & bright flares, affected by high background \\
 \noalign{\smallskip}
                 & 0679810401 & 2012-09-16 02:55 & 2012-09-16 07:30 & 15.0 & \citet{drave14} & rise to a flaring state \\
 \noalign{\smallskip}  
 \hline  
 \noalign{\smallskip} 
SAX\,J1818.6-1703 & 0693900101 & 2013-03-21 12:19 & 2013-03-21 21:56 & 33.3 & \citet{boon16} & bright flares, moderately affected by high background \\  
 \noalign{\smallskip} 
 \hline  
 \noalign{\smallskip} 
IGR\,J17354-3255 & 0693900201 & 2013-03-15 12:47 & 2013-03-15 22:22 & 33.2 & this work & flares, moderately affected by high background \\ 
 \noalign{\smallskip} 
                 & 0701230101 & 2013-03-29 11:31 & 2013-03-29 21:22 & 34.2 & this work & flares, moderately affected by high background \\
 \noalign{\smallskip} 
                 & 0701230701 & 2013-03-31 02:17 & 2013-03-31 08:41 & 21.7 & this work & several flares \\ 
 \noalign{\smallskip} 
 \hline  
   \noalign{\smallskip} 
IGR\,J16328-4726 & 0728560201 & 2014-08-24 19:17 & 2014-08-25 05:20 & 34.9 & \citet{fiocchi16} & moderately bright flares \\
                 & 0728560301 & 2014-08-26 18:44 & 2014-08-27 01:07 & 21.7 & \citet{fiocchi16} & moderately bright flares \\ 
 \noalign{\smallskip} 
  \hline  
  \end{tabular}
  \end{center}
  \end{table*}
 \begin{table*} 
 \scriptsize	
 \begin{center} 	
 \caption{Log of all \xmm\ observations of flaring/outbursting SFXTs published previously and for which an adaptively 
 rebinned HR-resolved spectral analysis was already carried out.} 	
 \label{tab:log2} 	
 \begin{tabular}{lllllll} 
 \hline 
 \hline 
 \noalign{\smallskip}  
 Source & Obs. ID. & Start & Stop & Exp. & Reference & Comments \\
 \noalign{\smallskip} 
        &          & (UTC) & (UTC) & (ks) & \\
 \hline 
 \noalign{\smallskip} 
IGR\,J17544-2619 & 0744600101 & 2015-03-20 05:00 & 2015-03-21 20:17 & 140.0 & \citet{bozzo16b} & outburst \\
  \noalign{\smallskip} 
 \hline  
 \noalign{\smallskip} 
IGR\,J18410-0535 & 0604820301 & 2010-03-15 13:09 & 2010-03-16 02:13 & 45.7 & \citet{bozzo11} & bright flare \\  
 \noalign{\smallskip}   
 \hline  
 \noalign{\smallskip} 
  \end{tabular}
  \end{center}
  \end{table*} 
  
It was originally proposed that largely eccentric orbits and extremely clumpy winds could 
explain the SFXT phenomenology \citep{zand05,negueruela10,walter07}. In a wide elongated orbit, the NS remains quiescent for a 
significant fraction of time, and the accretion of clumps endowed with an over-density of a factor of $\gtrsim$1000 compared 
to the surrounding wind could give rise to the brightest SFXT outbursts (especially when the NS approaches the companion). 
The problem with this interpretation 
is that a number of SFXTs were discovered to have short orbital periods (3-5~days) and studies of isolated supergiant 
winds showed that clumps are  typically only a factor of $\simeq$10 over-dense compared to a smooth wind  
\citep[see, e.g.][and references therein]{lutovinov13,bozzo15,nunez17}. 
\citet{bozzo08} and \citet{grebenev07} proposed that the inhibition of accretion caused by the neutron star rotating magnetosphere 
could combine with the presence of moderately dense clumps to boost the X-ray 
variability of the SFXTs compared to classical systems. An alternative way to inhibit accretion and widen the dynamic range 
of the SFXTs involve the combined presence of a long-lasting ``subsonic settling accretion regime'' and a significantly magnetized 
supergiant wind \citep{shakura12,shakura14}.  
 
Although the theoretical interpretation of the SFXT behavior is still matter of debate, 
convincing evidence has been reported in the literature about the presence of massive structures in the  
stellar winds around the NSs in these systems. Observationally, the role of these structures is two-fold. A sufficiently large and dense 
structure simply passing in front of the NS (without being accreted) causes source dimming or even obscuration. 
Its presence can be revealed by the signature of the photoelectric absorption, even when there is no  
significant increase of the source X-ray emission.   
Structures that instead are intercepted by the NS lead to a temporarily larger mass accretion rate, giving rise to  
X-ray flares or outbursts characterized by an enhanced local absorption  
that is proportional to the lateral size of the accreted structure \citep[see, e.g., the discussion in][]{bozzo13}. 
Events of source dimming have been reported in several observations of SFXTs  
and tentatively associated with clumps passing in front of the compact object  
\citep[see, e.g.,][]{rampy09,drave13}. X-ray flares and outbursts in SFXTs have also been associated  
to accretion of clumps in several publications, but finding direct proofs of the existence of these 
structures through X-ray observations has proved challenging. The reason is that SFXT flare and outbursts occur at unpredictable times.  
Their detailed spectroscopic analysis requires monitorings with a large effective area instrument able to give access to integrations as short 
as the local dynamical timescales (hundred of seconds) with the necessary signal-to-noise ratio (S/N) and energy resolution.

So far, the most convincing 
evidence for the ``ingestion'' of a massive clump by a NS hosted in a SFXT was reported by \citet{bozzo11} using 
an \xmm\ observation that luckily caught IGR\,J18410-0535 undergoing a bright X-ray flare. 
In that occasion, an increase by a factor of $\sim$10 of the absorption column density in the direction of the source preceding the flare 
revealed that a massive and dense structure was rapidly approaching the NS, giving first rise to a large increase in flux 
(a factor of $\sim$1000) before obscuring the compact object and finally moving away from the observer line of sight. 
\citet{bozzo16b} also fortuitously caught with \xmm\ an episode of strongly enhanced 
X-ray activity from the SFXT IGR\,J17544-2619 that reached the typical luminosity of an X-ray outburst 
but displayed only a marginal increase in the local absorption column density (a factor of $\lesssim$2). 
Estimating the physical properties of the clumps in a reliable way from these observations have been proved complicated, 
due to our poor knowledge of the spin period and magnetic field of the NS in the SFXTs. This prevents us 
from correctly quantifying the effect of the mechanisms inhibiting accretion and contributing to produce the total variation 
of the X-ray luminosity \citep[see also the detailed discussion in][]{bozzo11}. 

In this paper, we analyze all publicly available \xmm\ observations of confirmed SFXTs in which sufficiently bright 
X-ray flares/outbursts are detected to perform a detailed spectral analysis.   
We take advantage of the large collecting area of the EPIC cameras 
on-board \xmm\ \citep{jansen01,struder01,turner01} 
to investigate spectral variations during flares that can be associated to 
the presence of structures in the wind of the supergiant companion. An adaptively rebinned hardness-ratio resolved spectral 
analysis is homogeneously carried out for a total of 12 observations performed in the direction of 5 out of the $\sim$10 known SFXTs 
(IGR\,J18450-0435, IGR\,J17544-2619, SAX\,J1818.6-1703, IGR\,J17354-3255, and IGR\,J16328-4726).  
Our dataset includes both data that were published previously but not searched for rapid spectral variations, as well as 
5 unpublished observations. 
The unique capability of \xmm\ for these studies is to provide a sufficiently large effective area in the soft X-ray domain 
(0.5-12~keV) to perform spectroscopic investigations down to timescales at short as a few hundred of seconds, thus allowing us to resolve 
changes in the photoelectric absorption in the direction of the source during the rise and decays of the flares 
(as well as in the quiescent periods between them).  
Based on our experience, we consider sufficiently bright flares those in which 
the source count-rate achieves at least 2-3 cts~s$^{-1}$ in the EPIC cameras\footnote{The only exception in this paper is the 
observation ID.~0679810401, in which we could reveal a modest (but still significant) variation in the absorption column density during a faint 
X-ray flare from IGR\,J17544-2619.}.  

Our final goal is to test if all observed flares/outbursts are characterized by a spectral variability that can be associated with the 
presence of a massive clump around the compact object or if also less massive structures are able to trigger these phenomena. 
We summarize in Sect.~\ref{sec:data} the data analysis procedure and describe in the associated sub-sections the results obtained 
for each source and each observation. An overview of all the results is provided in Sect.~\ref{sec:overview}. The corresponding 
discussion is presented in Sect.~\ref{sec:discussion}, together with our conclusions.

\section{\xmm\ data reduction and analysis}
\label{sec:data}

For all sources considered in this paper, we produced calibrated event lists by using the Science Analysis System (SAS) v.15.0. 
EPIC-pn, EPIC-MOS, and RGS event files were generated with standard analysis performed with the 
{\sc epproc}, {\sc emproc}, and {\sc rgsproc} tasks. 
EPIC-pn and EPIC-MOS data were filtered in the 0.5-12~keV and 0.5-10~keV energy range, respectively. 
Extraction regions for the lightcurves and spectra of all sources and background  
were selected by following recommendations in the latest SAS on-line analysis 
threads\footnote{http://xmm.esac.esa.int/sas/current/documentation/threads.}.  
The difference in extraction areas between source and background was accounted for by using the SAS {\sc
backscale} task for the spectra and {\sc lccorrr} for the lightcurves. 
EPIC spectra were rebinned before fitting in order to have 15-25 counts per bin (depending on the available 
statistics) and, at the same 
time, to prevent oversampling of the energy resolution by more than a factor of three. 
RGS spectra were also rebinned accordingly, whenever required.    
The timing analysis of all EPIC data was performed on barycenter-corrected event lists 
(we used the SAS {\em barycen} tool). 
Details on specific issues regarding the observations of the different sources are provided 
in the dedicated following sections. 
Throughout this paper, uncertainties are given at 90\%~c.l., unless stated otherwise.   

For all sources, we searched for rapid spectral changes in their X-ray emission by performing a 
hardness ratio (HR) resolved spectral analysis. The HR is computed in all cases from the adaptively-rebinned 
energy-resolved lightcurves and calculated as the ratio of the 
source count-rate in the soft to hard energy band versus time. The adaptive rebinning and 
the choice of the soft and hard energy bands are performed differently for each source, depending on the data statistics 
and the averaged spectral properties. 
The HR-resolved spectral analysis we employ in this paper 
is better suited to reveal rapid spectral variations during the flares/outbursts of the SFXTs 
compared to the more ``standard'' intensity or time-resolved spectral analyses 
carried out before for 7 (out of 12) X-ray observations included in our study. The HR-resolved spectral analysis 
probes {\it directly} when a source is undergoing a significant spectral variation, without the need to assume {\it a priori} 
that such changes can occur either during periods of lower/higher X-ray activity or in specific time intervals.  
This technique was already exploited in two of our previous publications and 
allowed us to prove that the bright flare observed by \xmm\ from the SFXT IGR\,J18410-0535 was most 
likely related to the ingestion of a massive clump by the NS hosted in this system \citep{bozzo11}, while the  
outburst caught from the SFXT prototype IGR\,J17544-2619 was shown to require the simultaneous inter-play of additional 
mechanisms to give rise to the measured dynamic range in the X-ray luminosity \citep{bozzo16b}. The conclusion in the case of IGR\,J17544-2619 
was mainly derived from the fact that no particularly large increase of the accretion column density was recorded during this 
event, although it achieved an X-ray luminosity a factor of $\gtrsim$10 larger than that of the X-ray flare caught from 
IGR\,J18410-0535 and the latter displayed evidence for an increase in the local $N_{\rm H}$ by a 
factor of $\sim$10 preceding the X-ray flare. Note that for all sources analyzed in this paper the estimated distance is 
of $\sim$2-4~kpc \citep[][and references therein]{nunez17}, and thus all episodes of enhanced X-ray emission reported in the next sections 
are compatible with being flares as the peak luminosity is in all cases $\lesssim$10$^{36}$~erg~s$^{-1}$. 

The soft X-ray spectra of almost all sources considered here are well described by using an absorbed power-law model, 
as it is commonly the case for SFXTs and other HMXBs. The cut-off of the power-law at high energies is usually measurable 
only when a broad-band spectrum extending up to $\gtrsim$20-30~keV is available \citep[see, e.g.,][and references therein]{walter15}. 
The 1-10~keV X-ray flux was measured by using the {\sc pegpow} model in {\sc xspec}, in order to 
have values not dependent from the largely variable absorption column densities and the uncertainties that arise when 
the spectra are de-absorbed after the fit in {\sc xspec}.

SFXTs are characterized (on average) by a relatively large absorption column density ($\gg$10$^{22}$~cm$^{-2}$), 
well in excess to the galactic extinction in the source direction. However, the generally low statistics of the data at 
energies below 2-3~keV do not allow us to unambiguously separate the contribution of the Galactic absorption and the additional 
absorption local to the source. The former contribution can be rougly estimated by using an online 
tool\footnote{http://heasarc.gsfc.nasa.gov/cgi-bin/Tools/w3nh/w3nh.pl?} based on the results published 
by \citet{dickey90} or by more recent works considering also the contribution of molecular hydrogen \citep{dame01}.  
The density of absorbing material located close to the source, that is generally largely in excess of the expected Galactic value, 
can be obtained by measuring an excess with respect to the expected Galactic value using a simple {\sc phabs} spectral component in {\sc xspec}.
This is a common practice in the case of HMXBs for which high statistics X-ray observations 
with grating spectrometers are not available \citep[see, e.g.,][]{grinberg15}. In a few observations used  
in this work, slightly more complex spectral models were also considered, including a partial covering absorber 
({\sc pcfabs} in {\sc xspec}).  
The latter is also widely used to describe the soft X-ray emission of HMXBs, and it is usually ascribed to 
localized absorbers that partially intercept the radiation from the NS \citep[see, e.g.,][and references therein]{tomsick09b}.  
The X-ray emission that escapes this extra absorption is only affected by the Galactic column density.  
The fits to the spectra of a few sources also required the addition of a thin Gaussian line with a centroid energy compatible 
with that expected from neutral iron at 6.4~keV. Similar lines in HMXBs are known to occur due to the fluorescence 
of the X-ray photons released by the NS onto the surrounding stellar wind material \citep[see, e.g.,][]{torrejon10b,gimenez16}.  
We refer the reader to the following sub-sections dedicated to each source for further details. 

The log of all \xmm\ observations used in this paper is reported in Table~\ref{tab:log}. 
For completeness, we also add in Table~\ref{tab:log2} the \xmm\ observations that we 
published previously for which an adaptively rebinned HR-resolved spectral analysis 
of the SFXT flares/outbursts was already carried out. These observations are used in Sect.~\ref{sec:discussion} 
to provide a more exhaustive discussion and conclusions of all results.

\subsection{IGR\,J18450-0435}
\label{sec:AXJ1845}

\subsubsection{Observation ID.~0306170401}
\label{sec:ax401} 

\begin{figure}
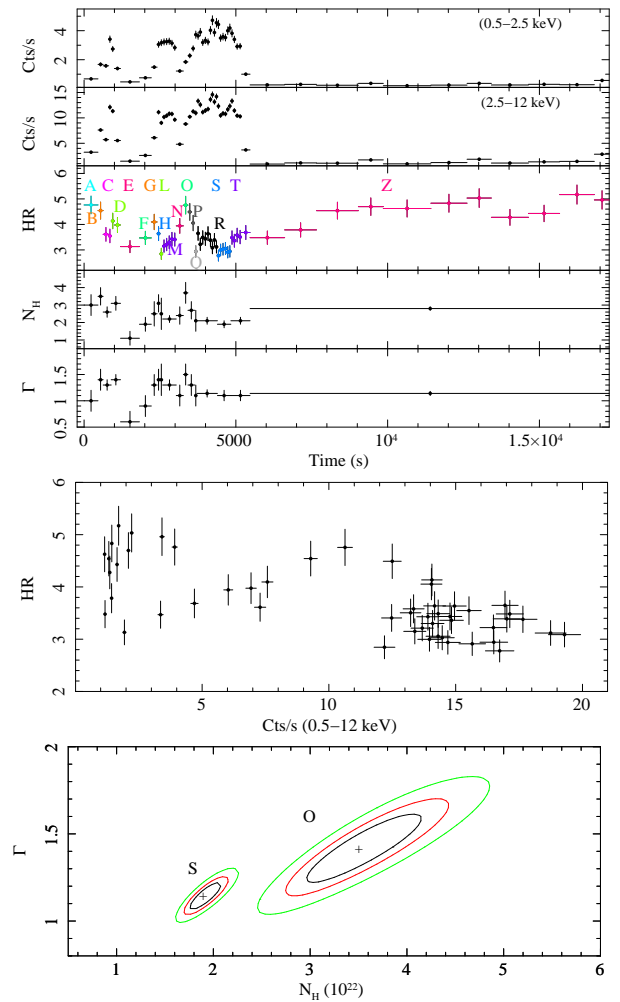

\centering
\includegraphics[scale=0.35,angle=-90]{figures/AXJ1845/lcurve.ps}
\includegraphics[scale=0.35,angle=-90]{figures/AXJ1845/HR.ps}
\includegraphics[scale=0.35,angle=-90]{figures/AXJ1845/contour.ps}
\caption{Results obtained from the observation ID.~0306170401 of IGR\,J18450-0435. 
{\it Top}: EPIC-pn lightcurve of the source in two energy bands and the corresponding HR 
(an adaptive rebinning has been used to achieve a minimum S/N$>$15 in each soft time bin). 
Letters in the middle panel indicate the different HR resolved spectra reported in 
Table~\ref{tab:j1845_fit2}. The last two panels show the results obtained from the spectral fits to the HR-resolved spectral 
analysis (the measured absorption column density in units of 10$^{22}$~cm$^{-2}$ and the  
photon index $\Gamma$). {\it Middle}: Hardness intensity diagram (HID). 
{\it Bottom}: Contour plot showing the ranges spanned by the power-law photon index and the absorption column density 
measured from the most relevant time intervals for the spectral variations marked in the second plot from the top 
(in this case the intervals ``O'' and ``S'').}
\label{fig:j1845} 
\end{figure}
\begin{figure}
\centering
\includegraphics[scale=0.35,angle=-90]{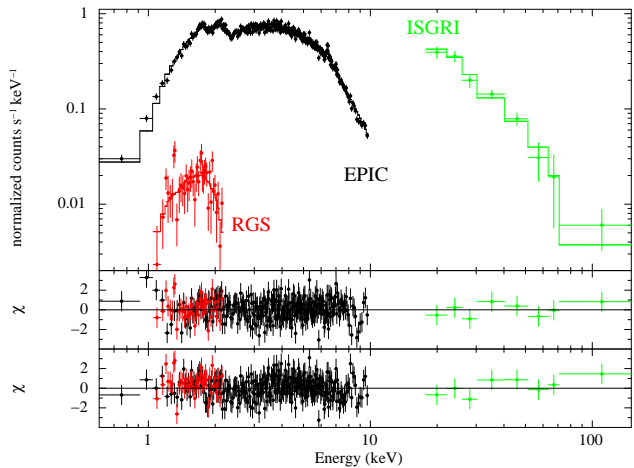}
\caption{Broad-band X-ray spectrum of IGR\,J18450-0435 obtained by summing together the data from the 
three \xmm\ EPIC cameras in the observation ID.~0306170401, combining the data from the 2 RGSs in the same observation, 
and using the INTEGRAL data described in the text. The best fit is 
obtained with a partial covering model comprising a cut-off power-law and an iron line. The middle panel shows the 
residuals from this fit. The bottom panel shows the residuals obtained by fitting the spectrum with the 
model proposed by \citet{zurita09}.}
\label{fig:j1845_broad} 
\end{figure}

This observation was first analyzed by \citet{zurita09}. It was not affected by high flaring background interval and 
thus the entire exposure time can be retained for the scientific analysis.  
By using the {\sc epatplot} tool, we found that the source X-ray emission was moderately affected by pile-up at 
count-rates $\gtrsim$7~cts/s both on the EPIC-pn and the MOS cameras. We verified that using an annular extraction region with a 
inner radius of 100~px for the EPIC-pn and 150~px for the MOS could remove the effect of pile-up. 
Figure~\ref{fig:j1845} (top) shows the source energy-resolved lightcurves, together with the corresponding HR.  
The latter is also shown in the same figure as a function of the total source 
intensity (hardness intensity diagram, hereafter HID).  
No apparent correlation emerges between the source count-rate and the HR, but some specific intervals of count-rates 
are visible where the variations of the HR are more pronounced.   
We extracted the pile-up corrected EPIC spectra and also made use of the 
RGS1 and RGS2 average spectra summed together with the SAS tool {\sc rgscombine} to improve the S/N. 

In order to carry out a comparison with previously reported results, we also extracted the spectrum 
from the IBIS/ISGRI instrument on-board \inte\ during the 7 pointings considered by \citet[][see their Table~2]{zurita09}. 
IBIS/ISGRI data were analyzed by using standard procedures and the version 10.2 of the OSA software distributed 
by the ISDC \citep{courvoisier03}.  
  
In Fig.~\ref{fig:j1845_broad} we show the averaged EPIC-pn and MOS spectra summed 
together by following the SAS online threads\footnote{http://xmm.esac.esa.int/sas/current/documentation/threads/} 
in order to achieve the highest possible S/N. This spectrum was grouped in order to have  
at least 150 photons per energy bin. We verified {\it a posteriori} that results obtained from the fit 
to this spectrum were compatible with those obtained by using only the EPIC-pn spectrum. 
A fit to the summed EPIC spectrum was carried out simultaneously with the RGS and the IBIS/ISGRI spectra first by using the same spectral 
model considered by \citet{zurita09}. This model comprises an emission from optically thin plasma at the lower energies 
({\sc mekal} in {\sc Xspec}), a cut-off power-law to fit the data above $\sim$4~keV, and an edge at $\sim$7.9~keV. 
We added to this model a gaussian line at $\sim$6.4~keV, that went previously unnoticed, and the inter-calibration constants 
for IBIS/ISGRI and the RGS 
with respect to the summed EPIC cameras (these constants also account for the source variability as the IBIS/ISGRI and 
\xmm\ data are not strictly simultaneous). This model gave a reasonable description of the data and parameter values  
in agreement with those previously reported by \citet{zurita09}.  
We found that a simpler model comprising a partial covering absorber, 
a cut-off power-law and an iron line could describe the data 
equally well ($\chi^2_{\rm red}$/d.o.f.=1.14/325). We measured in this case $N_{\rm H}^{\rm los}$=(1.8$\pm$0.1)$\times$10$^{22}$~cm$^{-2}$, 
$N_{\rm H}$=(5.5$^{+1.2}_{-1.0}$)$\times$10$^{22}$~cm$^{-2}$, $f$=5.6$\pm$0.4, $\Gamma$=1.3$\pm$0.1, 
$E_{\rm cut}$31.7$^{+16.9}_{-9.2}$~keV, $E_{\rm Fe}$=6.42$\pm$0.02~keV, $EQW_{\rm Fe}$=0.05$^{+0.02}_{-0.01}$, 
C$_{\rm RGS}$=1.1$\pm$0.1, and C$_{\rm IBIS/ISGRI}$=4.8$\pm$1.0. Here $N_{\rm H}^{\rm los}$ is the Galactic column density, 
$N_{\rm H}$ is the column density of the partial covering absorber, $f$ the covering fraction, $E_{\rm cut}$ is the cut-off energy 
of the cut-off power-law component and $\Gamma$ the corresponding photon index. $E_{\rm Fe}$ is the centroid energy of the iron line, 
and $EQW_{\rm Fe}$ its equivalent width. We fixed the normalization constant of the combined EPIC cameras spectrum to unity and indicated with 
C$_{\rm RGS}$ and C$_{\rm IBIS/ISGRI}$ the values of the normalization constants obtained for the RGS and the IBIS/ISGRI spectrum, respectively. 
The effective exposure time of the spectrum was of 38.2~ks for the EPIC, 12.0~ks for the RGS, and 17.6~ks for IBIS/ISGRI. The averaged flux 
is 4.3$\times$10$^{-11}$~erg~cm$^{-2}$~s$^{-1}$ (1-10~keV). 
We note that the value of $N_{\rm H}^{\rm los}$ in the partial covering model  
is compatible with the expected Galactic value in the direction of the source \citep{dickey90}.  
 
To investigate possible spectral changes during the \xmm\ observation, we extracted different EPIC-pn and EPIC-MOS spectra of the source 
by choosing the time intervals corresponding to the most noticeable variations of the HR. 
The different spectral intervals are labelled with letters A-Z and marked with different colors in Fig.~\ref{fig:j1845} (top). 
Spectra A-T were characterized by relatively low statistics and could be well fit by using a simple absorbed power-law model. 
The spectrum Z was extracted by using a significantly longer exposure time and could be fit only by using a partial covering 
cut-off power-law model. We fixed the cut-off energy at 31.7~keV (based on the \inte\ results) and measured 
$N_{\rm H}^{\rm los}$=(2.4$\pm$0.2)$\times$10$^{22}$~cm$^{-2}$, $N_{\rm H}$=(9.5$\pm$2.2)$\times$10$^{22}$~cm$^{-2}$, f=0.57$\pm$0.05, 
$\Gamma$=1.4$\pm$0.1, and an average flux of 1.4$\times$10$^{-11}$~erg~cm$^2$~s$^{-1}$.    
For each selected time interval (A to Z), the EPIC-pn and EPIC-MOS spectra were fit simultaneously and  
the estimated values of the absorption column density, power-law photon index, and flux are summarized 
in Table~\ref{tab:j1845_fit2}. Some results are also plot in Fig.~\ref{fig:j1845} (top). 
These fits revealed only a moderate change of the absorption column density (a factor of $\sim$2) 
and power-law photon index, despite the remarkable variation in the X-ray flux (a factor of $\gtrsim$10).

\subsubsection{Observation ID.~0728370801}

\begin{figure}
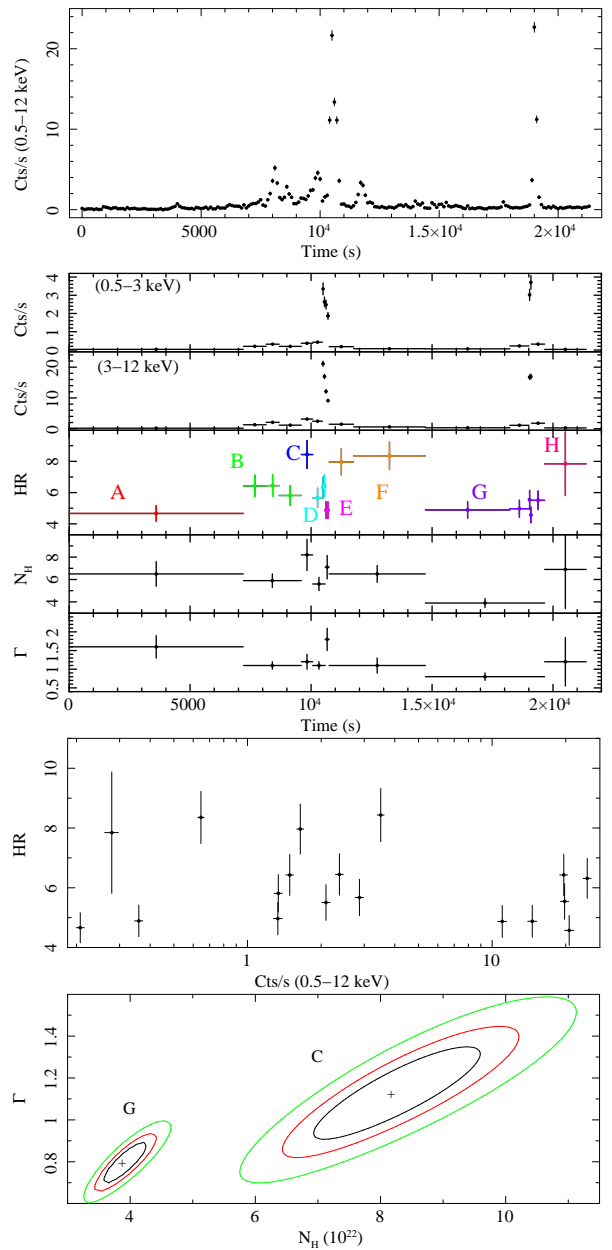

\centering
\includegraphics[scale=0.35,angle=-90]{figures/AXJ1845/801/lcurve.ps}
\includegraphics[scale=0.35,angle=-90]{figures/AXJ1845/801/ax801_HR.ps}
\includegraphics[scale=0.35,angle=-90]{figures/AXJ1845/801/hid.ps}
\includegraphics[scale=0.35,angle=-90]{figures/AXJ1845/801/ax801_contours.ps}
\caption{Same as Fig.~\ref{fig:j1845}, but in case of the IGR\,J18450-0435 observation ID.~0728370801. 
As this observation was not reported previously in the literature, the source EPIC-pn 
lightcurve in the full energy range (0.5-12~keV) is also shown on the top plot (time bin 100~s). 
An adaptive rebinning achieving S/N$>$10 in each soft time bin has been used to compute the HR and HID.}      
\label{fig:ax801} 
\end{figure}

This observation was not affected by flaring background intervals and 
thus the full exposure time could be retained for the scientific analysis. The pn was operated in small window mode, while the two MOS 
were set in large window mode. As this observation was not reported elsewhere yet, 
we show the full-band EPIC-pn lightcurve in Fig.~\ref{fig:ax801}, together with the energy-resolved lightcurves, the HR, and the HID. 

The source displayed two separated periods of flaring activity. During the first one, a bright episode is accompanied  
by lower intensity peaks, lasting roughly 5~ks in total. The second flare appears isolated and lasted less than 1~ks. We extracted 
first the average source spectra for all EPIC cameras using the entire exposure time available. We verified with the 
{\sc epaplot} tool that the pn was not affected by pile-up and thus we used the flux measured by this instrument in order to 
correct the pile-up on the two MOS. Due to the high extinction measured in the direction of the source (see below), the 
reduction of the two RGSs data did not result in usable scientific products. 
The average spectrum of the source cannot be satisfactorily fit by using a single 
power-law model ($\chi^2_{\rm red}$=1.39/318). As in the observation ID.~0306170401 (see Sect.~\ref{sec:ax401}), 
we improved the results by using a model comprising a cut-off power-law (the cut-off energy was fixed to 31.7~keV) 
and a partial absorber ({\sc pcfabs} in {\sc xspec}).  
Using the same notations as in Sect.~\ref{sec:ax401}, we measured $N_{\rm H}^{\rm los}$=(2.7$^{+1.0}_{-1.3}$)$\times$10$^{22}$~cm$^{-2}$, 
$N_{\rm H}$=(6.6$^{+1.2}_{-0.8}$)$\times$10$^{22}$~cm$^{-2}$, $f$=0.83$^{+0.09}_{-0.13}$, $\Gamma$=1.1$\pm$0.1, 
$F$=1.1$\times$10$^{-11}$~erg~cm$^{-2}$~s$^{-1}$, $C_{\rm MOS1}$=1.12$\pm$0.04, 
$C_{\rm MOS2}$=1.05$\pm$0.03, and $\chi^2_{\rm red}$/d.o.f=1.11/316. 

Following the HR variations, we also extracted the 8 different spectra indicated in Fig.~\ref{fig:ax801} (second plot from the top). 
The HID does not show any significant correlation, but similarly 
to the case of the observation ID.~0306170401, it is possible to identify intervals of count-rates corresponding 
to larger/smaller HR values than average. Due to the lower statistics, all the HR-resolved spectra could be well fit by using a 
simple absorbed power-law model and we report all the results in Table~\ref{tab:j1845_fit2}. Some of the results are also plot in 
Fig.~\ref{fig:ax801} (second plot from the top).
The largest variation in the column density is found between the spectra C and G, which correspond to the rise to the 
first large flare and the time interval before the second. The contour plots in Fig.~\ref{fig:ax801} (bottom) shows that the variation 
is highly significant and cannot be associated with the degeneracy between the spectral slope and the absorption column density.

\subsubsection{Observation ID.~0728371001}

\begin{figure}
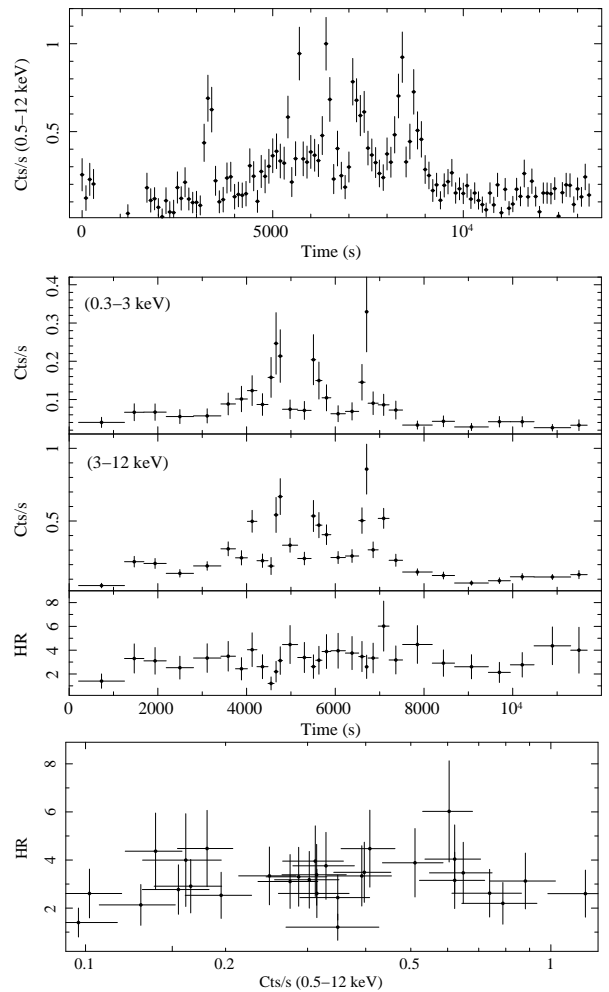

\centering
\includegraphics[scale=0.35,angle=-90]{figures/AXJ1845/1001/lcurve.ps}
\includegraphics[scale=0.35,angle=-90]{figures/AXJ1845/1001/ax1001_HR.ps}
\includegraphics[scale=0.35,angle=-90]{figures/AXJ1845/1001/ax1001_HID.ps}
\caption{Same as Fig.~\ref{fig:ax801}, but in case of the IGR\,J18450-0435 observation ID.~0728371001. 
An adaptive rebinning achieving S/N$>$3 in each soft time bin has been used to compute the HR and HID 
(note that the first $\sim$1.5~ks of observation visible in the top lightcurve have been excluded for the computation 
of the HR due to the low S/N and the fragmentation related to the filtering of the high background time intervals).}      
\label{fig:ax1001} 
\end{figure}

In this observation the same instrument set-up as in ID.~0728370801 was used, 
and the presence of solar flares led to a reduced effective exposure time of 
8.5~ks for the EPIC-pn and 6.9~ks for the two MOS. 
The observation was not reported elsewhere yet and thus we show the EPIC-pn lightcurve in the full available energy band 
in Fig.~\ref{fig:ax1001}, together with the energy-resolved lightcurves and the corresponding HR/HID. 

The source displayed low level X-ray flaring activity,  
but no significant HR variation could be revealed (see Fig.~\ref{fig:ax1001}). This is also confirmed by the HID 
in which all points are consistent with each other (to within the uncertainties). Consequently, we  
report for completeness on the average X-ray spectrum extracted from the three EPIC cameras, but we did not perform any HR-resolved 
spectral analysis. The best fit to the average spectra could 
be achieved by using a simple absorbed power-law model. We measured an absorption column density 
of (3.9$\pm$0.6)$\times$10$^{22}$~cm$^{-2}$ and a power-law photon index of 1.4$\pm$0.2. The normalization constant of the MOS1 
(MOS2) with respect to the pn was of 1.1$\pm$0.1 (0.96$\pm$0.09). The average 1-10~keV X-ray flux measured from the spectral fit is  
2.5$\times$10$^{-12}$~erg~cm$^{-2}$~s$^{-1}$ (not corrected for absorption).

\subsection{IGR\,J17544-2619}
\label{sec:J17544}

\subsubsection{Observation ID.~0148090501}

\begin{figure}
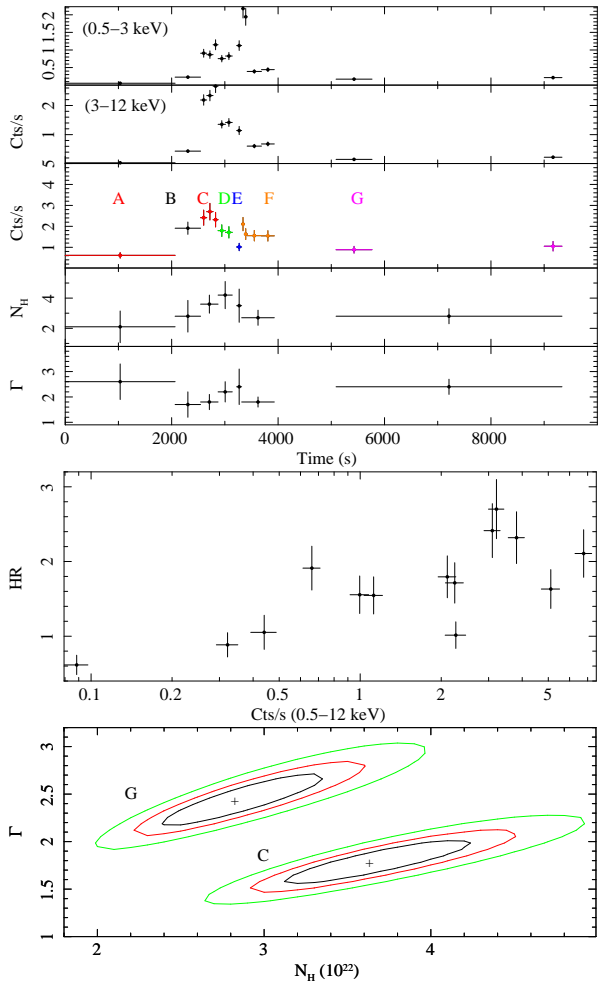

\centering
\includegraphics[scale=0.35,angle=-90]{figures/J17544/0148090501/lcurve.ps}
\includegraphics[scale=0.35,angle=-90]{figures/J17544/0148090501/HR.ps}
\includegraphics[scale=0.35,angle=-90]{figures/J17544/0148090501/contours.ps}
\caption{Same as Fig.~\ref{fig:j1845}, but in the case of the IGR\,J17544-2619 observation ID.~0148090501. 
An adaptive rebinning achieving S/N$>$10 in each soft time bin has been used to compute the HR and HID. 
The most relevant time intervals for the spectral variations are in this case ``C'' and ``G''.}     
\label{fig:J17544_lcurve} 
\end{figure}

This observation was originally reported by \citet{riestra04}, who retained for the scientific analysis the entire exposure time available.  
During our re-analysis of the data, we found that about $\sim$40\% of the total 
exposure time was significantly affected by a heavily flaring background and thus, based on the presently available suggestions in the \xmm\ data analysis threads, 
we discarded the corresponding time intervals for further analysis. One structured flare was recorded from the source during a time interval where 
the background was low and stable. The energy-resolved lightcurves of IGR\,J17544-2619 are shown in Fig.~\ref{fig:J17544_lcurve}, together 
with the corresponding HR and HID. 

Due to the relatively low S/N below 3~keV, no RGS spectra could be extracted from this observation. The summed EPIC-pn, MOS1, and MOS2 
average spectra could be well fit ($\chi^2_{\rm red}$=1.0/150) by using a simple absorbed power-law 
model. We measured $N_{\rm H}$=(2.9$\pm$0.2)$\times$10$^{22}$~cm$^{-2}$ and $\Gamma$=1.95$\pm$0.11. The absorbed 1-10 keV flux is 
5.5$\times$10$^{-12}$~erg~cm$^{-2}$~s$^{-1}$ (effective exposure time 6.6~ks).   

To investigate the source spectral variability during the flare, we extracted 6 spectra corresponding 
to the HR changes in Fig.~\ref{fig:J17544_lcurve}. 
They could all be well fit with a simple absorbed power-law model and we report the results in Table~\ref{tab:J17544_fit}. 
The largest change in the source spectral parameters was observed between the rise to the flare (interval C) and the 
post flare spectrum (interval G). The latter has a similar HR compared to the pre-flare interval A but also a higher statistics. 
This result is also confirmed by the HID, which shows that the largest HR is not achieved at the highest intensity. 
The contour plots of the spectra C and G suggested that 
the power-law photon index was harder during the rise to the flare, accompanied by a likely increase of the absorption 
column density. We note that the flux of the source before the flare was a factor 
of $\sim$5 lower that that achieved at the end of the flare. The relatively poor statistics of the A spectrum do not allow for a more 
refined assessment of possible spectral changes before and after the flare. 

\subsubsection{Observation ID.~0154750601}
 
The EPIC-pn and the MOS1 were operated in timing mode, while the MOS2 was in full frame. 
The total exposure available was of $\sim$8~ks for the two MOS cameras, and 2.5~ks for the EPIC-pn (the EPIC-pn 
started observing the target about 5.5~ks after the two MOS cameras). The data were first reported by \citet{riestra04}, 
who retained for the scientific analysis the entire exposure time available.  
During our re-analysis of the data, we found that the observation was heavily affected by the flaring background. 
The background filtering technique suggested in the presently available \xmm\ data analysis threads,
would result in a nearly complete rejection of all data, although the source displayed 
relatively bright flares during this observation. 
\begin{figure}
\centering
\includegraphics[scale=0.38,angle=-90]{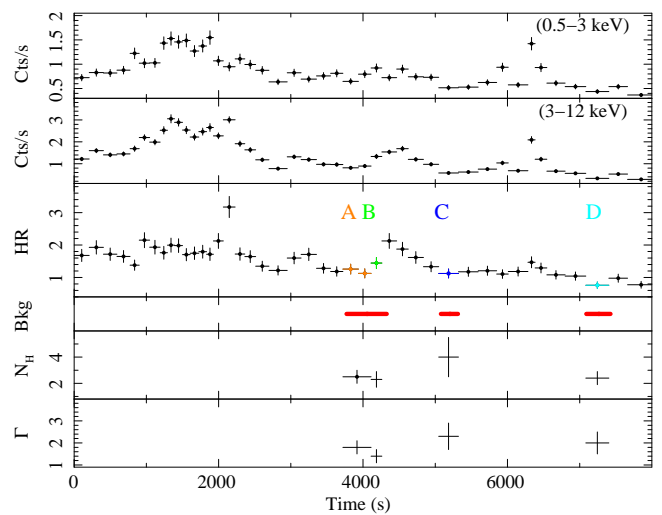}
\caption{Same as Fig.~\ref{fig:j1845}, but in the case of the IGR\,J17544-2619 observation ID.~0154750601. 
An adaptive rebinning achieving S/N$>$11 in each soft time bin has been used to compute the HR before filtering 
the lightcurves for the high flaring background. Only the time intervals marked with a red solid line in the fourth panel 
from the top could be used for the spectral analysis due to the high flaring background that affected the rest of 
the observation.}     
\label{fig:J17544b_lcurve} 
\end{figure} 
As the flaring background in the EPIC cameras is due to Solar protons arriving onto the X-ray detectors with hardly predicable 
spectral properties, it is not possible to correct the distortion in energy of the source spectra extracted during the affected 
time intervals\footnote{http://xmm.esac.esa.int/external/xmm\_user\_support/ documentation/uhb/epicextbkgd.html}. 
The only data that can be safely used for the scientific analysis are thus  
those collected during the time intervals marked with a red solid line in Fig.~\ref{fig:J17544b_lcurve} (bottom panel). 
We carried out the spectral analysis only during these time intervals.  
Due to the fragmented exposure, we did not compute the HID for this observation. 
The average MOS1 and MOS2 spectra (effective exposure time 1.8~ks) fit simultaneously 
with an absorbed power-law model provided the most accurate description of the source average X-ray emission. 
We measured $N_{\rm H}$=(2.2$\pm$0.2)$\times$10$^{22}$~cm$^{-2}$ and $\Gamma$=1.7$\pm$0.1  
($\chi_{\rm red}^{2}$/d.o.f.=1.25/138). The absorbed 1-10~keV flux is  
2.4$\times$10$^{-11}$~erg~cm$^{-2}$~s$^{-1}$. The fit to the average EPIC-pn spectrum (exposure time 1.2~ks) 
gave a consistent absorption column density ($N_{\rm H}$=(2.6$\pm$0.4)$\times$10$^{22}$~cm$^{-2}$) and a slightly steeper 
power-law photon index of $\Gamma$=2.1$\pm$0.2 ($\chi_{\rm red}^{2}$/d.o.f.=1.28/64). 
This is in agreement with the lower HR visible in 
Fig.~\ref{fig:J17544b_lcurve} toward the end of the observation (where the EPIC-pn data are available).   

We then extracted 4 source spectra during the usable time intervals, following the HR variations    
as shown in Fig.~\ref{fig:J17544b_lcurve} (the pn data were only available during the time interval D). 
We report the results of the fits to the spectra A-D in Table~\ref{tab:J17544_fit} (some of them are also 
plot on Fig.~\ref{fig:J17544b_lcurve}). No significant variations of the spectral parameters could be 
revealed due to the large associated uncertainties.

\subsubsection{Observation ID.~0679810401}

\begin{figure}
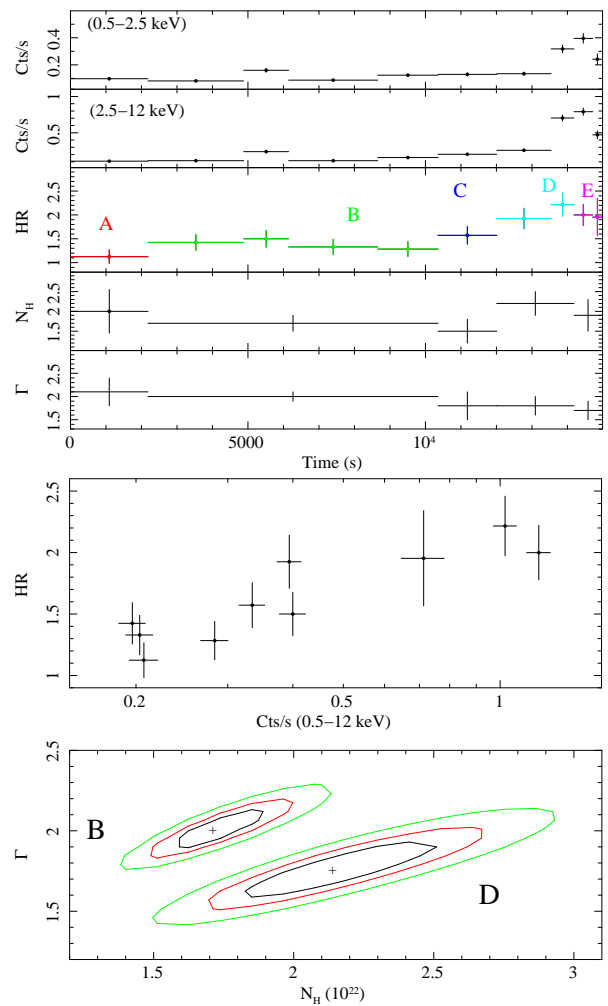

\centering
\includegraphics[scale=0.35,angle=-90]{figures/J17544/0679810401/lcurve.ps}
\includegraphics[scale=0.35,angle=-90]{figures/J17544/0679810401/hid.ps}
\includegraphics[scale=0.35,angle=-90]{figures/J17544/0679810401/contours.ps}
\caption{Same as Fig.~\ref{fig:j1845}, but in the case of the IGR\,J17544-2619 observation ID.~0679810401. 
An adaptive rebinning achieving S/N$>$11 in each soft time bin has been used to compute the HR and HID.}     
\label{fig:J17544c_lcurve} 
\end{figure}

The two MOS were operated in full frame, while the EPIC-pn was set in small window mode. 
No flaring background intervals were recorded. These data were firstly presented by \citet{drave14}. 
The energy-resolved EPIC-pn lightcurves of the source are shown in Fig.~\ref{fig:J17544c_lcurve}, together with the corresponding 
HR and HID. The source displayed only a marginal flaring activity toward the end of the observation. 
We summed-up the three EPIC spectra and found that the average spectrum 
of the source could not be well described by a single absorbed power-law model ($\chi^2_{\rm red}$=1.47/111). We adopted a partial covering 
model for the spectral fit and we measured $N_{\rm H}^{\rm los}$=(1.3$\pm$0.2)$\times$10$^{22}$~cm$^{-2}$, 
$N_{\rm H}$=(3.4$_{-0.8}^{1.0}$)$\times$10$^{22}$~cm$^{-2}$, $f$=0.73$\pm$0.08, and $\Gamma$=2.3$\pm$0.2.  
The average (absorbed) 1-10~keV X-ray flux is 2.1$\times$10$^{-12}$~erg~cm$^{-2}$~s$^{-1}$. 

We extracted five spectra following the HR variations in Fig.~\ref{fig:J17544c_lcurve}  
and report the results of the fits to these spectra in Table~\ref{tab:J17544_fit}. 
The relevant contour plots obtained from the spectral fits are shown in Fig.~\ref{fig:J17544c_lcurve} (bottom). 
We found an indication for a slightly increase in the absorption column density and hardening of the power law during the rise to the faint flare 
(spectra B and D). This is also confirmed by the HID (Fig.~\ref{fig:J17544c_lcurve} middle) which shows that the source emission 
becomes harder at count-rates $\gtrsim$0.5~cts/s. 

\subsection{SAX\,J1818.6-1703}

\subsubsection{Observation ID.~0693900101}
\label{sec:SAXJ1818} 

This is the only available observation where the source was pointed and detected by \xmm\ \citep[see also][]{bozzo08c,bozzo12}.  
A first analysis of these data was reported by \citet{boon16}. During our re-analysis of the data, we noticed that the observation  
was affected by a high background during the last 8.3~ks and thus we discarded these data for the 
spectral analysis. The EPIC-pn was operated in small window, while the two MOS were operated in full frame.  
Given the relatively high count-rate of the source after the initial $\sim$15~ks, the MOS suffered significant pile-up and thus we 
discarded these data (the quality of the pn data was high enough to carry out all necessary investigations). 
The average EPIC-pn spectrum (effective exposure time 15.6 ks) 
could be fit with a simple absorbed power-law model. We measured $N_{\rm H}$=(27.0$\pm$1.1)$\times$10$^{22}$~cm$^{-2}$, 
$\Gamma$=0.48$\pm$0.06, and a flux of 3.5$\times$10$^{-11}$~erg~cm$^{-2}$~s$^{-1}$ in the 1-10~keV energy range. 
As the source is heavily absorbed, we did not attempt an extraction of the RGS data. 

\begin{figure}
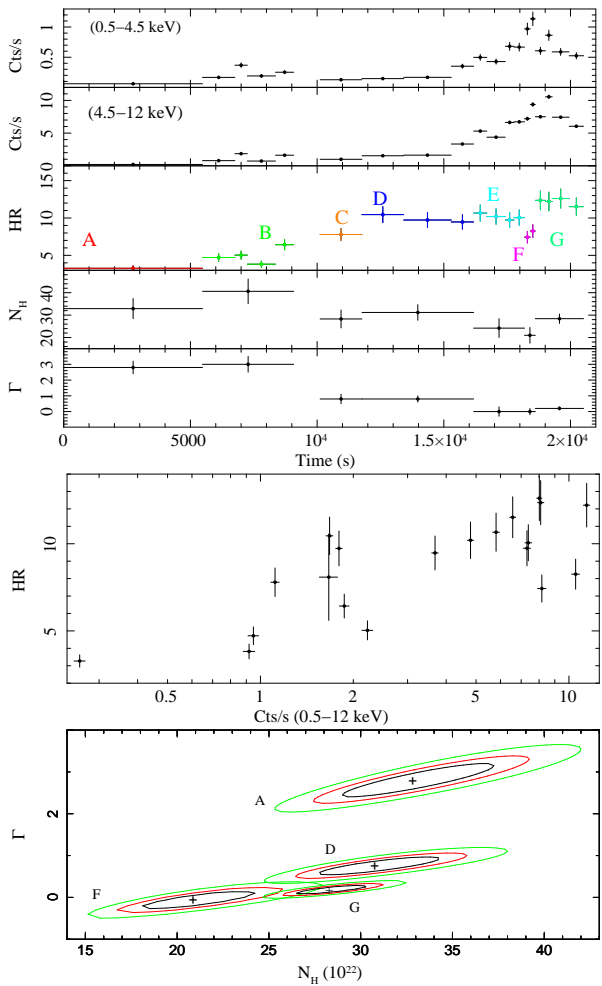

\centering
\includegraphics[scale=0.35,angle=-90]{figures/J1818/J1818_results.ps}
\includegraphics[scale=0.35,angle=-90]{figures/J1818/J1818_hid.ps}
\includegraphics[scale=0.35,angle=-90]{figures/J1818/J1818_contours.ps}
\caption{Same as Fig.~\ref{fig:j1845}, but in the case of the SAX\,J1818.6-1703 observation ID.~0693900101. 
An adaptive rebinning achieving S/N$>$12 in each soft time bin has been used to compute the HR and HID.}     
\label{fig:J1818_results} 
\end{figure}

We extracted the adaptively rebinned EPIC-pn lightcurves of SAX\,J1818.6-1703 and computed the source HR/HID shown  
in Fig.~\ref{fig:J1818_results}. We extracted a number of HR resolved spectra following the HR variations. 
Simultaneous MOS1 and MOS2 spectra were only extracted for the time intervals 
A, B, C, and D to avoid pile-up issues. 
We fit all EPIC spectra with an absorbed power-law model, which provided in all cases a good description of the data. The results are 
summarized in Table~\ref{tab:J1818_fit} and partly plot in Fig.~\ref{fig:J1818_results}. 
We recorded a progressive hardening of the X-ray emission during the rise to the flare (time intervals A, B, C, D, and E) 
and a significant decrease of the absorption column density at the peak of the event (time interval F). The data also show some evidence for a re-increase of the 
$N_{\rm H}$ following the peak of the flare. 
Remarkable changes were also observed in the power-law spectral index, as suggested by the HID (where the source emission becomes harder at higher count-rates). 
In order to further confirm these findings, we also show in 
Fig.~\ref{fig:J1818_results} the relevant contour plots of the fit parameters derived from the 
analysis of the HR-resolved spectra.

\subsection{IGR\,J17354-3255}
\label{sec:J17354}

\subsubsection{Observation ID.~0693900201}

During this observation, not reported yet in other works, the EPIC-pn was operated in small window mode but unfortunately was pointed to 
a different close-by source and thus the collected data are not usable.  
The two MOS cameras were instead operated in full frame and thus included IGR\,J17354-3255 in their FoV. In the following,  
we only report on the MOS data analysis and we merge in all cases the MOS1 and MOS2 spectra 
by using the SAS tool {\sc epicspeccombine} to improve the statistics. 
The observation was affected by a moderately high background for the first $\sim$3~ks 
and by very high background during the last $\sim$6~ks. We retained the first 3~ks of the observation as bright flares from 
the source were also detected, and discarded the last part of the observation for the scientific analysis. 
The average spectrum of the source is shown in Fig.~\ref{fig:J17354_spe}, as an example (this looks relatively similar to the 
spectra extracted from the following two observations ID.~0701230101 and ID.~0701230701). An acceptable fit to this spectrum  
($\chi_{\rm red}^{2}$/d.o.f.=0.87/146) could be obtained by using an absorbed power-law and a neutral iron emission line. We measured 
$N_{\rm H}$=(10.2$\pm$0.2)$\times$10$^{22}$~cm$^{-2}$, $\Gamma$=1.34$\pm$0.04, $E_{\rm Fe}$=6.36$\pm$0.02~keV, 
$EQW_{\rm Fe}$=0.06~keV, and a flux of 3.2$\times$10$^{-11}$~erg~cm$^{-2}$~s$^{-1}$ in the 1-10~keV 
energy range (effective exposure time 27.6~ks).  

\begin{figure}
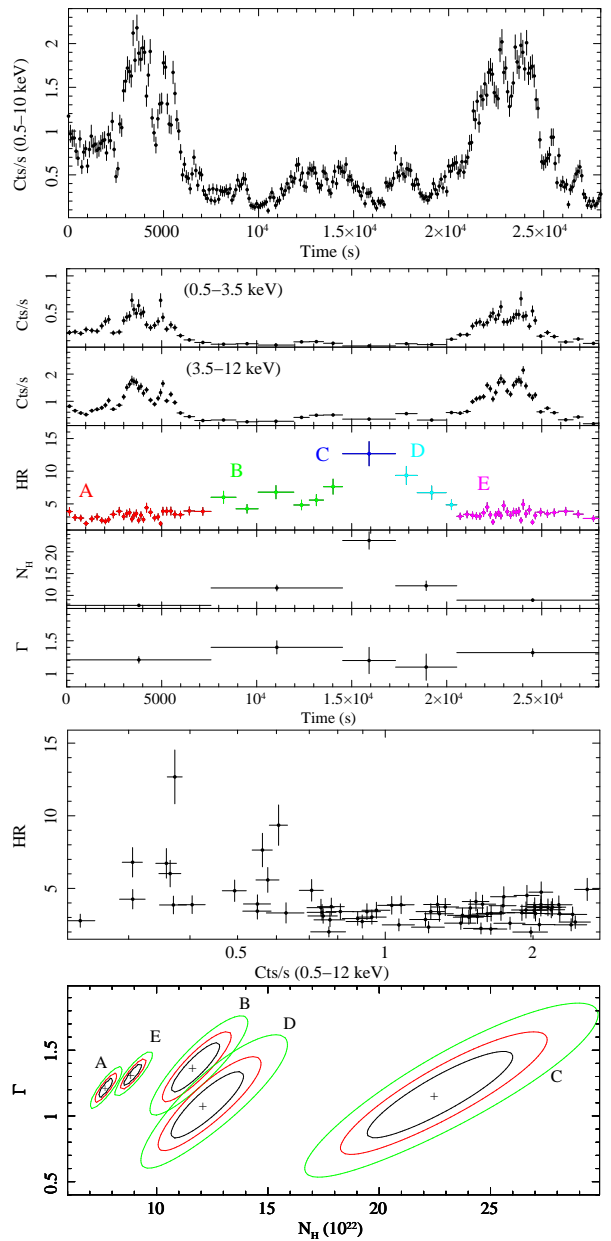

\centering
\includegraphics[scale=0.35,angle=-90]{figures/J17354/lcurve.ps}
\includegraphics[scale=0.35,angle=-90]{figures/J17354/J17354_lc.ps}
\includegraphics[scale=0.35,angle=-90]{figures/J17354/J17354_hid.ps}
\includegraphics[scale=0.35,angle=-90]{figures/J17354/J17354_contours.ps}
\caption{Same as Fig.~\ref{fig:ax801}, but in the case of the IGR\,J17354-3255 observation ID.~0693900201. 
An adaptive rebinning achieving S/N$>$7 in each soft time bin has been used to compute the HR and HID.}     
\label{fig:J17354_lc} 
\end{figure}

\begin{figure}
\centering
\includegraphics[scale=0.35,angle=-90]{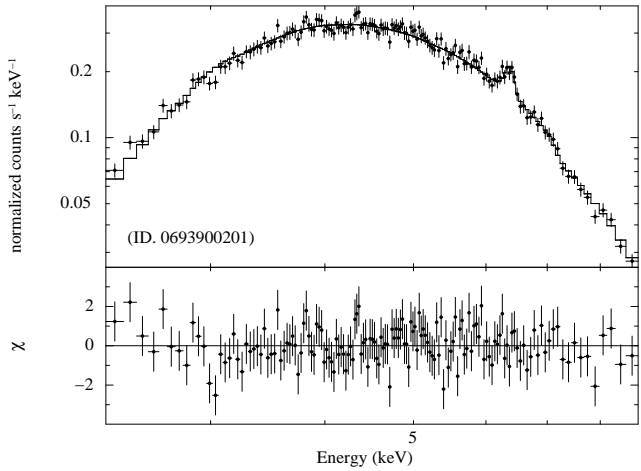}
\caption{Merged MOS spectra of IGR\,J17354-3255 in the observation ID.~0693900201.  
The best fit is obtained with an absorbed power-law plus a neutral 
iron emission line (see text for details). The residuals from the fit are shown in the bottom panel.}      
\label{fig:J17354_spe} 
\end{figure}

The energy-resolved lightcurves of the source are shown in Fig.~\ref{fig:J17354_lc}, together with the corresponding HR and HID. 
To investigate the variability suggested by the changes in the HR across the observation 
we extracted the HR-resolved spectra A-E indicated in the same figure. All spectra could be well fit with an 
absorbed power-law and we report all results in Table~\ref{tab:J17354_fit}. Some of the results obtained from the 
HR-resolved spectral analysis are also shown in Fig.~\ref{fig:J17354_lc}, together with the contour plots of the 
relevant spectral parameters determined from the fits. 
The most noticeable changes in the source spectral parameters are recorded during the lower X-ray intensity period 
(interval C) separating the flares at the beginning and at the end of the observation. This is also clearly visible from the HID, which shows  
that the highest HR values are achieved at the count-rate specifically recorded toward the middle of the low X-ray intensity period. 

\subsubsection{Observation ID.~0701230101}

During this observation, not yet reported in other literature works, 
the three EPIC cameras were operated in large window mode and several time intervals affected by 
a high background were recorded. The merged EPIC-pn and MOS spectrum extracted by using all available exposure time (17.1~ks) 
could be well fit ($\chi_{\rm red}^{2}$/d.o.f.=1.11/667) with 
an absorbed power-law model, plus a neutral iron emission line. 
We measured $N_{\rm H}$=(7.4$\pm$0.2)$\times$10$^{22}$~cm$^{-2}$, $\Gamma$=1.40$\pm$0.04, $E_{\rm Fe}$=6.43$\pm$0.02~keV, 
$EQW_{\rm Fe}$=0.04~keV, and a flux of 1.1$\times$10$^{-11}$~erg~cm$^{-2}$~s$^{-1}$ in the 1-10~keV 
energy range. 

\begin{figure}
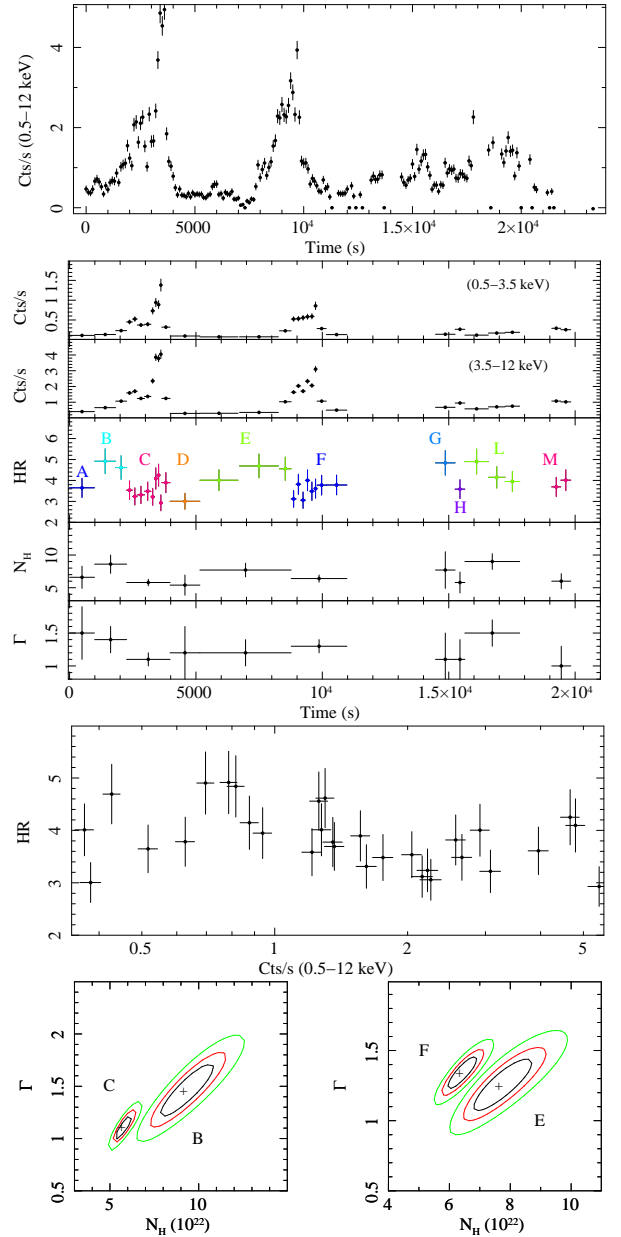

\centering
\includegraphics[scale=0.35,angle=-90]{figures/J17354/lcurve2.ps}
\includegraphics[scale=0.35,angle=-90]{figures/J17354/J17354_lc2.ps}
\includegraphics[scale=0.35,angle=-90]{figures/J17354/J17354_hid2.ps}
\includegraphics[scale=0.35,angle=-90]{figures/J17354/J17354_contours2a.ps}
\includegraphics[scale=0.35,angle=-90]{figures/J17354/J17354_contours2b.ps}
\caption{Same as Fig.~\ref{fig:ax801}, but in the case of the IGR\,J17354-3255 observation ID.~0701230101. 
An adaptive rebinning achieving S/N$>$10 in each soft time bin has been used to compute the HR and HID.}     
\label{fig:J17354_lc2} 
\end{figure}

After having excluded the affected time intervals, we obtained the energy-resolved lightcurves displayed in 
Fig.~\ref{fig:J17354_lc2}, together with the corresponding HR and HID. A number 
of moderately bright flares were recorded from the source. 
We extracted the source A-M spectra during the HR variations as indicated in Fig.~\ref{fig:J17354_lc2}
and fit them with an absorbed power-law model. The results 
are summarized in Table~\ref{tab:J17354_fit} and some plot also in Fig.~\ref{fig:J17354_lc2}. 
We recorded a significant increase in the absorption column density 
preceding the two flares (time intervals B and E) and a drop close to their peaks (time intervals C and F).  

\subsubsection{Observation ID.~0701230701}

During this observation, not reported yet in other works, the EPIC cameras were operated in large window mode and no high background 
time intervals were recorded. We could thus retain the entire exposure time available for the scientific analysis. 
The average spectrum obtained by merging all EPIC cameras (effective exposure time 19.1~ks)  
could be well fit ($\chi_{\rm red}^{2}$/d.o.f.=1.11/667) with an absorbed power-law model plus an iron emission line. 
We measured in this case $N_{\rm H}$=(6.0$\pm$0.1)$\times$10$^{22}$~cm$^{-2}$, 
$\Gamma$=1.28$\pm$0.03, $E_{\rm Fe}$=6.41$\pm$0.02~keV, $EQW_{\rm Fe}$=0.04~keV, and a 
flux of 1.0$\times$10$^{-11}$~erg~cm$^{-2}$~s$^{-1}$ in the 1-10~keV energy range. 

\begin{figure}
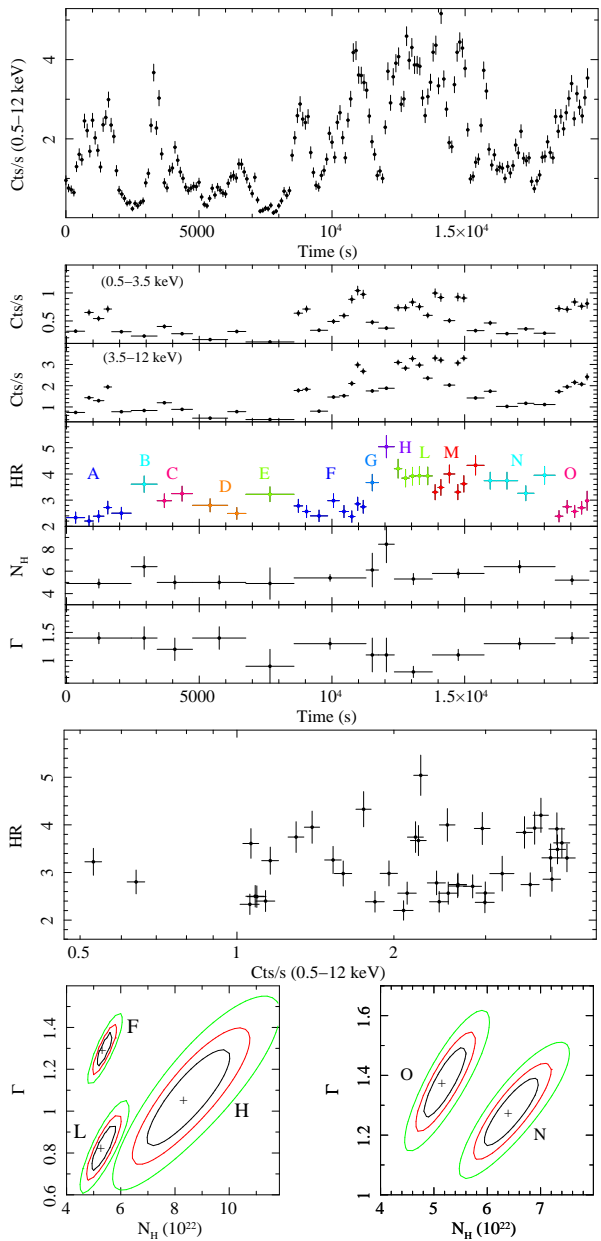

\centering
\includegraphics[scale=0.35,angle=-90]{figures/J17354/lcurve3.ps}
\includegraphics[scale=0.35,angle=-90]{figures/J17354/J17354_lc3.ps}
\includegraphics[scale=0.35,angle=-90]{figures/J17354/J17354_hid3.ps}
\includegraphics[scale=0.35,angle=-90]{figures/J17354/J17354_contours3a.ps}
\includegraphics[scale=0.35,angle=-90]{figures/J17354/J17354_contours3b.ps}
\caption{Same as Fig.~\ref{fig:ax801}, but in the case of the IGR\,J17354-3255 observation ID.~0701230701.  
An adaptive rebinning achieving S/N$>$12 in each soft time bin has been used to compute the HR and HID.}     
\label{fig:J17354_lc3} 
\end{figure}

The source displayed a remarkable variability, with moderately bright flares and short intervals of fainter X-ray emission. The  
energy-resolved lightcurves of the source are shown in Fig.~\ref{fig:J17354_lc3}, together with the corresponding HR and HID. 
At odds with the case of the observation ID.~0693900201, the most prominent HR variations 
were recorded during the flares. The HID shows that there is not a clear trend of the HR as a function of the X-ray intensity, 
but rather that the HR varies sporadically during some specific rise/decay phases of the flares.  
The results of the HR-resolved spectral analysis are presented in Table~\ref{tab:J17354_fit} and Fig.~\ref{fig:J17354_lc3}. 
We recorded significant increases in the absorption column density slightly before the rises of the different flares (time intervals H and N), 
accompanied by moderate power-law slope variations and $N_{\rm H}$ drops at the peak of those events (time intervals F, L, and O).  

\subsection{IGR\,J16328-4732}
\label{sec:J16328}

\subsubsection{Observation ID.~0679810201}

\begin{figure}
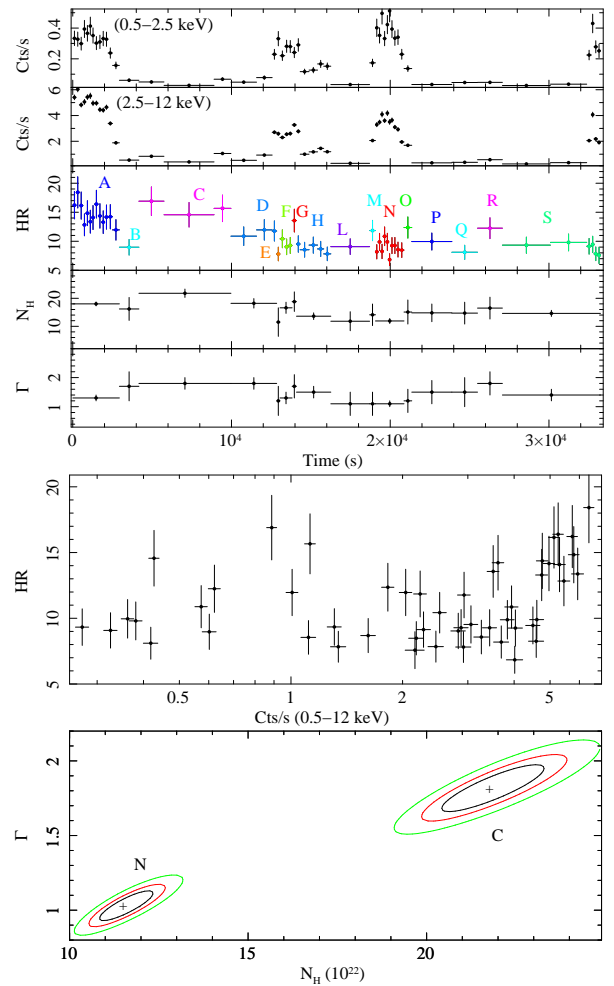

\centering
\includegraphics[scale=0.35,angle=-90]{figures/J16328/16328_201_lcurve.ps}
\includegraphics[scale=0.35,angle=-90]{figures/J16328/16328_hid.ps}
\includegraphics[scale=0.35,angle=-90]{figures/J16328/contours_c_n.ps}
\caption{Same as Fig.~\ref{fig:j1845}, but in the case of the IGR\,J16328-4732 observation ID.~0679810201. 
An adaptive rebinning achieving S/N$>$7 in each soft time bin has been used to compute the HR and HID.}     
\label{fig:J16328_lc} 
\end{figure}

These data were first analyzed by \citet{fiocchi16}, who found that no particularly high-flaring 
background time interval affected the data, and thus the entire exposure time could be used for the scientific analysis. 
The source displayed several relatively bright flares, 
achieving up to $\sim$6~cts~s$^{-1}$ in the EPIC-pn (0.5-12~keV). We verified that no pile-up was present 
in the data. The EPIC-pn energy-resolved lightcurves of the source are shown in 
Fig.~\ref{fig:J16328_lc}, together with the corresponding HR and HID. A total of 16 EPIC spectra 
were extracted following the HR variations evidenced in the figure (the MOS1 and MOS2 spectra were corrected for  
pile-up whenever required). The pn, MOS1, and MOS2 spectra were fit together with a simple absorbed power-law 
model. All results are summarized in Table~\ref{tab:J16328_fit} and some are also plot in Fig.~\ref{fig:J16328_lc}.
The largest spectral variations, especially in terms of absorption column densities, are recorded between the time interval C 
preceding the rise to the second flare and E or N (corresponding to the peaks of the two flares). The time interval L is also characterized 
by a relatively low absorption and corresponds to the separation between the second and the third flare. Note that the count-rate corresponding to these 
specific intervals are those for which the HID also shows a large increase/decrease of the HR (see Fig.~\ref{fig:J16328_lc}). 

\subsubsection{Observation ID.~0679810301}

\begin{figure}
\centering
\includegraphics[scale=0.35,angle=-90]{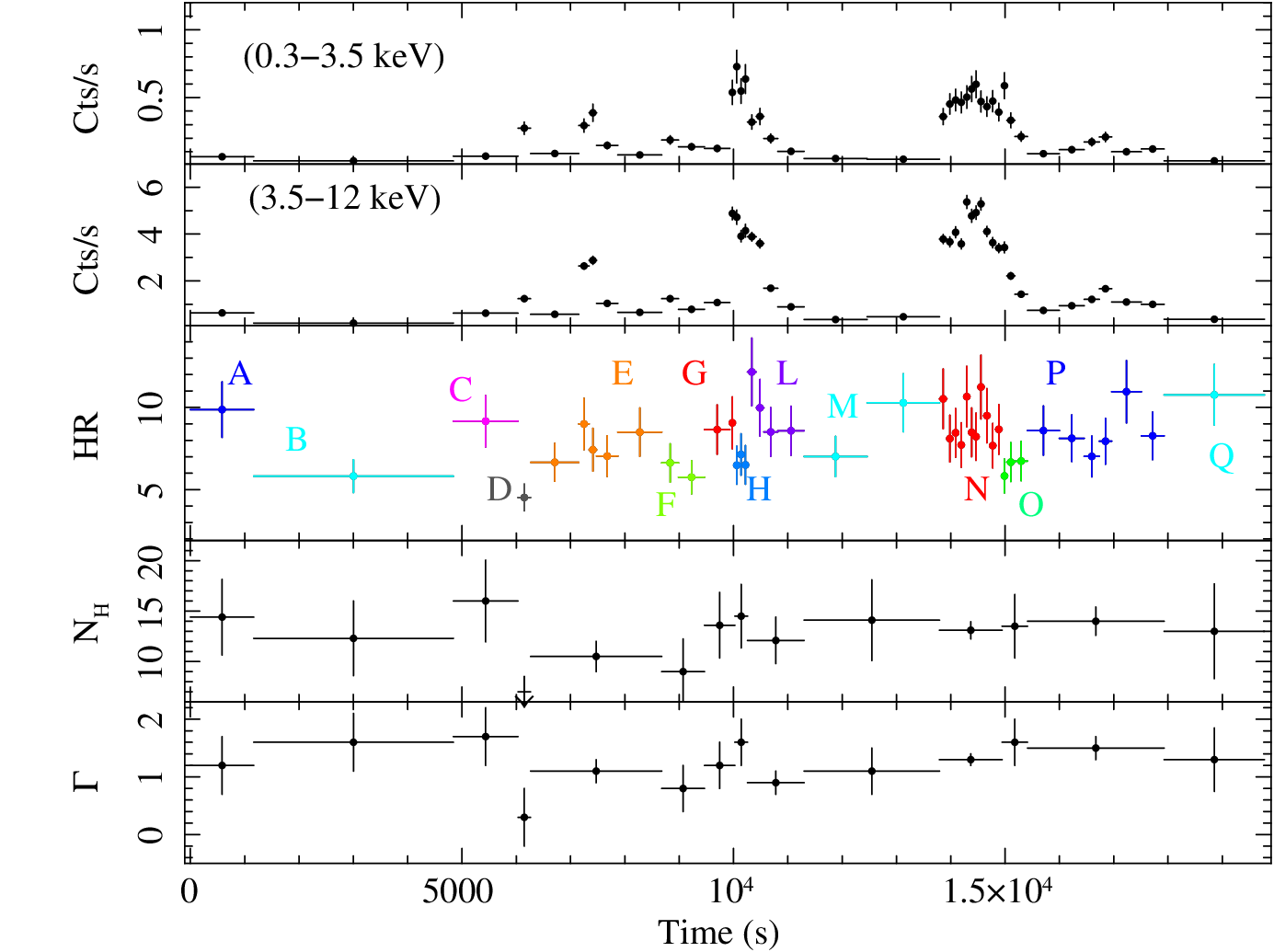}
\includegraphics[scale=0.35,angle=-90]{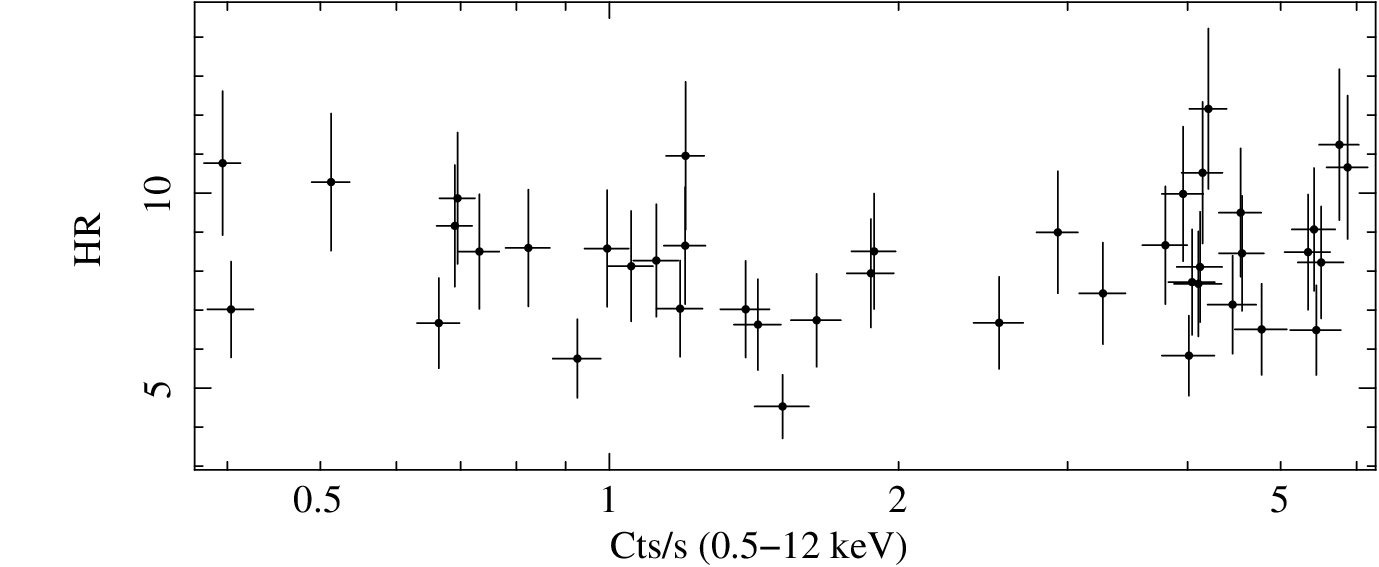}
\includegraphics[scale=0.35,angle=-90]{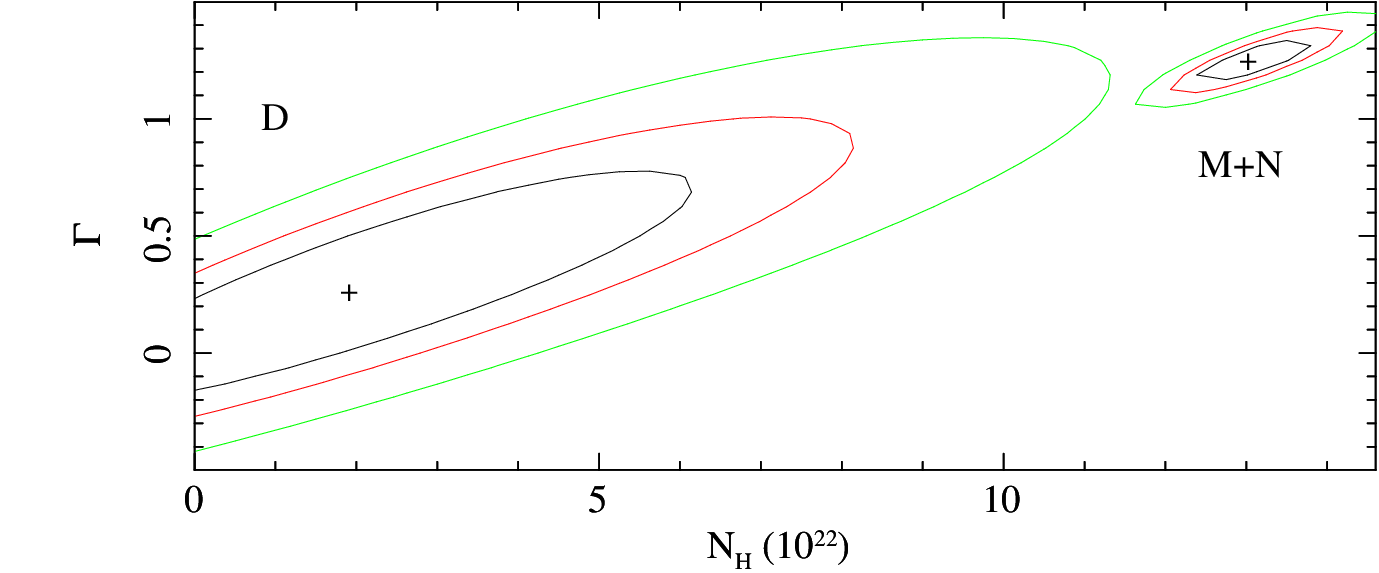}
\caption{Same as Fig.~\ref{fig:j1845}, but in the case of the IGR\,J16328-4732 observation ID.~0679810301. 
An adaptive rebinning achieving S/N$>$6 in each soft time bin has been used to compute the HR and HID.}     
\label{fig:J16328_lc2} 
\end{figure}

The analysis of these data was first discussed by \citet{fiocchi16}. We report the source EPIC-pn 
energy-resolved lightcurves in Fig.~\ref{fig:J16328_lc2}, together with the corresponding HR and HID. 
The time intervals indicated in this figure (top) were used to extract the HR-resolved spectra for the EPIC-pn, MOS1, and MOS2 cameras. 
All these spectra could be well fit with a simple absorbed power-law model and we summarize all results in Table~\ref{tab:J16328_fit} 
(some results are also plot in Fig.~\ref{fig:J16328_lc2}). 
Although the overall flaring behavior seems remarkably similar to that discussed for the observation ID.~0679810201, 
in the present case we measured more pronounced variations of the spectral slope (ranging from 0.3 to 1.7). 
The contour plots obtained from the different intervals showed that in most cases there is a large degeneracy between 
$\Gamma$ and $N_{\rm H}$, and isolating significant changes in the absorption column density proved to be more difficult 
(this is also confirmed by the relatively flat HID). 
The largest variation of the $N_{\rm H}$ was recorded between the intervals M/N and D, but we were forced to merge 
together the data of the M and N intervals in order to improve the S/N and obtain the results shown at the bottom of 
Fig.~\ref{fig:J16328_lc2}. The time interval D is particularly interesting because the corresponding 
spectra were characterized by a remarkably low absorption for which we could only set an upper limit 
of $<$7$\times$10$^{22}$~cm$^{-2}$ (at 90\% c.l.).

\section{Overview of the results}
\label{sec:overview}

In this section, we summarize the most relevant findings for each of the five sources before 
discussing any interpretation of these results in Sect.~\ref{sec:discussion}. 
\begin{itemize}

\item {\it IGR\,J18450-0435}: the source displayed relatively bright flares in all the three observations we reported on. 
The most prominent flares were recorded during the observation ID.~0306170401, being characterized by a flux increase 
$\gtrsim$10 with respect to the following quiescent period (the time interval Z). Despite this variability, 
we only measured a modest increase of the absorption column density (a factor of $\sim$2) between the flares and quiescence. 
We found evidence for an increase of the absorption during the rise to the main flares (time intervals A, B, G, and O) and for a drop of the absorption 
column density every time the source approached the peak of a flare (the most significant being during the interval S). 
The quiescent time interval Z is more difficult to characterize, but the fit with the partial covering model 
revealed a moderate but significant increase of the partial absorber column density compared to the average spectrum. 
A similar behavior was found during the observation ID.~0728370801. In this case, the increase in the absorption column 
density before the flare was recorded during the time interval C, while a (moderate) decrease was revealed from the fits to the 
intervals D and G. We found some evidence for a re-increase in the absorption column density during the decay 
of the flares, i.e. between the spectra C and D in the observation 
ID.~0306170401 and between the spectra G (D) and H (F) in the observation ID.~0728370801. 

\item {\it IGR\,J17544-2619}: the source showed moderately bright flares during the observation ID.~0148090501, achieving an overall dynamic range of 
$\gtrsim$10 in the X-ray flux. The HR increases during the rise to the flare (time interval B) and drops close to the peak of the event (time interval E). 
We found some evidence for an increase of the $N_{\rm H}$ during the rise (time interval C), but it seems that it is the combined effect of the variation in 
$N_{\rm H}$ and $\Gamma$ to drive the changes in most of the analyzed intervals. While we cannot firmly assess which spectral parameters are changing 
during the observation ID.~0154750601 due to the contamination from the flaring background, some other interesting HR variations were recorded during the observation
ID.~0679810401. We measured an increase in the absorption column density and hardening of the power law during the time 
intervals B and D that precede the peak of a faint flare of the source. 
Unfortunately, the observation ends before any data can be collected on the post-flare emission to verify a possible 
re-enhancement of the absorption column density with the decay of the X-ray flux. 

\item {\it SAX\,J1818.6-1703}: a progressive hardening has been measured  
in the only available observation during the rise to the peak of a relatively bright flare (time 
intervals A, B, C, D, and E). This is mostly due to the change in the power-law slope, with only a minor 
change in the absorption column density. However, we measured a sharp drop of the absorption column density slightly 
before the peak of the flare (time interval F). We also reported a marginal evidence for a re-enhancement of the local absorption 
following the peak of the flare (time interval G). 

\item {\it IGR\,J17354-3255}: during the observation ID.~0693900201, the source displayed little (if any) spectral 
variability during the rise and drop 
from the two brightest X-ray flares, but showed a prominent HR increase between them that corresponds to an enhancement  
of the local absorption column density by a factor of $\gtrsim$3. 
In the other two observations, the source displays an increase in the absorption column density preceding 
the source X-ray flares (time intervals B and E in the observation ID.~0701230101, and H and N in the observation 
ID.~0701230701) and decrease in the absorption column density during the brightest phases of the different flares (time intervals 
C and F in the observation ID.~0701230101, and F, L, and O in the observation ID.~0701230701). 

\item {\it IGR\,J16328-4732}: the observation ID.~0679810201 shows that there seems to be little spectral variability during the 
decay from the first flare (time intervals A and B), while a significant increase in the local absorption column density  
precedes the second X-ray flare (time interval C). A drop of the absorption column density is recorded during the peaks of the flares, 
as shown by the results of the fits to the spectra in the time intervals E and N. We also measured significant variations of the 
power-law slope, generally becoming softer during the time intervals corresponding to relatively low flux states (B, C, D, and R).  
During the observation ID.~0679810301, the variations of the power-law slope were more pronounced and this made it difficult to 
measure the changes in the absorption column density. In particular, we could not find clear indication for an increase of the $N_{\rm H}$ 
before the rise to more prominent flares and a decrease during the brightest phases of these events (as we measured in the observation ID.~0679810201). 
Quite peculiar are the cases of the time intervals C, H, and O, where a relatively large absorption column density is measured 
although the corresponding value of the HR is low due to a significant steeping of the power-law slope. 
An interesting case is also that of the time interval D, which is characterized 
by the lowest absorption column density measured during the observation and it is associated with the peak 
of a relatively faint X-ray flare. 

\end{itemize}

\section{Discussion}
\label{sec:discussion}

The goal of the present study is to use \xmm\ data to verify if X-ray flares from the SFXTs can be all associated to the presence of intervening 
dense clumps from the supergiant star winds. Accurate measurements of the absorption column 
densities during the flares are mandatory to achieve this goal. 
For the analysis of all data, we adopted a technique that we verified in previous publications 
to be particularly powerful to probe the accretion physics in wind-fed X-ray binaries. In particular, we adaptively rebin the energy resolved 
lightcurves of all sources and use the correspondingly derived HR to drive the selection of the different time intervals 
for the spectral extraction.  
\begin{figure*}
\centering
\includegraphics[scale=0.3]{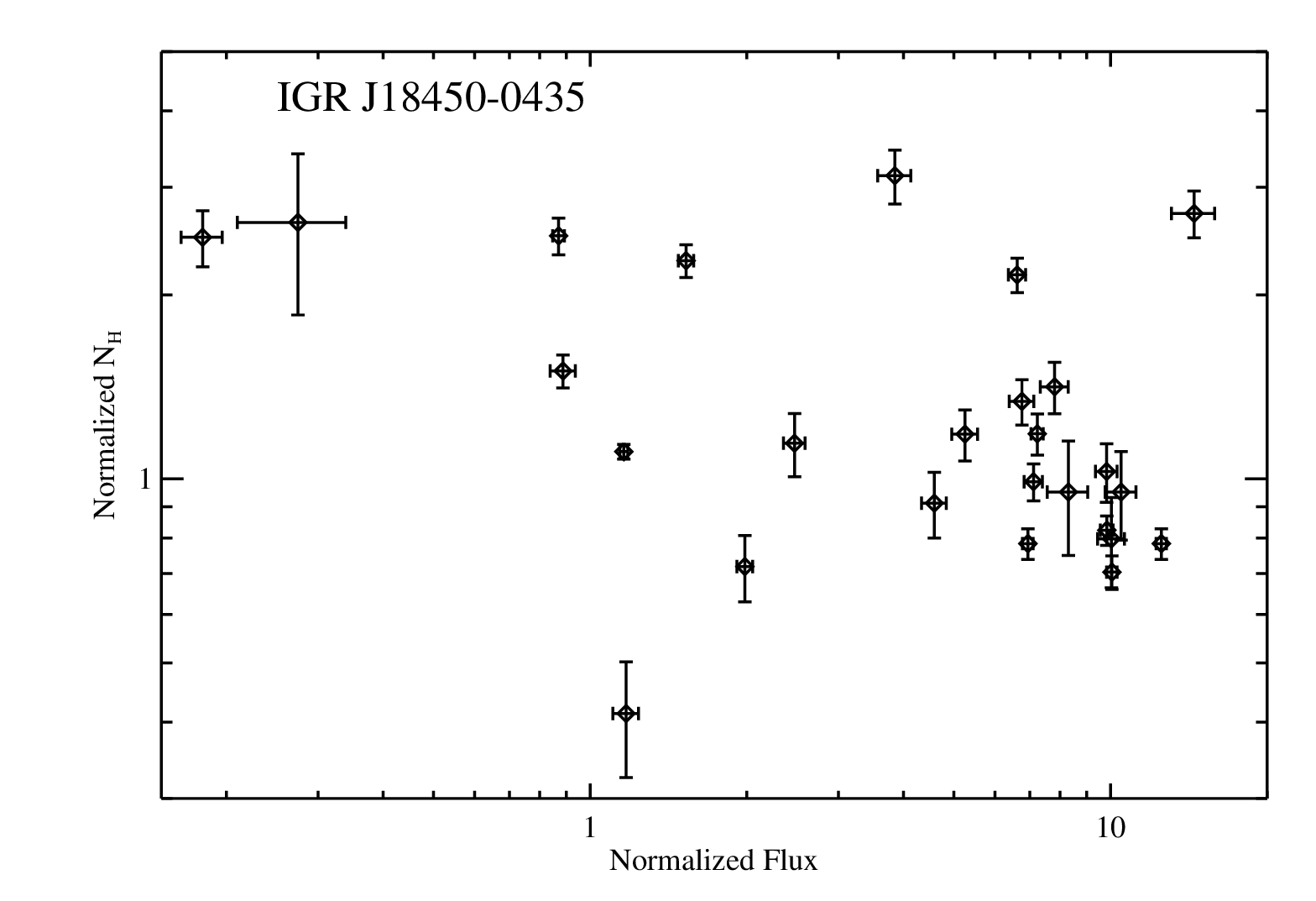}
\includegraphics[scale=0.3]{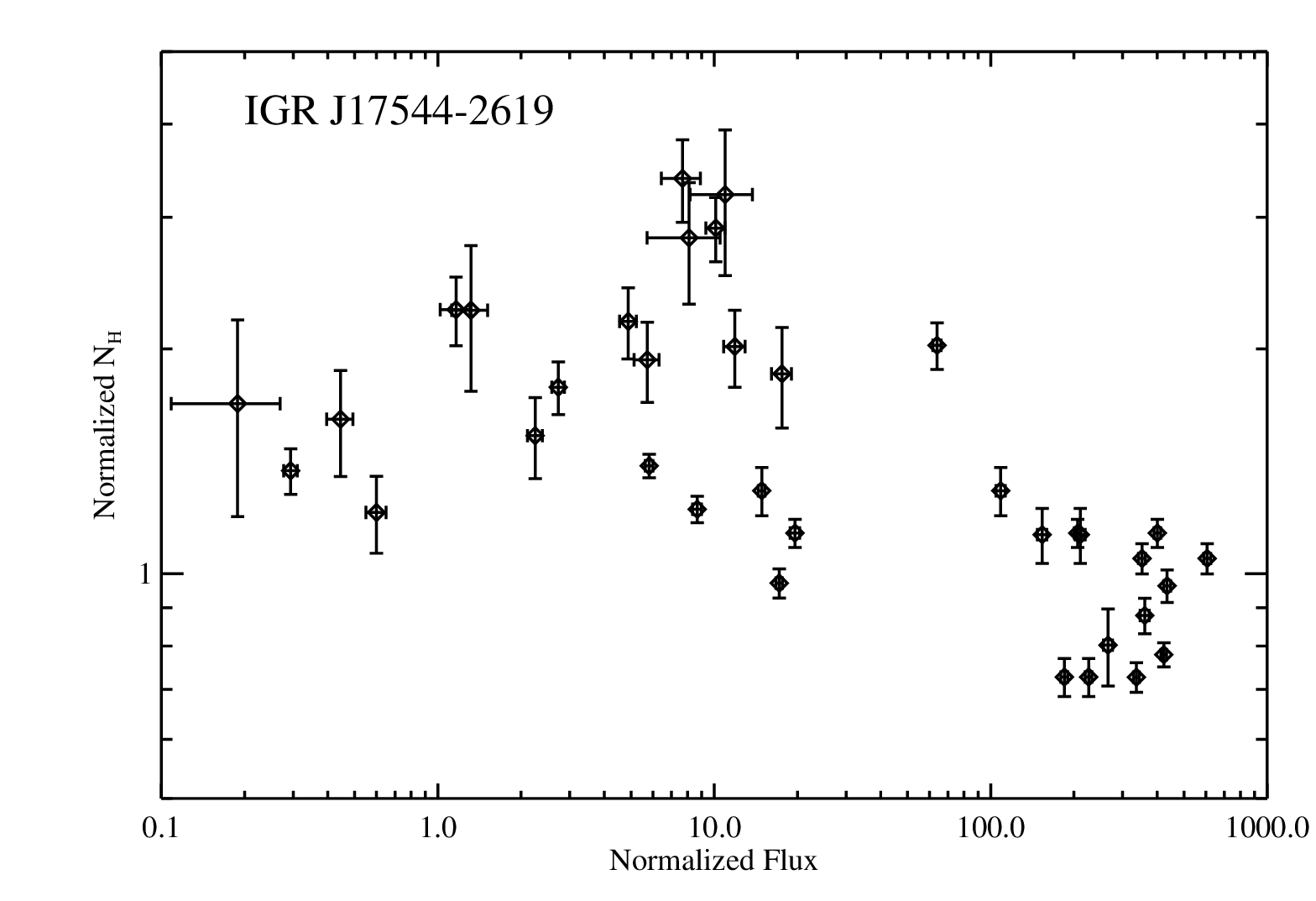}
\includegraphics[scale=0.3]{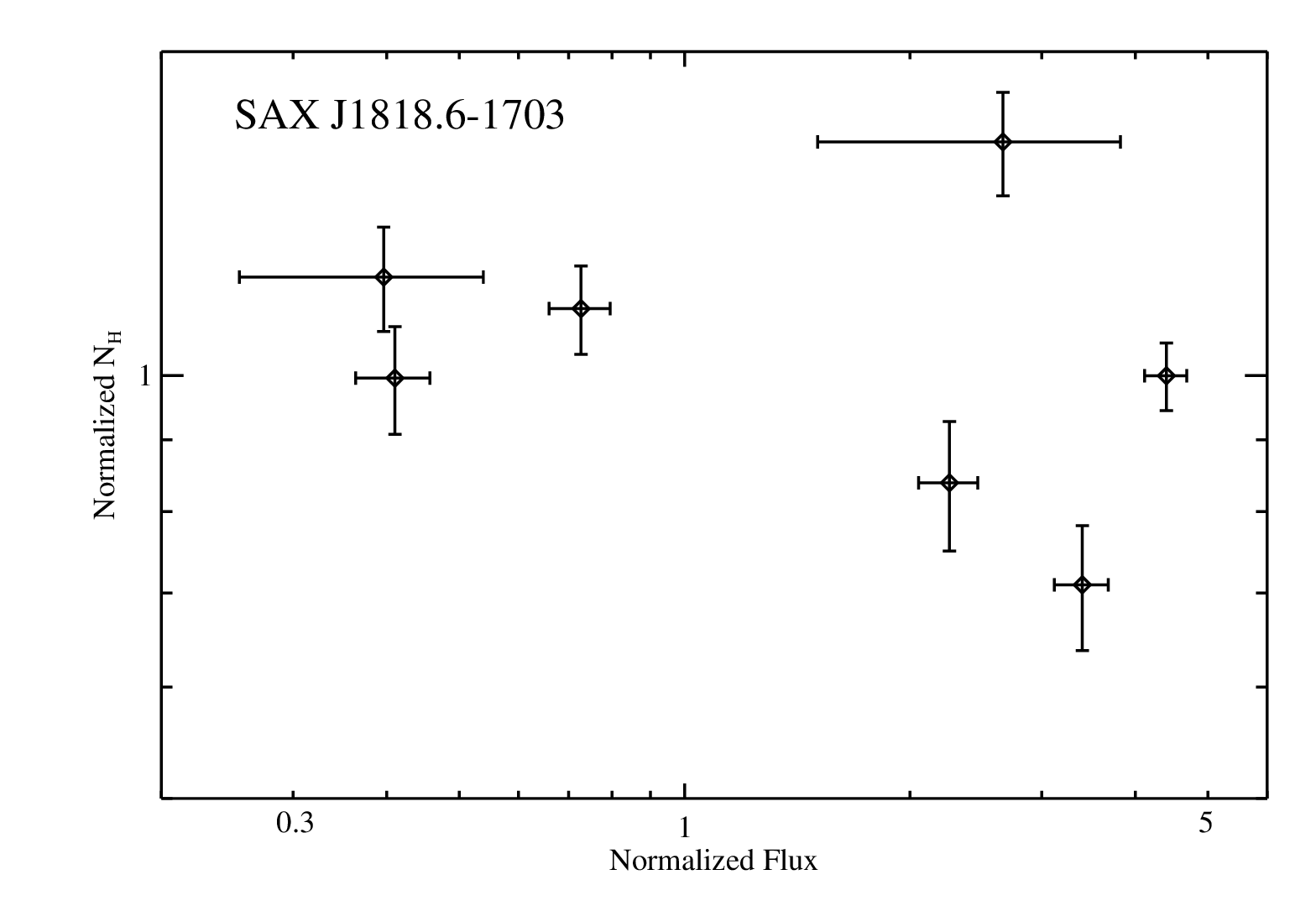}
\includegraphics[scale=0.3]{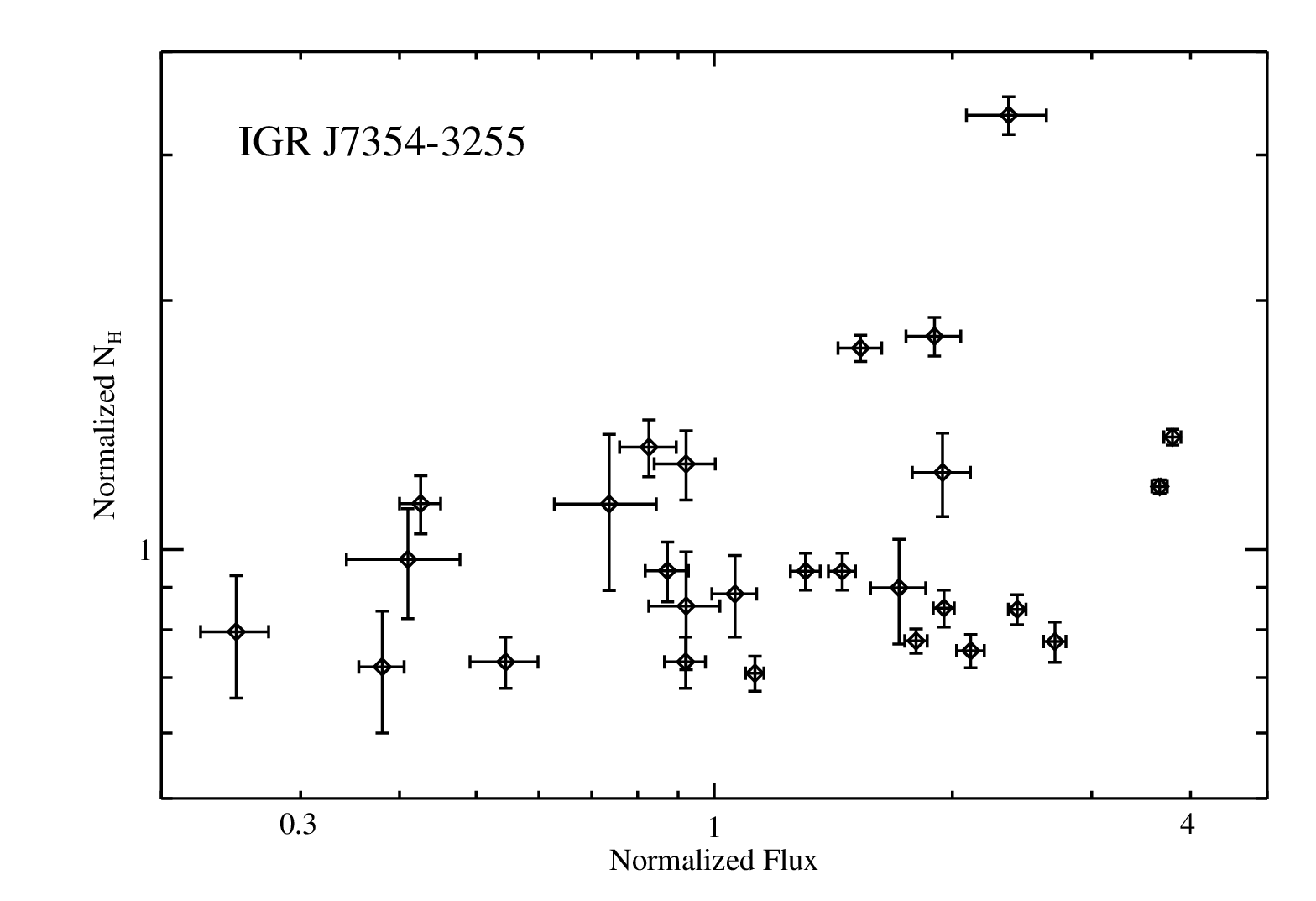}
\includegraphics[scale=0.3]{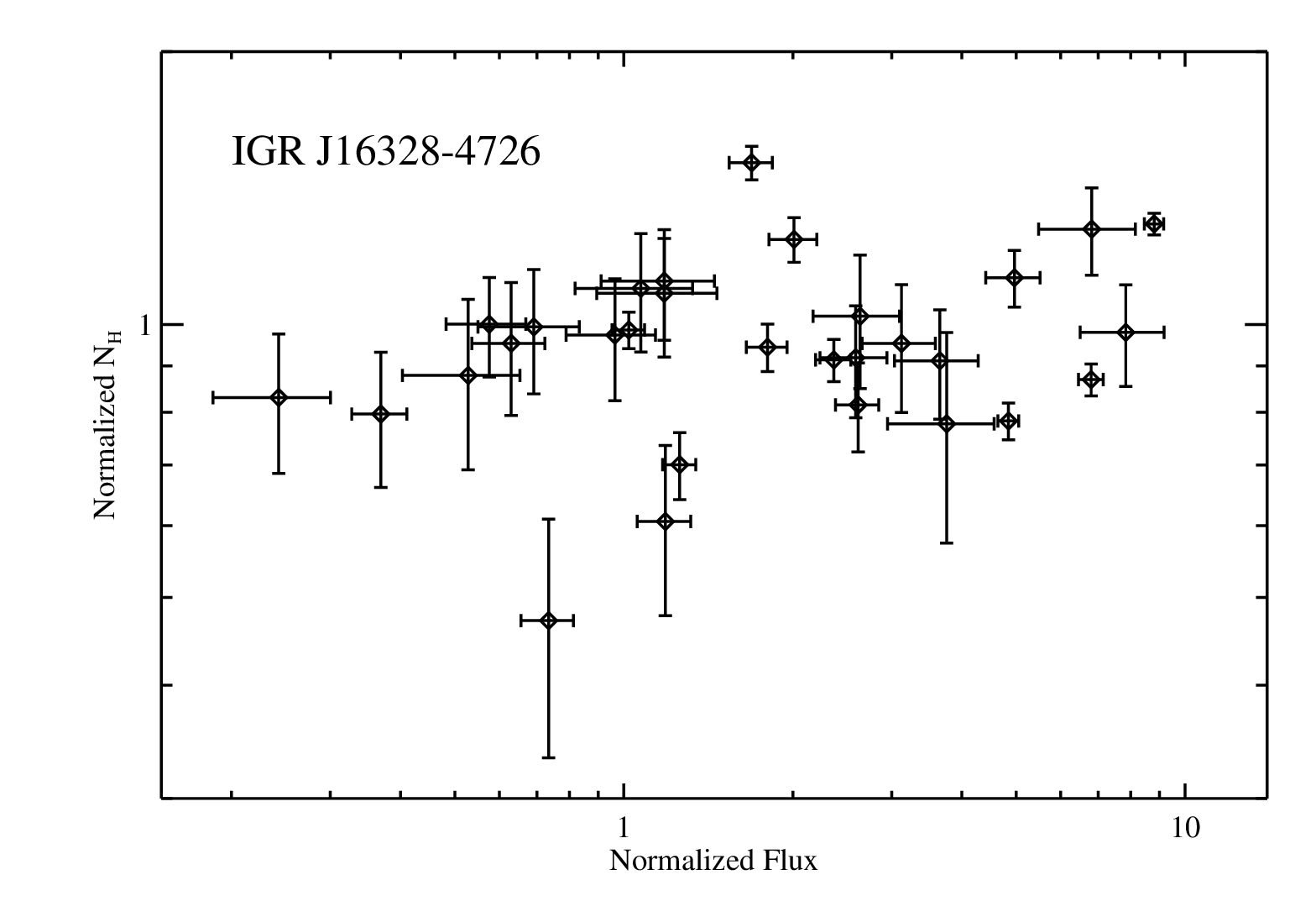}
\includegraphics[scale=0.3]{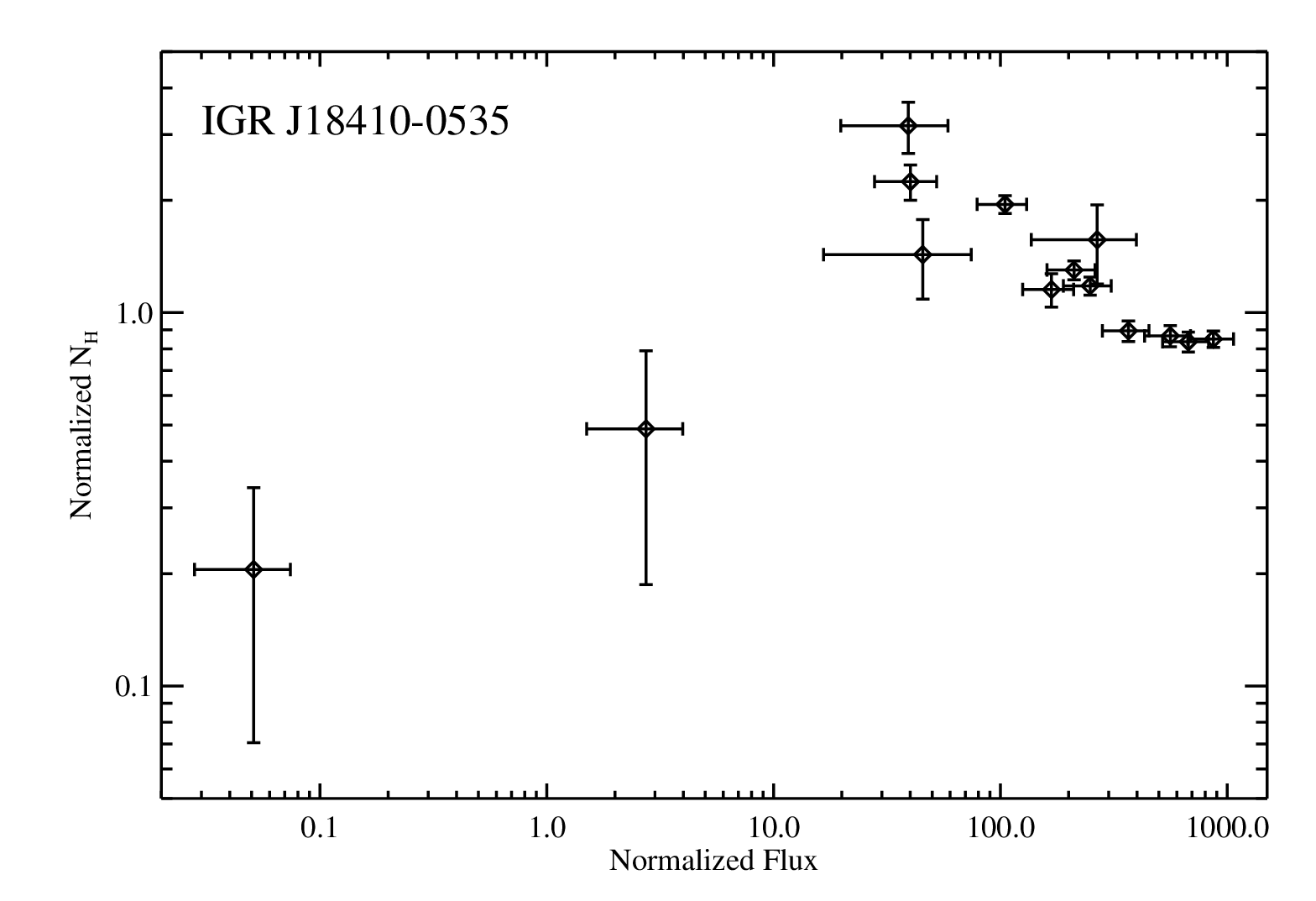}
\caption{Distributions of the normalized absorption column density as a function of the normalized flux (1-10~keV) measured from the HR-resolved 
spectral analysis for the 5 sources in Sect.~\ref{sec:data} (see text for further details). The previously published \xmm\ data on the outburst of IGR\,J17544-2619 
\citep{bozzo16} and the bright flare from IGR\,J18410-0535 \citep{bozzo11} are also included. All uncertainties are at 1$\sigma$ c.l.}     
\label{fig:last} 
\end{figure*} 

The analysis of the flares reported in Sect.~\ref{sec:data} and the corresponding results summarized in 
Sect.~\ref{sec:overview} evidenced that, generally, an increase in the local absorption column density is recorded  
during the rise of these events; a drop follows when the source reaches the peak flux. In some cases a re-increase in the absorption column 
density after the flare is also detected. This phenomenology 
was already pointed out during our previous analysis of the bright flare from IGR\,J18410-0535 \citep{bozzo11} 
and the outburst from IGR\,J17544-2619 \citep{bozzo16b}. Similar variations of the absorption column density are 
expected if the flares/outbursts are triggered by the presence of stellar wind clumps approaching the 
compact object. The drop of the $N_{\rm H}$ close to the peak of the X-ray activity can be most likely ascribed to the photoionization of the clump medium.  
A re-increase of the $N_{\rm H}$ is expected after the flare due to the recombination that is allowed by the decreasing X-ray flux. 
We also reported on a new detection of an enhanced absorption event not connected with an X-ray flare during the observation ID.~0693900201 
of the SFXT IGR\,J17354-3255. Few other similar cases were reported in the literature and associated 
with the passage of a clump along the observer line of sight to the source without being accreted. 
This is expected due to the two-fold role of the clumps mentioned in Sect.~\ref{sec:intro}. 

In order to carry out a statistically meaningful analysis of the results reported in Tables~\ref{tab:j1845_fit2}-\ref{tab:J16328_fit}, 
we also built for each source the distributions of the 
$N_{\rm H}$ as a function of the flux in Fig.~\ref{fig:last}. We plot for each observational time interval the measured 
absorption column density and flux normalized to the weighted average values (computed each time by excluding the reference time interval). We included 
the data from the previously published outburst of IGR\,J17544-2619 and the flare from IGR\,J18410-0535, as for these observations we had also exploited the 
HR-resolved spectral analysis technique. We used the time interval selection presented in Fig.~6 of \citet{bozzo16b} and Fig.~7 of \citet{bozzo11} and re-fit 
the corresponding spectra with the {\sc phabs*pegpow} model in order to have consistent measurements of the spectral parameters and fluxes for the 
entire data-set. The usage of a normalized $N_{\rm H}$ allows us to carry out an easier comparison between the results obtained for the 
different sources and draw some preliminarily conclusion for the SFXT as a class. 

Two main results emerge from Fig.~\ref{fig:last}. The first result is that there is no significant correlation between 
the dynamic range in the X-ray flux and the dynamic range in the absorption column density achieved by any of the sources. 
This means that, as an example, sources achieving a larger dynamic range 
in the X-ray flux do not necessarily display the largest swings in the absorption column density. 
The largest dynamic range in $N_{\rm H}$ was measured during the bright flare from IGR\,J18410-0535 (a factor of $\sim$15), which was also 
characterized by a dynamic range in the X-ray flux of a factor of $\sim$2000. In the case of IGR\,J17544-2619, 
which underwent several bright flares and one outburst with a total dynamic range in the X-ray flux of $\sim$3.5$\times$10$^3$, the total 
dynamic range in the $N_{\rm H}$ was limited to a factor of $\sim$5. A similar factor is measured in the case of IGR J18450-0435, which did not undergo an outburst 
but displayed several relatively bright flares (the total dynamic range in the X-ray flux was of $\lesssim$100). 
More limited dynamic ranges in the absorption column density ($\lesssim$3) have been recorded in the cases of 
IGR\,J16328-4726 and IGR\,J7354-3255, even though these sources displayed a  flux dynamic range similar to IGR\,J18450-0435  
(the only exception is the case of IGR\,J7354-3255 if the highest recorded $N_{\rm H}$ corresponding to the passage of a clump in front of the 
X-ray source is taken into account). This is in agreement with the general idea that clumps cannot be the only 
ingredient to explain the SFXT extreme X-ray variability. If this were the case, the dynamic range in the X-ray flux should be directly proportional 
to the density contrast of the clump compared to the intra-clump medium and thus to the observed dynamic range of the absorption column density. Instead, 
we measure similar $N_{\rm H}$ swings in different sources for X-ray dynamic ranges that differ by a factor of $\simeq$100. 
The presence of mechanisms inhibiting the accretion of 
the stellar wind material in SFXTs can explain this discrepancy, as they can boost the variability of these sources by switching between periods of quiescence and 
enhanced accretion (and {\it viceversa}) even when only moderate variations of the density of the stellar wind material take place in the surroundings of the compact object. 
This complication and the lack of a proper knowledge of the NS spin period and magnetic field intensity in the SFXTs prevent us to efficiently use the measurements 
of the column density variations in the these sources to estimate the physical properties of the stellar clumps (see Sect.~\ref{sec:intro}).    

The second result that emerges from Fig.~\ref{fig:last} is an indication 
that for the SFXT sources the absorption column densities measured at the highest fluxes are lower compared to those measured during the 
low/intermediate fluxes. Among all sources in the present sample, IGR\,J17544-2619 and IGR\,J18410-0535 provide the clearest evidence for such phenomenon. 
This is not surprising because these are also the only two sources that have been observed by \xmm\ in a sufficiently wide range of X-ray flux to draw 
a more corroborated conclusion. To quantify this conclusion at the best of our current possibilities, we calculated the weighted mean and the standard 
deviation of the normalized $N_{\rm H}$ in different normalized flux intervals for the two sources. The results are provided in Table~\ref{tab:last}. 
We note that for the SFXT sources a lower absoprtion column density at the highest fluxes compared to lower flux states  
is expected according to previous discussions in this section,  
as the largest increase in the absorption column density is measured during the rise of the flares/outbursts and a drop should occur above the threshold 
at which the clump material is significantly photoionized. Future \xmm\ observations of the confirmed SFXTs during quiescence, flares, and outbursts 
will permit us to improve this analysis and carry out more quantitative statistical investigations of the entire class of sources.   
 \begin{table}
   \scriptsize	
 \begin{center} 	
 \caption{Weighted mean of the normalized absorption column 
 density as a function of a chosen flux interval in the two SFXTs IGR\,J17544-2619 and 
 IGR\,J18410-0535 (all uncertainties are given at 1~$\sigma$ c.l.).} 
 \label{tab:last} 
 \begin{tabular}{@{}ll@{}} 
 \hline 
 \hline 
 \noalign{\smallskip}  
 Normalized flux range  & Normalized $N_{\rm H}$ (uncertainty) \\ 
 \noalign{\smallskip} 
 \hline 
       \multicolumn{2}{c}{IGR\,J17544-2619} \\
  \hline 
 \noalign{\smallskip} 
0.1-1.0 & 1.35 (0.08) \\
\noalign{\smallskip}
1.0-20.0 & 1.25 (0.02) \\
\noalign{\smallskip}
20.0-1000.0 & 0.90 (0.01) \\
\noalign{\smallskip}
 \hline
       \multicolumn{2}{c}{IGR\,J18410-0535} \\
 \hline 
\noalign{\smallskip} 
0.01-10.0 & 0.25 (0.12) \\
\noalign{\smallskip}
10.0-100.0 & 2.00 (0.09) \\
\noalign{\smallskip}
100.0-1000.0 & 0.95 (0.02) \\
\noalign{\smallskip}
    \hline 
     \hline 
  \end{tabular}
  \end{center}
  \end{table}

\subsection{Classification of SFXT flares and outbursts}
\label{sec:classification} 
  
Even though our poor understanding of the mechanisms inhibiting accretion in the SFXTs does not allow us to directly infer the properties 
of the stellar wind clumps, we can propose a preliminary classification of the enhanced accretion episodes from these 
sources to link the observations with the accretion physics and make prediction which could drive future investigations. Based on our current theoretical understanding 
of the SFXT accretion physics and the evidence collected so far from the X-ray observations, we propose to define the following four categories of flares/outbursts:  
\begin{enumerate}[label=\Alph*:]
\item flares ($L_{\rm X}$$\simeq$10$^{34}$-10$^{35}$~\erg) that are generated by the impact of moderately dense structures. The latter are not able to fully overcome 
all mechanisms inhibiting accretion, and thus accretion is only partly restored during the flare\footnote{It is generally believed that the inhibition of accretion 
is only completely removed during the outbursts (see Sect.~\ref{sec:intro}).}. 
As these structures are not extremely massive/dense, the increase on the 
local absorption can be limited ($\Delta$$N_{\rm H}$$\lesssim$2-3) or even hidden within the uncertainty of the $N_{\rm H}$ 
in the direction of the source. If flares in category A are observed from a source, it cannot be excluded that brighter flares in the following category B 
can also occur, as well as outbursts in the categories C and D below.   

\item flares ($L_{\rm X}$$\simeq$10$^{34}$-10$^{35}$~\erg) that are generated by the impact of sufficiently dense structures producing measurable  
increases in the local absorption ($\Delta$$N_{\rm H}$$\gg$3). If such flares are found in a system, it shall be expected that the 
mechanisms inhibiting accretion are more difficult to be overcame, as massively enough structures are requested to produce even moderately faint flares. Therefore, 
these sources should not show outbusts associated with modest or no increase in the absorption column density (category C below). It is still possible 
that flares belonging to the category A might occur at lower luminosities. 

\item outbursts ($L_{\rm X}$$\simeq$10$^{36}$-10$^{38}$~\erg) that are generated by the impact of moderately massive structures. 
The limited increase in the absorption during these events ($\Delta$$N_{\rm H}$$\lesssim$2-3) indicates that 
the inhibiting mechanisms are particularly easy to overcome. We should thus expect frequent outbursts 
from these systems and also fainter flares which do not display at all the signature of a measurable increased X-ray extinction 
(as the partial inhibition of accretion is even more easy to be overcame; category A). Flares belonging to the category B should 
not be expected from these systems.  

\item outbursts ($L_{\rm X}$$\simeq$10$^{36}$-10$^{38}$~\erg) that are generated by the impact of sufficiently dense structures 
to produce large increases in the absorption ($\Delta$$N_{\rm H}$$\gg$3). If the onset of an outburst require the presence of such structures, 
these systems should show less frequent outbursts and a larger number of fainter flares (as for them it is, on average, more difficult to completely overcome 
the inhibition mechanisms). Both flares from categories A and B can be expected, with those in category B achieving higher luminosities.  
\end{enumerate}

In the \xmm\ data available so far, we found episodes in three of the above categories (A, B, and C) but outbursts giving rise to very large variations of the 
accretion column density ($\gg$3, case D) are still to be observed (the only outburst observed by \xmm\ is the one from IGR\,J17544-2619 and it is of the C type). 
We can, however, use also all previously published results in the literature to preliminarly test the proposed classification and the applicability to different 
sources. 

In the case of the SFXT prototype IGR\,J17544-2619, large increases in the 
absorption column density were not detected during either the many outbursts observed with Swift\,/XRT \citep{romano08,romano11,romano14}, 
\xmm\ \citep{bozzo16b}, and \suzaku\ \citep{rampy09}, or during the fainter flares analyzed here. 
The spectral variability seems more related to changes in the power-law photon index  
and so to the emission mechanism intrinsic to the source \citep[see the exhaustive discussion in][and references therein]{bozzo16}, 
while the absorption column density does not change dramatically between quiescence and the very bright X-ray states \citep{zand05}. 
We should thus conclude that in the case of IGR\,J17544-2619 the mechanism(s) inhibiting the accretion can easily be 
overcome by moderate variations in the density of the accreting material (a factor of $\sim$2-3 at the most) in order to produce 
both flares and outbursts (cases A and C of the classification). 
The prediction for IGR\,J17544-2619 is that during its future outbursts there should not be large absorption variations and that 
during its X-ray flares variations should be even lower.  

In the case of IGR\,J18450-0435, we recorded variations in the $N_{\rm H}$ as large as a factor of $\sim$8 
during some bright flares (see Fig.~\ref{fig:last}) and we should thus conclude that overcoming the inhibition mechanisms for this source 
is not as easy as in the case of IGR\,J17544-2619. Large swings of the absorption column density would thus be expected during future flares and 
outbursts (cases B and D). 
In the sole outburst observed in the soft X-ray domain from this source with \swift/XRT, no variations of the $N_{\rm H}$  
were reported \citep{romano12,romano13,goossens13}. However, we note that some limitations in the observational capabilities of this instrument could hamper the 
searches for these variations within the required time scales and in the relevant phases of the event (see Sect.~\ref{sec:caveats}). 

The analysis of the flares from SAX\,J1818.6-1703 reported in this paper did not reveal very large variations of the absorption column 
density, but the average $N_{\rm H}$ was as large as 3$\times$10$^{23}$~cm$^{-2}$. Similar high values of the $N_{\rm H}$ 
have been measured by \swift/XRT during a number of previous outbursts/flares with swings down to a few 
times 10$^{22}$~cm$^{-2}$\citep{sidoli09,romano09d,romano09e,kennea14}. Even though we do not have a clear evidence of 
the absorption column density regularly increasing during the rise of these events and the properties of the source emission 
outside the outburst/flaring state are not known \citep[two deep \xmm\ observations carried out to this aim failed to detect the source in both 
occasions; see][]{bozzo08c,bozzo12}, we would preliminarly conclude that the mechanisms inhibiting accretion in 
SAX\,J1818.6-1703 cannot be overcame without the intervention of a substantially dense structure. 
We thus expect to detect with  \xmm\ much larger $N_{\rm H}$ variations during the rising phase of an outburst 
than during a flare (case D of the classification). 

For IGR\,J17354-3255 our analysis did not reveal large HR variations corresponding 
to the rise and decay of the flares. 
At present, we might thus conclude that for this source overcoming the 
mechanisms inhibiting the accretion should not be very difficult, as noticeable X-ray flares 
are associated with modest increase of the absorption column density (a factor of $\lesssim$2;  
case A of the classification). This does not preclude the existence of more dense structures  
in the stellar wind that could pass in front of the compact object, causing its  
obscuration outside flares and outbursts (see also Sect.~\ref{sec:intro}). In fact, 
we detected an increase in the $N_{\rm H}$ between two flares most likely due to the 
passage of a clump along the observer line of sight.  
More observations of this source in outburst with large area instruments are definitively needed in order 
to confirm this conclusion, as so far no outbursts from IGR\,J17354-3255 could be studied in the soft 
X-ray domain \citep[see also][and references therein]{ducci13}. Our prediction is that for this source no 
outbursts with large variations of the $N_{\rm H}$ should be observed (case C of the classification). 

In the case of IGR\,J16328-4726,  
the flares analyzed here showed that there is usually some limited variability (a factor of $\sim$2-3) in the absorption 
column density during these events (case A of the classification). 
It could thus be initially concluded that this source is similar to IGR\,J17354-3255. However, we note that the outburst that was followed-up with 
\beppo/MECS was characterized by large swings in the absorption column density, reaching factors as high as $\sim$7 \citep{fiocchi13}. 
We can thus predict that an observation of the source with \xmm\ during an outburst will most likely reveal large swings in the 
absorption column density (case D of the classification) and thus that this source is more similar to IGR\,J18450-0435. 
Brighter and less frequent flares associated to large variations of the $N_{\rm H}$ (category B) should also be observable. 

In case of IGR\,J18410-0535, only one single bright flare has been 
observed by \xmm\ with variations in the absorption column density up to a factor of $\gtrsim$15 (case B of the classification). 
We should thus expect to eventually observe only outbursts with large changes in the $N_{\rm H}$ from this system.  
No other evidence of large increases in the absorption column 
density have been reported in the literature, but we note that the observations performed in the direction of the source 
with {\em ASCA} and {\em SUZAKU} were endowed with a limited statistics to perform any meaningful HR-resolved spectral analysis 
and \swift/XRT observed the source thousand of seconds after the onset of all three detected outbursts 
\citep{bamba01,depasquale10,romano10,romano10b,mangano11,romano12b,romano12c,kawabata12}.  

There are a number of SFXTs which have not been discussed in this paper due to the lack of reasonable S/N data and 
for which it is difficult to understand if they follow or not the proposed classification and predictions above. 
XTE\,J1739-302 and IGR\,J16479-4514 have been 
observed in outbursts many times with \swift/XRT and thus several measurements of the local 
absorption column density have been reported in the literature. 
For XTE\,J1739-302 an enhanced $N_{\rm H}$ at the on-set of the outbursts compared to quiescent states  
has clearly been detected in several cases \citep{sidoli08,sidoli09b,sidoli09c}, even though the usually 
fragmented \swift/XRT lightcurves do not allow to carry out an in depth analysis of the events. 
The only observations of the source available with \xmm\ caught the system in quiescence and the flares 
detected are too faint to investigate the presence of any spectral variability \citep[see][]{bozzo10,bodaghee11}. 
Large swings of the absorption column density have been detected during the outbursts of the SFXT IGR\,J08408-4503, 
which seems to display on average a much harder emission during the active X-ray states compared to quiescence 
\citep{romano09t,bozzo10,sidoli10b,mangano13}. 
At present, we would thus preliminarly conclude that XTE\,J1739-302 and IGR\,J08408-4503 are similar to IGR\,J18450-0435, SAX\,J1818.6-1703, and 
IGR\,J16328-4726 for what concerns the flares/outbursts classification and the accretion inhibition mechanisms, 
but this conclusion needs to be corroborated by additional X-rays observations during flaring or outburst states. 
We should conclude the opposite for IGR\,J16479-4514, which displayed evidence for changes in the 
local absorption column density especially during the eclipses \citep{sidoli08,bozzo08b,sidoli13}, but most likely not during the rise to the few 
outbursts observed with \swift/XRT \citep{romano09f,romano08b,parola09}.

\subsubsection{Caveats to the flares/outbursts classification}
\label{sec:caveats}

Even though it seems that all the SFXTs that could be carefully checked so far qualitatively follow the proposed 
flares/outbursts classification and the predictions presented in Sect.~\ref{sec:classification}, there are two caveats that need 
to be taken into account: 
\begin{itemize}
\item {\it Impact of geometrical effects}. The values of the absorption column density measured during the rising or decaying stages 
of an SFXT flare or outburst are likely affected by the projection effects related to the viewing angle between the observer line of sight 
and the direction from which the clump approaches the compact object. These effects have been discussed in theoretical and observational 
papers \citep{walter07,ducci10,bozzo11}. They are known to give rise to large uncertainties in the estimate of the clump size from the duration 
and intensity of the X-ray flare or outburst. In this respect, when no variation in the absorption column density is observed, 
it is difficult to conclude if this is due to the lack of dense enough structures gathering around the NS or if 
there is a projection effect that might limit our ability to reveal those structures. Clearly, in all those cases in which a significant 
$N_{\rm H}$ increase is measured, any possible geometrical effect would only further enhance it.  
\item {\it Observability of the variation of the absorption column density during  the relevant stages of the flare/outburst}. 
\xmm\ results have shown that the absorption column density 
might only increase during the rising or the final phases of the flare/outburst, while it can be comparable to the value measured during quiescence 
in other phases of the same event. As most of the SFXT flares/outbursts were observed with \swift/XRT hundred to thousand of seconds 
after the onset of the event, it is possible that in most of the cases we missed the large swings in the absorption column density.  
Uninterrupted observations of many more flares and outbursts with \xmm\ are thus needed in order to further test 
our proposed classification and predictions.   
\end{itemize}

\subsection{Stellar wind clumps and corotating interaction regions}  

Since the early discovery of the SFXTs, it has always been assumed that the density variations in the material around the  
compact object triggering the onset of flares/outbursts were associated with the presence of clumps. This assumption followed 
especially from the need of having sufficiently dense and compact structures to trigger the SFXT fast X-ray flares within the required 
short timescales (few hundred of seconds) and explain the entire observed dynamic range in the SFXT X-ray luminosity only 
through the density contrast between the clump and the intra-clump medium. This need has been largely alleviated by the observation that  
all SFXTs are largely sub-luminous compared to classical systems and that some mechanisms should be at work to inhibit accretion 
in these systems. These mechanisms act to boost the dynamical range in the X-ray luminosity achievable by the SFXTs, 
which in turns is no longer bound to the sole density ratio between the structure being accreted onto the NS and the surrounding 
stellar wind material. 

Studies of isolated supergiants have shown that, beside clumps, 
at least another kind of large scale structures can be formed within the wind of an OB supergiant star, i.e. the so-called 
Corotating Interaction Regions \citep[CIRs;][]{mullan84}. The presence and properties 
of these structures are mainly inferred from the detection of Discrete Absorption Components in the UV spectra of many 
OB supergiants \citep[DACs;][]{bates90,howarth92}, and used in several cases to interpret the X-ray variability of these objects 
\citep{oskinova01,massa14,naze13}.  
DACs are revealed in the form of optical depth enhancements in the absorption troughs of unsaturated 
UV P Cygni profiles and were directly observed to accelerate towards the blue wing of the line profile, becoming narrower as they 
approach the limiting value of the stellar wind asymptotic velocity. The timescales on which these features move along the line profile 
is of a few days, and it thus seems more plausible that they are connected with the star rotational period rather than with the 
intrinsic acceleration of the stellar wind (occurring on much short timescales if a relatively standard terminal velocity of few 
thousands km~s$^{-1}$ is assumed for a typical OB supergiant). The widely agreed hypothesis is that DACs are the outcomes of slowly 
evolving velocity and/or density perturbations along the main stellar wind stream, originating from a substantial fraction of 
the stellar surface \citep[see, e.g.,][and references therein]{prinja92,prinja94}. Hydrodynamical simulations have shown that 
photospheric disturbances on the stellar surface can generate streams of enhanced density along the 
profile of an unperturbed wind, with a significant velocity drop at the edge of these streams where they 
interact with the unperturbed wind \citep{feldmeier00,feldmeier02}. Even though the details on the formation and evolution of these 
structures are still under study \citep{hamann01,lobel08} and the overall picture of large scale structures 
in the winds of OB supergiant could be complicated by the later discovery of the so-called rotational modulations 
\citep[see, e.g.,][and references therein]{lobel13}, it seems that the typical 
density and/or velocity contrasts of these structures with respect to the surrounding wind could be as large as a factor 
of $\sim$3 \citep{cranmer96}. The presence of these large and mildly dense structure could thus provide 
the required changes in the physical properties of the accretion medium that in turn can overcome the accretion inhibition mechanisms and trigger 
those SFXT flares/outbursts in which large variations in the local accretion column densities are not measured (cases A and C of our classification). 
It is not clear at present if CIRs or other similar large scale structures could survive in the heavily clumpy wind of an OB supergiant 
(the instability usually adopted to generate clumps in the hydrodynamical simulations of their winds is not included 
in the simulations developed to produce the CIRs), but we propose here that 
many of the enhanced X-ray emission episodes that we detect from SFXTs could be associated to these structures rather than to 
more confined and dense clumps. 
As the CIRs are supposed to be somehow connected with the rotational velocity of the star, there is the 
interesting possibility that they could induce a periodical modulation of the X-ray luminosity in classical SgXBs and in SFXTs 
on much longer timescales than those of the flares/outbursts or the binary revolution \citep[e.g., giving rise to the 
observed super-orbital modulation; see][]{corbet13}. The analysis of this possibility and a full exploitation on the expected effect of 
CIRs on the X-ray emission of SFXTs and classical SgXBs is beyond the scope of the current paper and 
will be reported in a forthcoming publication (Bozzo et al. 2017, in preparation).

\section*{Acknowledgments}

We thank the anonymous referee for constructive comments which helped to improve the paper. 
This publication was motivated by a team meeting
sponsored by the International Space Science Institute in Bern,
Switzerland. EB and LO thank ISSI for the financial
support during their stay in Bern. EB is grateful for the 
hospitality of the New York University Abu Dhabi 
during part of this work. PR and EB 
acknowledge financial contribution from 
the agreement ASI-INAF I/037/12/0. 
PR acknowledges grant ASI-INAF I/004/11/0. 
LO acknowledges support by the DLR grant 50\,OR\,1302 and partial
support by the Russian Government Program of Competitive 
Growth of Kazan Federal University.

\bibliographystyle{aa}
\bibliography{wind}

\begin{thebibliography}{84}
\expandafter\ifx\csname natexlab\endcsname\relax\def\natexlab#1{#1}\fi

\bibitem[{{Bamba} {et~al.}(2001){Bamba}, {Yokogawa}, {Ueno}, {Koyama}, \&
  {Yamauchi}}]{bamba01}
{Bamba}, A., {Yokogawa}, J., {Ueno}, M., {Koyama}, K., \& {Yamauchi}, S. 2001,
  \pasj, 53, 1179

\bibitem[{{Bates} \& {Gilheany}(1990)}]{bates90}
{Bates}, B. \& {Gilheany}, S. 1990, \mnras, 243, 320

\bibitem[{{Bodaghee} {et~al.}(2011){Bodaghee}, {Tomsick}, {Rodriguez}, {Chaty},
  {Pottschmidt}, {Walter}, \& {Romano}}]{bodaghee11}
{Bodaghee}, A., {Tomsick}, J.~A., {Rodriguez}, J., {et~al.} 2011, \apj, 727, 59

\bibitem[{{Boon} {et~al.}(2016){Boon}, {Bird}, {Hill}, {Sidoli}, {Sguera},
  {Goossens}, {Fiocchi}, {McBride}, \& {Drave}}]{boon16}
{Boon}, C.~M., {Bird}, A.~J., {Hill}, A.~B., {et~al.} 2016, \mnras, 456, 4111

\bibitem[{{Bozzo} {et~al.}(2016{\natexlab{a}}){Bozzo}, {Bhalerao}, {Pradhan},
  {Tomsick}, {Romano}, {Ferrigno}, {Chaty}, {Oskinova}, {Manousakis}, {Walter},
  {Falanga}, {Campana}, {Stella}, {Ramolla}, \& {Chini}}]{bozzo16b}
{Bozzo}, E., {Bhalerao}, V., {Pradhan}, P., {et~al.} 2016{\natexlab{a}}, \aap,
  596, A16

\bibitem[{{Bozzo} {et~al.}(2008{\natexlab{a}}){Bozzo}, {Campana}, {Stella},
  {Falanga}, {Israel}, {Rampy}, {Smith}, \& {Negueruela}}]{bozzo08c}
{Bozzo}, E., {Campana}, S., {Stella}, L., {et~al.} 2008{\natexlab{a}}, The
  Astronomer's Telegram, 1493

\bibitem[{{Bozzo} {et~al.}(2008{\natexlab{b}}){Bozzo}, {Falanga}, \&
  {Stella}}]{bozzo08}
{Bozzo}, E., {Falanga}, M., \& {Stella}, L. 2008{\natexlab{b}}, \apj, 683, 1031

\bibitem[{{Bozzo} {et~al.}(2011){Bozzo}, {Giunta}, {Cusumano}, {Ferrigno},
  {Walter}, {Campana}, {Falanga}, {Israel}, \& {Stella}}]{bozzo11}
{Bozzo}, E., {Giunta}, A., {Cusumano}, G., {et~al.} 2011, \aap, 531, A130

\bibitem[{{Bozzo} {et~al.}(2016{\natexlab{b}}){Bozzo}, {Oskinova}, {Feldmeier},
  \& {Falanga}}]{bozzo16}
{Bozzo}, E., {Oskinova}, L., {Feldmeier}, A., \& {Falanga}, M.
  2016{\natexlab{b}}, \aap, 589, A102

\bibitem[{{Bozzo} {et~al.}(2012){Bozzo}, {Pavan}, {Ferrigno}, {Falanga},
  {Campana}, {Paltani}, {Stella}, \& {Walter}}]{bozzo12}
{Bozzo}, E., {Pavan}, L., {Ferrigno}, C., {et~al.} 2012, \aap, 544, A118

\bibitem[{{Bozzo} {et~al.}(2015){Bozzo}, {Romano}, {Ducci}, {Bernardini}, \&
  {Falanga}}]{bozzo15}
{Bozzo}, E., {Romano}, P., {Ducci}, L., {Bernardini}, F., \& {Falanga}, M.
  2015, Advances in Space Research, 55, 1255

\bibitem[{{Bozzo} {et~al.}(2013){Bozzo}, {Romano}, {Ferrigno}, {Esposito}, \&
  {Mangano}}]{bozzo13}
{Bozzo}, E., {Romano}, P., {Ferrigno}, C., {Esposito}, P., \& {Mangano}, V.
  2013, Advances in Space Research, 51, 1593

\bibitem[{{Bozzo} {et~al.}(2010){Bozzo}, {Stella}, {Ferrigno}, {Giunta},
  {Falanga}, {Campana}, {Israel}, \& {Leyder}}]{bozzo10}
{Bozzo}, E., {Stella}, L., {Ferrigno}, C., {et~al.} 2010, \aap, 519, A6

\bibitem[{{Bozzo} {et~al.}(2008{\natexlab{c}}){Bozzo}, {Stella}, {Israel},
  {Falanga}, \& {Campana}}]{bozzo08b}
{Bozzo}, E., {Stella}, L., {Israel}, G., {Falanga}, M., \& {Campana}, S.
  2008{\natexlab{c}}, \mnras, 391, L108

\bibitem[{{Corbet} \& {Krimm}(2013)}]{corbet13}
{Corbet}, R.~H.~D. \& {Krimm}, H.~A. 2013, \apj, 778, 45

\bibitem[{{Courvoisier} {et~al.}(2003){Courvoisier}, {Walter}, {Beckmann},
  {Dean}, {Dubath}, {Hudec}, {Kretschmar}, {Mereghetti}, {Montmerle},
  {Mowlavi}, {Paltani}, {Preite Martinez}, {Produit}, {Staubert}, {Strong},
  {Swings}, {Westergaard}, {White}, {Winkler}, \& {Zdziarski}}]{courvoisier03}
{Courvoisier}, T.~J.-L., {Walter}, R., {Beckmann}, V., {et~al.} 2003, \aap,
  411, L53

\bibitem[{{Cranmer} \& {Owocki}(1996)}]{cranmer96}
{Cranmer}, S.~R. \& {Owocki}, S.~P. 1996, \apj, 462, 469

\bibitem[{{Dame} {et~al.}(2001){Dame}, {Hartmann}, \& {Thaddeus}}]{dame01}
{Dame}, T.~M., {Hartmann}, D., \& {Thaddeus}, P. 2001, \apj, 547, 792

\bibitem[{{de Pasquale} {et~al.}(2010){de Pasquale}, {Barthelmy},
  {Baumgartner}, {Burrows}, {Holland}, {Kennea}, {Mangano}, {Palmer}, {Romano},
  {Sbarufatti}, {Starling}, \& {Vetere}}]{depasquale10}
{de Pasquale}, M., {Barthelmy}, S.~D., {Baumgartner}, W.~H., {et~al.} 2010, The
  Astronomer's Telegram, 2661

\bibitem[{{Dickey} \& {Lockman}(1990)}]{dickey90}
{Dickey}, J.~M. \& {Lockman}, F.~J. 1990, \araa, 28, 215

\bibitem[{{Drave} {et~al.}(2014){Drave}, {Bird}, {Sidoli}, {Sguera}, {Bazzano},
  {Hill}, \& {Goossens}}]{drave14}
{Drave}, S.~P., {Bird}, A.~J., {Sidoli}, L., {et~al.} 2014, \mnras, 439, 2175

\bibitem[{{Drave} {et~al.}(2013){Drave}, {Bird}, {Sidoli}, {Sguera}, {McBride},
  {Hill}, {Bazzano}, \& {Goossens}}]{drave13}
{Drave}, S.~P., {Bird}, A.~J., {Sidoli}, L., {et~al.} 2013, \mnras, 433, 528

\bibitem[{{Ducci} {et~al.}(2013){Ducci}, {Romano}, {Esposito}, {Bozzo},
  {Krimm}, {Vercellone}, {Mangano}, \& {Kennea}}]{ducci13}
{Ducci}, L., {Romano}, P., {Esposito}, P., {et~al.} 2013, \aap, 556, A72

\bibitem[{{Ducci} {et~al.}(2010){Ducci}, {Sidoli}, \& {Paizis}}]{ducci10}
{Ducci}, L., {Sidoli}, L., \& {Paizis}, A. 2010, \mnras, 408, 1540

\bibitem[{{Feldmeier} \& {Shlosman}(2000)}]{feldmeier00}
{Feldmeier}, A. \& {Shlosman}, I. 2000, \apjl, 532, L125

\bibitem[{{Feldmeier} \& {Shlosman}(2002)}]{feldmeier02}
{Feldmeier}, A. \& {Shlosman}, I. 2002, \apj, 564, 385

\bibitem[{{Fiocchi} {et~al.}(2013){Fiocchi}, {Bazzano}, {Bird}, {Drave},
  {Natalucci}, {Persi}, {Piro}, \& {Ubertini}}]{fiocchi13}
{Fiocchi}, M., {Bazzano}, A., {Bird}, A.~J., {et~al.} 2013, \apj, 762, 19

\bibitem[{{Fiocchi} {et~al.}(2016){Fiocchi}, {Bazzano}, {Natalucci},
  {Ubertini}, {Sguera}, {Bird}, {Boon}, {Persi}, \& {Piro}}]{fiocchi16}
{Fiocchi}, M., {Bazzano}, A., {Natalucci}, L., {et~al.} 2016, \apj, 829, 125

\bibitem[{{Gim{\'e}nez-Garc{\'{\i}}a}
  {et~al.}(2016){Gim{\'e}nez-Garc{\'{\i}}a}, {Shenar}, {Torrej{\'o}n},
  {Oskinova}, {Mart{\'{\i}}nez-N{\'u}{\~n}ez}, {Hamann}, {Rodes-Roca},
  {Gonz{\'a}lez-Gal{\'a}n}, {Alonso-Santiago}, {Gonz{\'a}lez-Fern{\'a}ndez},
  {Bernabeu}, \& {Sander}}]{gimenez16}
{Gim{\'e}nez-Garc{\'{\i}}a}, A., {Shenar}, T., {Torrej{\'o}n}, J.~M., {et~al.}
  2016, \aap, 591, A26

\bibitem[{{Gonz{\'a}lez-Riestra} {et~al.}(2004){Gonz{\'a}lez-Riestra},
  {Oosterbroek}, {Kuulkers}, {Orr}, \& {Parmar}}]{riestra04}
{Gonz{\'a}lez-Riestra}, R., {Oosterbroek}, T., {Kuulkers}, E., {Orr}, A., \&
  {Parmar}, A.~N. 2004, \aap, 420, 589

\bibitem[{{Goossens} {et~al.}(2013){Goossens}, {Bird}, {Drave}, {Bazzano},
  {Hill}, {McBride}, {Sguera}, \& {Sidoli}}]{goossens13}
{Goossens}, M.~E., {Bird}, A.~J., {Drave}, S.~P., {et~al.} 2013, \mnras, 434,
  2182

\bibitem[{{Grebenev} \& {Sunyaev}(2007)}]{grebenev07}
{Grebenev}, S.~A. \& {Sunyaev}, R.~A. 2007, Astronomy Letters, 33, 149

\bibitem[{{Grinberg} {et~al.}(2015){Grinberg}, {Leutenegger}, {Hell},
  {Pottschmidt}, {B{\"o}ck}, {Garc{\'{\i}}a}, {Hanke}, {Nowak}, {Sundqvist},
  {Townsend}, \& {Wilms}}]{grinberg15}
{Grinberg}, V., {Leutenegger}, M.~A., {Hell}, N., {et~al.} 2015, \aap, 576,
  A117

\bibitem[{{Hamann} {et~al.}(2001){Hamann}, {Brown}, {Feldmeier}, \&
  {Oskinova}}]{hamann01}
{Hamann}, W.-R., {Brown}, J.~C., {Feldmeier}, A., \& {Oskinova}, L.~M. 2001,
  \aap, 378, 946

\bibitem[{{Howarth}(1992)}]{howarth92}
{Howarth}, I.~D. 1992, in Astronomical Society of the Pacific Conference
  Series, Vol.~22, Nonisotropic and Variable Outflows from Stars, ed.
  L.~{Drissen}, C.~{Leitherer}, \& A.~{Nota}, 155

\bibitem[{{in 't Zand}(2005)}]{zand05}
{in 't Zand}, J.~J.~M. 2005, \aap, 441, L1

\bibitem[{{Jansen} {et~al.}(2001){Jansen}, {Lumb}, {Altieri}, {Clavel}, {Ehle},
  {Erd}, {Gabriel}, {Guainazzi}, {Gondoin}, {Much}, {Munoz}, {Santos},
  {Schartel}, {Texier}, \& {Vacanti}}]{jansen01}
{Jansen}, F., {Lumb}, D., {Altieri}, B., {et~al.} 2001, \aap, 365, L1

\bibitem[{{Kawabata} {et~al.}(2012){Kawabata}, {Nobukawa}, {Tsuru}, \&
  {Koyama}}]{kawabata12}
{Kawabata}, K., {Nobukawa}, M., {Tsuru}, T.~G., \& {Koyama}, K. 2012, \pasj,
  64, 99

\bibitem[{{Kennea} {et~al.}(2014){Kennea}, {Krimm}, {Romano}, {Evans},
  {Barthelmy}, {Burrows}, {Mangano}, {Sbarufatti}, {Esposito}, {Gehrels}, \&
  {Vercellone}}]{kennea14}
{Kennea}, J.~A., {Krimm}, H.~A., {Romano}, P., {et~al.} 2014, The Astronomer's
  Telegram, 5980

\bibitem[{{La Parola} {et~al.}(2009){La Parola}, {Romano}, {Sidoli}, {Evans},
  {Kennea}, {Cusumano}, {Mangano}, {Vercellone}, {Krimm}, {Burrows}, \&
  {Gehrels}}]{parola09}
{La Parola}, V., {Romano}, P., {Sidoli}, L., {et~al.} 2009, The Astronomer's
  Telegram, 1929

\bibitem[{{Lobel}(2013)}]{lobel13}
{Lobel}, A. 2013, in Massive Stars: From alpha to Omega

\bibitem[{{Lobel} \& {Blomme}(2008)}]{lobel08}
{Lobel}, A. \& {Blomme}, R. 2008, \apj, 678, 408

\bibitem[{{Lutovinov} {et~al.}(2013){Lutovinov}, {Revnivtsev}, {Tsygankov}, \&
  {Krivonos}}]{lutovinov13}
{Lutovinov}, A.~A., {Revnivtsev}, M.~G., {Tsygankov}, S.~S., \& {Krivonos},
  R.~A. 2013, \mnras, 431, 327

\bibitem[{{Mangano} {et~al.}(2011){Mangano}, {Barthelmy}, {Romano}, {Sakamoto},
  {Chester}, {Siegel}, {Evans}, {Esposito}, {Vercellone}, {Burrows},
  {Cusumano}, {Farinelli}, {Kennea}, {Krimm}, {La Parola}, \&
  {Gehrels}}]{mangano11}
{Mangano}, V., {Barthelmy}, S.~D., {Romano}, P., {et~al.} 2011, The
  Astronomer's Telegram, 3453

\bibitem[{{Mangano} {et~al.}(2013){Mangano}, {Romano}, {Ceccobello}, \&
  {Farinelli}}]{mangano13}
{Mangano}, V., {Romano}, P., {Ceccobello}, C., \& {Farinelli}, R. 2013,
  \memsai, 84, 604

\bibitem[{{Mart{\'{\i}}nez-N{\'u}{\~n}ez}
  {et~al.}(2017){Mart{\'{\i}}nez-N{\'u}{\~n}ez}, {Kretschmar}, {Bozzo},
  {Oskinova}, {Puls}, {Sidoli}, {Sundqvist}, {Blay}, {Falanga}, {F{\"u}rst},
  {G{\'{\i}}menez-Garc{\'{\i}}a}, {Kreykenbohm}, {K{\"u}hnel}, {Sander},
  {Torrej{\'o}n}, \& {Wilms}}]{nunez17}
{Mart{\'{\i}}nez-N{\'u}{\~n}ez}, S., {Kretschmar}, P., {Bozzo}, E., {et~al.}
  2017, ArXiv e-prints

\bibitem[{{Massa} {et~al.}(2014){Massa}, {Oskinova}, {Fullerton}, {Prinja},
  {Bohlender}, {Morrison}, {Blake}, \& {Pych}}]{massa14}
{Massa}, D., {Oskinova}, L., {Fullerton}, A.~W., {et~al.} 2014, \mnras, 441,
  2173

\bibitem[{{Mullan}(1984)}]{mullan84}
{Mullan}, D.~J. 1984, \apj, 283, 303

\bibitem[{{Naz{\'e}} {et~al.}(2013){Naz{\'e}}, {Oskinova}, \&
  {Gosset}}]{naze13}
{Naz{\'e}}, Y., {Oskinova}, L.~M., \& {Gosset}, E. 2013, \apj, 763, 143

\bibitem[{{Negueruela}(2010)}]{negueruela10}
{Negueruela}, I. 2010, in Astronomical Society of the Pacific Conference
  Series, Vol. 422, High Energy Phenomena in Massive Stars, ed.
  {J.~Mart{\'{\i}}, P.~L.~Luque-Escamilla, \& J.~A.~Combi}, 57

\bibitem[{{Oskinova} {et~al.}(2001){Oskinova}, {Clarke}, \&
  {Pollock}}]{oskinova01}
{Oskinova}, L.~M., {Clarke}, D., \& {Pollock}, A.~M.~T. 2001, \aap, 378, L21

\bibitem[{{Prinja}(1994)}]{prinja94}
{Prinja}, R.~K. 1994, in IAU Symposium, Vol. 162, Pulsation; Rotation; and Mass
  Loss in Early-Type Stars, ed. L.~A. {Balona}, H.~F. {Henrichs}, \& J.~M. {Le
  Contel}, 507

\bibitem[{{Prinja} \& {Smith}(1992)}]{prinja92}
{Prinja}, R.~K. \& {Smith}, L.~J. 1992, \aap, 266, 377

\bibitem[{{Rampy} {et~al.}(2009){Rampy}, {Smith}, \& {Negueruela}}]{rampy09}
{Rampy}, R.~A., {Smith}, D.~M., \& {Negueruela}, I. 2009, \apj, 707, 243

\bibitem[{{Romano} {et~al.}(2009{\natexlab{a}}){Romano}, {Barthelmy}, {Sidoli},
  {Vercellone}, {Burrows}, {Chester}, {Esposito}, {Evans}, {Gehrels}, {Kennea},
  {Krimm}, \& {La Parola}}]{romano09d}
{Romano}, P., {Barthelmy}, S., {Sidoli}, L., {et~al.} 2009{\natexlab{a}}, The
  Astronomer's Telegram, 2191

\bibitem[{{Romano} {et~al.}(2012{\natexlab{a}}){Romano}, {Barthelmy},
  {Chester}, {Oates}, {Burrows}, {Esposito}, {Evans}, {Kennea}, {Krimm},
  {Mangano}, {Vercellone}, \& {Gehrels}}]{romano12}
{Romano}, P., {Barthelmy}, S.~D., {Chester}, M.~M., {et~al.}
  2012{\natexlab{a}}, The Astronomer's Telegram, 4095

\bibitem[{{Romano} {et~al.}(2012{\natexlab{b}}){Romano}, {Barthelmy}, {Kennea},
  {Esposito}, {Evans}, {Mangano}, {Palmer}, {Burrows}, {Chester}, {Krimm},
  {Ukwatta}, {Vercellone}, \& {Gehrels}}]{romano12c}
{Romano}, P., {Barthelmy}, S.~D., {Kennea}, J.~A., {et~al.} 2012{\natexlab{b}},
  The Astronomer's Telegram, 4176

\bibitem[{{Romano} {et~al.}(2015){Romano}, {Bozzo}, {Mangano}, {Esposito},
  {Israel}, {Tiengo}, {Campana}, {Ducci}, {Ferrigno}, \& {Kennea}}]{romano15}
{Romano}, P., {Bozzo}, E., {Mangano}, V., {et~al.} 2015, \aap, 576, L4

\bibitem[{{Romano} {et~al.}(2010){Romano}, {Cusumano}, {Baumgartner}, {Krimm},
  {Sakamoto}, {de Pasquale}, {Barthelmy}, {Burrows}, {Chester}, {Kennea},
  {Evans}, {Esposito}, {Gehrels}, {Palmer}, {La Parola}, \&
  {Vercellone}}]{romano10}
{Romano}, P., {Cusumano}, G., {Baumgartner}, W.~H., {et~al.} 2010, The
  Astronomer's Telegram, 2662

\bibitem[{{Romano} {et~al.}(2008{\natexlab{a}}){Romano}, {Cusumano}, {Sidoli},
  {Vercellone}, {Kuulkers}, {Kennea}, {Mangano}, {Krimm}, {Burrows}, \&
  {Gehrels}}]{romano08}
{Romano}, P., {Cusumano}, G., {Sidoli}, L., {et~al.} 2008{\natexlab{a}}, The
  Astronomer's Telegram, 1697

\bibitem[{{Romano} {et~al.}(2012{\natexlab{c}}){Romano}, {Krimm}, {Sbarufatti},
  {Sakamoto}, {Burrows}, {Barthelmy}, {Chester}, {Kennea}, {Esposito}, {Evans},
  {Gehrels}, {Mangano}, {Palmer}, \& {Vercellone}}]{romano12b}
{Romano}, P., {Krimm}, H., {Sbarufatti}, B., {et~al.} 2012{\natexlab{c}}, The
  Astronomer's Telegram, 4276

\bibitem[{{Romano} {et~al.}(2014){Romano}, {Krimm}, {Palmer}, {Ducci},
  {Esposito}, {Vercellone}, {Evans}, {Guidorzi}, {Mangano}, {Kennea},
  {Barthelmy}, {Burrows}, \& {Gehrels}}]{romano14}
{Romano}, P., {Krimm}, H.~A., {Palmer}, D.~M., {et~al.} 2014, \aap, 562, A2

\bibitem[{{Romano} {et~al.}(2011{\natexlab{a}}){Romano}, {La Parola},
  {Vercellone}, {Cusumano}, {Sidoli}, {Krimm}, {Pagani}, {Esposito},
  {Hoversten}, {Kennea}, {Page}, {Burrows}, \& {Gehrels}}]{romano11}
{Romano}, P., {La Parola}, V., {Vercellone}, S., {et~al.} 2011{\natexlab{a}},
  \mnras, 410, 1825

\bibitem[{{Romano} {et~al.}(2011{\natexlab{b}}){Romano}, {Mangano}, {Cusumano},
  {Esposito}, {Evans}, {Kennea}, {Vercellone}, {La Parola}, {Krimm}, {Burrows},
  \& {Gehrels}}]{romano10b}
{Romano}, P., {Mangano}, V., {Cusumano}, G., {et~al.} 2011{\natexlab{b}},
  \mnras, 412, L30

\bibitem[{{Romano} {et~al.}(2013){Romano}, {Mangano}, {Ducci}, {Esposito},
  {Vercellone}, {Bocchino}, {Burrows}, {Kennea}, {Krimm}, {Gehrels},
  {Farinelli}, \& {Ceccobello}}]{romano13}
{Romano}, P., {Mangano}, V., {Ducci}, L., {et~al.} 2013, Advances in Space
  Research, 52, 1593

\bibitem[{{Romano} {et~al.}(2009{\natexlab{b}}){Romano}, {Sidoli}, {Cusumano},
  {Evans}, {Ducci}, {Krimm}, {Vercellone}, {Page}, {Beardmore}, {Burrows},
  {Kennea}, {Gehrels}, {La Parola}, \& {Mangano}}]{romano09t}
{Romano}, P., {Sidoli}, L., {Cusumano}, G., {et~al.} 2009{\natexlab{b}},
  \mnras, 392, 45

\bibitem[{{Romano} {et~al.}(2009{\natexlab{c}}){Romano}, {Sidoli}, {Krimm},
  {Chester}, {Evans}, {Barthelmy}, {Vercellone}, {La Parola}, {Mangano},
  {Kennea}, {Burrows}, \& {Gehrels}}]{romano09e}
{Romano}, P., {Sidoli}, L., {Krimm}, H.~A., {et~al.} 2009{\natexlab{c}}, The
  Astronomer's Telegram, 2044

\bibitem[{{Romano} {et~al.}(2009{\natexlab{d}}){Romano}, {Sidoli}, {Mangano},
  {Kennea}, {La Parola}, {Cusumano}, {Vercellone}, {Page}, {Burrows}, \&
  {Gehrels}}]{romano09f}
{Romano}, P., {Sidoli}, L., {Mangano}, V., {et~al.} 2009{\natexlab{d}}, The
  Astronomer's Telegram, 1920

\bibitem[{{Romano} {et~al.}(2008{\natexlab{b}}){Romano}, {Sidoli}, {Mangano},
  {Vercellone}, {Kennea}, {Cusumano}, {Krimm}, {Burrows}, \&
  {Gehrels}}]{romano08b}
{Romano}, P., {Sidoli}, L., {Mangano}, V., {et~al.} 2008{\natexlab{b}}, \apjl,
  680, L137

\bibitem[{{Shakura} {et~al.}(2012){Shakura}, {Postnov}, {Kochetkova}, \&
  {Hjalmarsdotter}}]{shakura12}
{Shakura}, N., {Postnov}, K., {Kochetkova}, A., \& {Hjalmarsdotter}, L. 2012,
  \mnras, 420, 216

\bibitem[{{Shakura} {et~al.}(2014){Shakura}, {Postnov}, {Sidoli}, \&
  {Paizis}}]{shakura14}
{Shakura}, N., {Postnov}, k., {Sidoli}, L., \& {Paizis}, A. 2014,
  arXiv/1405.5707

\bibitem[{{Sidoli} {et~al.}(2010){Sidoli}, {Esposito}, \& {Ducci}}]{sidoli10b}
{Sidoli}, L., {Esposito}, P., \& {Ducci}, L. 2010, \mnras, 409, 611

\bibitem[{{Sidoli} {et~al.}(2013){Sidoli}, {Esposito}, {Sguera}, {Bodaghee},
  {Tomsick}, {Pottschmidt}, {Rodriguez}, {Romano}, \& {Wilms}}]{sidoli13}
{Sidoli}, L., {Esposito}, P., {Sguera}, V., {et~al.} 2013, \mnras, 429, 2763

\bibitem[{{Sidoli} {et~al.}(2009{\natexlab{a}}){Sidoli}, {Romano}, {Ducci},
  {Paizis}, {Cusumano}, {Mangano}, {Krimm}, {Vercellone}, {Burrows}, {Kennea},
  \& {Gehrels}}]{sidoli09b}
{Sidoli}, L., {Romano}, P., {Ducci}, L., {et~al.} 2009{\natexlab{a}}, \mnras,
  397, 1528

\bibitem[{{Sidoli} {et~al.}(2009{\natexlab{b}}){Sidoli}, {Romano}, {Esposito},
  {Parola}, {Kennea}, {Krimm}, {Chester}, {Bazzano}, {Burrows}, \&
  {Gehrels}}]{sidoli09}
{Sidoli}, L., {Romano}, P., {Esposito}, P., {et~al.} 2009{\natexlab{b}},
  \mnras, 400, 258

\bibitem[{{Sidoli} {et~al.}(2009{\natexlab{c}}){Sidoli}, {Romano}, {Mangano},
  {Cusumano}, {Vercellone}, {Kennea}, {Paizis}, {Krimm}, {Burrows}, \&
  {Gehrels}}]{sidoli09c}
{Sidoli}, L., {Romano}, P., {Mangano}, V., {et~al.} 2009{\natexlab{c}}, \apj,
  690, 120

\bibitem[{{Sidoli} {et~al.}(2008){Sidoli}, {Romano}, {Mangano}, {Pellizzoni},
  {Kennea}, {Cusumano}, {Vercellone}, {Paizis}, {Burrows}, \&
  {Gehrels}}]{sidoli08}
{Sidoli}, L., {Romano}, P., {Mangano}, V., {et~al.} 2008, \apj, 687, 1230

\bibitem[{{Str{\"u}der} {et~al.}(2001){Str{\"u}der}, {Briel}, {Dennerl},
  {Hartmann}, {Kendziorra}, {Meidinger}, {Pfeffermann}, {Reppin}, {Aschenbach},
  {Bornemann}, {Br{\"a}uninger}, {Burkert}, {Elender}, {Freyberg}, {Haberl},
  {Hartner}, {Heuschmann}, {Hippmann}, {Kastelic}, {Kemmer}, {Kettenring},
  {Kink}, {Krause}, {M{\"u}ller}, {Oppitz}, {Pietsch}, {Popp}, {Predehl},
  {Read}, {Stephan}, {St{\"o}tter}, {Tr{\"u}mper}, {Holl}, {Kemmer}, {Soltau},
  {St{\"o}tter}, {Weber}, {Weichert}, {von Zanthier}, {Carathanassis}, {Lutz},
  {Richter}, {Solc}, {B{\"o}ttcher}, {Kuster}, {Staubert}, {Abbey}, {Holland},
  {Turner}, {Balasini}, {Bignami}, {La Palombara}, {Villa}, {Buttler},
  {Gianini}, {Lain{\'e}}, {Lumb}, \& {Dhez}}]{struder01}
{Str{\"u}der}, L., {Briel}, U., {Dennerl}, K., {et~al.} 2001, \aap, 365, L18

\bibitem[{{Tomsick} {et~al.}(2009){Tomsick}, {Chaty}, {Rodriguez}, {Walter},
  {Kaaret}, \& {Tovmassian}}]{tomsick09b}
{Tomsick}, J.~A., {Chaty}, S., {Rodriguez}, J., {et~al.} 2009, \apj, 694, 344

\bibitem[{{Torrej{\'o}n} {et~al.}(2010){Torrej{\'o}n}, {Schulz}, {Nowak}, \&
  {Kallman}}]{torrejon10b}
{Torrej{\'o}n}, J.~M., {Schulz}, N.~S., {Nowak}, M.~A., \& {Kallman}, T.~R.
  2010, \apj, 715, 947

\bibitem[{{Turner} {et~al.}(2001){Turner}, {Abbey}, {Arnaud}, {Balasini},
  {Barbera}, {Belsole}, {Bennie}, {Bernard}, {Bignami}, {Boer}, {Briel},
  {Butler}, {Cara}, {Chabaud}, {Cole}, {Collura}, {Conte}, {Cros}, {Denby},
  {Dhez}, {Di Coco}, {Dowson}, {Ferrando}, {Ghizzardi}, {Gianotti}, {Goodall},
  {Gretton}, {Griffiths}, {Hainaut}, {Hochedez}, {Holland}, {Jourdain},
  {Kendziorra}, {Lagostina}, {Laine}, {La Palombara}, {Lortholary}, {Lumb},
  {Marty}, {Molendi}, {Pigot}, {Poindron}, {Pounds}, {Reeves}, {Reppin},
  {Rothenflug}, {Salvetat}, {Sauvageot}, {Schmitt}, {Sembay}, {Short},
  {Spragg}, {Stephen}, {Str{\"u}der}, {Tiengo}, {Trifoglio}, {Tr{\"u}mper},
  {Vercellone}, {Vigroux}, {Villa}, {Ward}, {Whitehead}, \& {Zonca}}]{turner01}
{Turner}, M.~J.~L., {Abbey}, A., {Arnaud}, M., {et~al.} 2001, \aap, 365, L27

\bibitem[{{Walter} {et~al.}(2015){Walter}, {Lutovinov}, {Bozzo}, \&
  {Tsygankov}}]{walter15}
{Walter}, R., {Lutovinov}, A.~A., {Bozzo}, E., \& {Tsygankov}, S.~S. 2015,
  \aapr, 23, 2

\bibitem[{{Walter} \& {Zurita Heras}(2007)}]{walter07}
{Walter}, R. \& {Zurita Heras}, J. 2007, \aap, 476, 335

\bibitem[{{Zurita Heras} \& {Walter}(2009)}]{zurita09}
{Zurita Heras}, J.~A. \& {Walter}, R. 2009, \aap, 494, 1013

\end{thebibliography}

\newpage

\appendix

\section{Detailed results of the fits to the HR-resolved spectral analysis}

We report in this section all tables containing the detailed results of the HR-resolved spectral analysis 
performed for all sources and all observations. 
 \begin{table}[h!]
 \scriptsize
 \begin{center} 	
 \caption{Results of the fits to the A-Z spectra of IGR\,J18450-0435 in the observation ID.~0306170401 (see Fig.~\ref{fig:j1845}, top) 
 and the A-H spectra of the same source in the observation ID.~0728370801 (see Fig.~\ref{fig:ax801}, second plot from the top). 
 The absorption column density is measured in units of 10$^{22}$~cm$^{-2}$, while the 1-10~keV flux is given in units of 
 10$^{-11}$~erg~cm$^{-2}$~s$^{-1}$.} 	
 \label{tab:j1845_fit2} 
	\resizebox{\columnwidth}{!}{%
 \begin{tabular}{@{}llllllll@{}} 
 \hline 
 \hline 
 \noalign{\smallskip}  
 S.  & $N_{\rm H}$ & $\Gamma$ & Flux & C$_{\rm MOS1}$ & C$_{\rm MOS2}$ & $\chi^2_{\rm red}$/ & Exp  \\ 
            &  &  &  & & & d.o.f. & (ks)      \\
 \noalign{\smallskip}
  \hline 
 \multicolumn{8}{c}{Observation ID.~0306170401} \\
 \hline 
 \noalign{\smallskip} 
A$^a$	&	3.0$\pm$0.6 &	1.0$\pm$0.2 &	4.4$^{+0.4}_{-0.3}$ &	1.1$\pm$0.2 & 1.1$\pm$0.1 & 1.40/69	& 0.32 \\
\noalign{\smallskip}
B	&	3.5$\pm$0.5 &	1.4$\pm$0.2 &	12.1$^{+1.2}_{-1.0}$ &	1.1$\pm$0.1 & 1.0$\pm$0.1 & 1.19/81	& 0.17	\\
\noalign{\smallskip}
C	&	2.6$\pm$0.3 &	1.3$\pm$0.1 &	12.7$^{+0.9}_{-0.8}$ &	1.2$\pm$0.2 & 1.2$\pm$0.1 & 0.96/82	& 0.25	\\
\noalign{\smallskip}
D	&	3.1$\pm$0.4 &	1.4$\pm$0.1 &	12.8$^{+1.0}_{-0.9}$ &	1.2$\pm$0.1 & 0.9$\pm$0.1 & 1.18/116 & 0.27	\\
\noalign{\smallskip}
E	&	1.1$\pm$0.4 &	0.6$\pm$0.2 &	2.1$\pm$0.2 &	1.0$\pm$0.2 & 0.9$\pm$0.1 & 1.20/61	& 0.60 \\
\noalign{\smallskip}
F	&	1.9$\pm$0.4 &	0.9$\pm$0.2 &	3.5$\pm$0.2 &	1.0$\pm$0.2 & 0.9$\pm$0.1 & 0.79/64	& 0.35 \\
\noalign{\smallskip}
G	&	3.1$\pm$0.5 & 1.4$\pm$0.2 &	9.4$^{+1.0}_{-0.8}$ & 1.3$\pm$0.2 & 1.4$\pm$0.2 & 0.90/52	& 0.20	\\
\noalign{\smallskip}
H	&	2.5$\pm$0.7 & 1.3$\pm$0.2 &	18.8$^{+2.3}_{-2.0}$ & 1.2$\pm$0.2 & 1.1$\pm$0.2 & 0.90/37	& 0.09	\\
\noalign{\smallskip}
L$^a$	&	2.5$\pm$0.9 & 1.4$\pm$0.3 & 14.9$^{+2.5}_{-2.0}$ & 1.3$\pm$0.2 & 1.1$\pm$0.2 & 1.47/20	& 0.09	\\		
\noalign{\smallskip}
M	&	2.2$\pm$0.2 & 1.3$\pm$0.1 & 17.5$\pm$0.8 & 0.97$\pm$0.06 & 0.93$\pm$0.06 & 1.10/176	& 0.35	\\		
\noalign{\smallskip}
N$^a$	&	2.4$\pm$0.5 & 1.1$\pm$0.2 & 8.2$^{+0.8}_{-0.7}$ & 1.1$\pm$0.1 & 0.9$\pm$0.1 & 1.23/58	& 0.26	\\		
\noalign{\smallskip}
O	&	3.7$\pm$0.6 & 1.5$\pm$0.2 &	14.0$^{+1.6}_{-1.3}$ & 1.3$\pm$0.2 & 1.3$\pm$0.2 & 0.97/58	& 0.15	\\
\noalign{\smallskip}
P	&	2.7$\pm$0.5 & 1.3$\pm$0.2 &	17.6$^{+1.5}_{-1.3}$ & 1.1$\pm$0.2 & 1.2$\pm$0.2 & 0.98/74	& 0.18	\\
\noalign{\smallskip}
Q$^a$	&	2.1$\pm$0.6 & 1.1$\pm$0.2 &	18.0$^{+1.9}_{-1.7}$ & 1.1$\pm$0.1 & 1.0$\pm$0.1 & 1.26/42	& 0.10	\\
\noalign{\smallskip}
R	&	2.1$\pm$0.2 & 1.14$\pm$0.07 &	22.2$\pm$0.8 & 1.15$\pm$0.06 & 1.0$\pm$0.06 & 1.18/219	& 0.62	\\
\noalign{\smallskip}
S	&	1.9$\pm$0.2 & 1.1$\pm$0.1 &	17.8$^{+0.7}_{-0.6}$ & 1.1$\pm$0.1 & 1.0$\pm$0.1 & 1.15/166 & 0.33	\\
\noalign{\smallskip}
T	&	2.1$\pm$0.2 & 1.1$\pm$0.1 &	12.2$^{+0.5}_{-0.4}$ & 1.1$\pm$0.1 & 1.1$\pm$0.1 & 1.11/208 & 0.62	\\
\noalign{\smallskip}
Z$^b$	&	2.8$\pm$0.1 & 1.14$\pm$0.04 &	1.94$\pm$0.04 & 1.05$\pm$0.03 & 1.00$\pm$0.03 & 1.50/357 & 11.72	\\
\noalign{\smallskip} 
\hline
   \multicolumn{8}{c}{Observation ID.~0728370801} \\
 \hline 
\noalign{\smallskip}
A	&	6.5$_{-1.1}^{+1.2}$ & 1.6$\pm$0.3 & 0.39$^{+0.07}_{-0.05}$ &	1.1$\pm$0.1 & 1.0$\pm$0.1 & 1.01/78	& 7.1 \\
\noalign{\smallskip}
B	&	5.9$\pm$0.6 & 1.1$\pm$0.1 & 2.7$^{+0.2}_{-0.1}$ & 0.98$\pm$0.07 & 0.97$\pm$0.07 & 1.08/170	& 2.4 	\\
\noalign{\smallskip}
C	&	8.2$_{-1.3}^{+1.5}$ & 1.2$\pm$0.2 & 6.9$^{+1.0}_{-0.7}$ &	1.0$\pm$0.1 & 1.1$\pm$0.1 & 0.86/80	& 0.5  \\
\noalign{\smallskip}
D	&	5.6$\pm$0.6 & 1.1$\pm$0.1 & 11.8$^{+0.8}_{-0.7}$ &	1.09$\pm$0.08 & 0.98$\pm$0.08 & 0.97/131	& 0.5 	\\
\noalign{\smallskip}
E	&	7.1$_{-1.0}^{+1.1}$ & 1.8$\pm$0.3 & 26.0$^{+4.9}_{-3.5}$ &	1.1$\pm$0.1 & 1.0$\pm$0.1 & 1.08/78	& 0.2 \\		
\noalign{\smallskip}
F	&	6.5$_{-0.7}^{+0.8}$ &  1.1$\pm$0.2 & 1.60$^{+0.07}_{-0.06}$ & 1.08$\pm$0.08 & 1.07$\pm$0.08 & 1.14/161	& 3.9 	\\
\noalign{\smallskip}
G	&	3.9$\pm$0.4 &	0.8$\pm$0.1 & 1.6$^{+0.2}_{-0.1}$ &	0.92$\pm$0.05 & 0.92$\pm$0.05 & 1.08/215	& 4.9 	\\
\noalign{\smallskip}
H	&	6.9$_{-2.8}^{+4.1}$ & 1.2$_{-0.6}^{+0.7}$ & 0.5$^{+0.3}_{-0.1}$ &	1.2$\pm$0.2 & 1.0$\pm$0.2 & 0.46/17	& 1.5 	\\
\noalign{\smallskip} 
  \hline
  \hline
  \end{tabular}
  }
  \tablefoot{$^a$: In these cases we verified that the slightly large $\chi^2_{\rm red}$ 
  were only due to the noise in reduced statistics spectra rather than evident trends in the residuals 
  (that could have suggested that a more refined model was needed to fit the data). 
  $^b$: This spectrum could not be well fit by using a simple absorbed power-law model. 
  See Sect.~\ref{sec:AXJ1845} for details.}
  \end{center}
  \end{table} 

 \begin{table}[h!]
   \scriptsize	
 \begin{center} 	
 \caption{Same as Table~\ref{tab:j1845_fit2}, but in the case of the IGR\,J17544-2619 
 observation ID.~0148090501 (Fig.~\ref{fig:J17544_lcurve}), ID.~0154750601 
 (Fig.~\ref{fig:J17544b_lcurve}), and ID.~0679810401 (Fig.~\ref{fig:J17544c_lcurve}).} 	
 \label{tab:J17544_fit} 	
\resizebox{\columnwidth}{!}{%
 \begin{tabular}{@{}llllllll@{}} 
 \hline 
 \hline 
 \noalign{\smallskip}  
 S.   & $N_{\rm H}$$^a$ & $\Gamma$ & Flux$^b$ & C$_{\rm MOS1}$ & C$_{\rm MOS2}$ & $\chi^2_{\rm red}$/ & Exp  \\ 
            &  &  &  & & & d.o.f. & (ks)      \\
 \noalign{\smallskip} 
 \hline 
\multicolumn{8}{c}{Observation ID.~0148090501} \\
 \hline 
 \noalign{\smallskip} 
A	&	2.1$_{-0.8}^{1.3}$ & 2.6$_{-0.6}^{+0.8}$ &	0.09$_{-0.02}^{+0.11}$ & 1.2$\pm$0.4 & 1.3$\pm$0.6 & 0.60/8	& 2.05 \\
\noalign{\smallskip}
B	&	2.8$_{-0.9}^{+1.2}$ &	1.7$\pm$0.5 &	0.6$_{-0.1}^{+0.2}$ & 1.1$\pm$0.3 & 1.1$\pm$0.4 & 1.05/51	& 0.42	\\
\noalign{\smallskip}
C	&	3.6$\pm$0.6 &	1.8$\pm$0.3 &	4.6$_{-0.5}^{+0.7}$ & 1.0$\pm$0.1 & 1.4$\pm$0.2 & 1.02/62	& 0.32	\\
\noalign{\smallskip}
D	&	4.2$_{-0.8}^{+1.0}$ &	2.2$\pm$0.4 &	3.5$_{-0.7}^{+1.2}$ & 1.0$\pm$0.2 & 1.2$\pm$0.3 & 0.97/30	& 0.27 \\
\noalign{\smallskip}
E	&	3.5$\pm$1.1 & 2.4$\pm$0.7 & 3.7$_{-1.1}^{+2.6}$ &	1.2$\pm$0.3 & 1.4$\pm$0.5 & 0.60/10	& 0.10	\\
\noalign{\smallskip}
F	&	2.7$\pm$0.5 & 1.8$\pm$0.2 &	2.2$_{-0.2}^{+0.3}$ & 1.1$\pm$0.1 & 1.5$\pm$0.3 & 0.95/69	& 0.60	\\		
\noalign{\smallskip}
G	&	2.8$\pm$0.5 & 2.4$\pm$0.3 &	0.53$_{-0.09}^{+0.13}$ & 1.2$\pm$0.2 & 1.4$\pm$0.3 & 1.14/53	& 1.72	\\
\noalign{\smallskip}
  \hline
  \multicolumn{8}{c}{Observation ID.~0154750601} \\
    \hline 
 \noalign{\smallskip} 
A$^a$	&	2.5$\pm$0.5 &	1.8$\pm$0.3 &	5.4$_{-0.7}^{+0.9}$ & 1.0 & 1.0$\pm$0.1 & 1.11/52	& 0.32 \\
\noalign{\smallskip}
B$^a$	&	2.3$_{-0.5}^{+0.7}$ & 1.4$\pm$0.3 & 8.0$_{-1.0}^{+1.2}$ &	1.0 & 1.0$\pm$0.1 & 0.95/33	& 0.15	\\
\noalign{\smallskip}
C$^a$	&	4.0$\pm$1.5 & 2.3$\pm$0.6 &	5.0$_{-1.4}^{+2.9}$ & 1.0 & 1.0$\pm$0.2 & 1.50/19	& 0.20	\\		
\noalign{\smallskip}
D	&	2.4$_{-0.4}^{0.6}$ & 2.0$\pm$0.5 & 2.6$_{-0.4}^{+0.5}$ & 0.9$\pm$0.1 & 0.9$\pm$0.1 & 0.82/30	& 0.30 \\
\noalign{\smallskip}
  \hline
    \multicolumn{8}{c}{Observation ID.~0679810401} \\
 \hline 
 \noalign{\smallskip} 
A	&	2.0$_{-0.5}^{0.6}$ & 2.1$\pm$0.3 & 0.22$\pm$0.04 & 0.9$\pm$0.2 & 0.9$\pm$0.2 & 0.94/27	& 1.40 \\
\noalign{\smallskip}
B	&	1.7$\pm$0.2 & 2.0$\pm$0.1 & 0.22$\pm$0.02 & 0.97$\pm$0.09 & 0.99$\pm$0.09 & 1.01/130	& 8.07 \\
\noalign{\smallskip}
C	&	1.5$\pm$0.3 &	1.8$\pm$0.3 &	0.29$\pm$0.04 & 0.9$\pm$0.1 & 1.0$\pm$0.1 & 1.17/33	& 1.15	\\
\noalign{\smallskip}
D	&	2.2$\pm$0.3 &	1.8$\pm$0.2 &	1.2$\pm$0.1 & 1.0$\pm$0.1 & 1.1$\pm$0.1 & 0.92/80	& 0.87 \\
\noalign{\smallskip}
E	&	1.9$\pm$0.4 & 1.7$\pm$0.2 & 1.0$\pm$0.1 & 1.1$\pm$0.2 & 1.1$\pm$0.2 & 0.87/45	& 0.56 \\
\noalign{\smallskip}
  \hline
\hline
  \end{tabular}
  }
  \tablefoot{$a$: For these spectra 
  no EPIC-pn data were available and thus the normalization constant of the MOS1 was fixed to unity 
  and only that of the MOS2 was left free to vary in the fit.}
  \end{center}
  \end{table} 
  
 \begin{table}[h!]
   \scriptsize	
 \begin{center} 	
 \caption{Same as Table~\ref{tab:j1845_fit2}, but in the case of the SAX\,J1818.6-1703 observation ID.~0679810501.} 	
 \label{tab:J1818_fit} 	
 \resizebox{\columnwidth}{!}{%
 \begin{tabular}{@{}llllllll@{}} 
 \hline 
 \hline 
 \noalign{\smallskip}  
 S.   & $N_{\rm H}$ & $\Gamma$ & Flux & C$_{\rm MOS1}$ & C$_{\rm MOS2}$ & $\chi^2_{\rm red}$/ & Exp  \\ 
            &  &  &  & & & d.o.f.& (ks)      \\
 \noalign{\smallskip} 
 \hline 
 \noalign{\smallskip} 
A	&	32.9$\pm$4.5 & 2.8$\pm$0.4 & 3.3$_{-1.3}^{+2.7}$ & 0.97$\pm$0.1 & 1.0$\pm$0.1 & 1.05/85	& 3.8 \\
\noalign{\smallskip}
B  &  40.6$\pm$5.5 & 3.0$\pm$0.5 & 21.2$_{-9.5}^{+21.7}$ & 1.0$\pm$0.1 & 1.0$\pm$0.1 & 0.94/69	& 0.9 \\
\noalign{\smallskip}
C  &  28.3$\pm$4.0 & 0.8$\pm$0.3 & 4.4$_{-0.7}^{+0.9}$ & 1.0$\pm$0.1 & 1.0$\pm$0.1 & 0.86/93	& 1.2 \\
\noalign{\smallskip}
D  &  31.2$\pm$3.5 & 0.8$\pm$0.2 & 6.4$_{-0.8}^{+1.0}$ & 0.93$\pm$0.08 & 0.97$\pm$0.09 & 1.11/123	& 1.1 \\
\noalign{\smallskip}
E  &  24.2$\pm$4.2 & 0.0$\pm$0.3 & 16.9$_{-2.0}^{+2.6}$ & --- & --- & 0.84/66	& 0.4 \\
\noalign{\smallskip}
F  &  21.0$\pm$3.5 & 0.0$\pm$0.2 & 25.3$_{-2.7}^{+3.4}$ & --- & --- & 1.14/75	& 0.3\\
\noalign{\smallskip}
G  &  28.4$\pm$2.1 & 0.2$\pm$0.1 & 28.4$_{-1.9}^{+2.1}$ & --- & --- & 1.05/114 & 1.3\\
\noalign{\smallskip}
  \hline
  \end{tabular}
  }
  \end{center}
  \end{table}
  
 \begin{table}[h!]
   \scriptsize	
 \begin{center} 	
 \caption{Same as Table~\ref{tab:j1845_fit2}, but in the case of the IGR\,J17354-3255 observation 
 ID.~0693900201 (see Fig.~\ref{fig:J17354_lc}), ID.~0701230101 (see Fig.~\ref{fig:J17354_lc2}), 
 and ID.~0701230701 (see Fig.~\ref{fig:J17354_lc3}).} 	
 \label{tab:J17354_fit}
   \resizebox{\columnwidth}{!}{%
 \begin{tabular}{@{}llllllll@{}} 
 \hline 
 \hline 
 \noalign{\smallskip}  
 S.   & $N_{\rm H}$$^a$ & $\Gamma$ & Flux$^b$ & C$_{\rm MOS1}$ & C$_{\rm MOS2}$ & $\chi^2_{\rm red}$/ & Exp  \\ 
            &  &  &  & &  & d.o.f.& (ks)      \\
 \noalign{\smallskip} 
 \hline 
       \multicolumn{8}{c}{Observation ID.~0693900201} \\
  \hline 
 \noalign{\smallskip} 
A$^a$	&	7.7$\pm$0.2 & 1.21$\pm$0.05 & 7.5$^{+0.3}_{-0.2}$ & --- & --- & 0.94/673	& 7.5 \\
\noalign{\smallskip}
B$^a$  &  11.7$\pm$0.7 & 1.4$\pm$0.1 & 3.3$^{+0.4}_{-0.3}$ & --- & ---  & 1.00/217	& 5.9 \\
\noalign{\smallskip}
C$^a$  &  22.6$\pm$2.0 & 1.2$\pm$0.2 & 5.1$^{+1.2}_{-0.8}$ & --- & ---  & 1.04/108	& 2.8 \\
\noalign{\smallskip}
D$^a$  &  12.2$\pm$1.1 & 1.1$\pm$0.2 & 4.1$^{+0.6}_{-0.5}$ & --- & ---  & 0.96/123	& 2.6 \\
\noalign{\smallskip}
E$^a$  &  8.9$\pm$0.3 & 1.32$\pm$0.06 & 7.9$\pm$0.3 & --- & ---  & 0.98/692	& 7.9 \\
\noalign{\smallskip}
  \hline
      \multicolumn{8}{c}{Observation ID.~0701230101} \\
  \hline 
 \noalign{\smallskip} 
A	&	6.6$^{+1.9}_{-1.5}$ & 1.5$\pm$0.4 & 0.9$^{+0.3}_{-0.2}$ & 1.1$\pm$0.2 & 0.9$\pm$0.2  & 0.94/27	& 0.9 \\
\noalign{\smallskip}
B  &  8.6$^{+1.5}_{-1.3}$ & 1.4$\pm$0.2 & 2.0$\pm$0.3 & 1.1$\pm$0.1 & 1.0$\pm$0.1 & 1.04/76	& 1.2 \\
\noalign{\smallskip}
C  &  5.8$\pm$0.5 & 1.1$\pm$0.1 & 4.1$\pm$0.2 & 1.05$\pm$0.07 & 1.00$\pm$0.06  & 0.82/172	& 1.6 \\
\noalign{\smallskip}
D  &  5.4$^{+1.7}_{-1.4}$ & 1.2$\pm$0.4 & 0.6$\pm$0.1 & 1.0$\pm$0.2 & 1.0$\pm$0.2  & 0.59/22	& 1.1 \\
\noalign{\smallskip}
E  &  7.7$^{+1.1}_{-1.0}$ & 1.2$\pm$0.2 & 1.0$\pm$0.1 & 1.0$\pm$0.1 & 1.0$\pm$0.1  & 0.79/97	& 3.3 \\
\noalign{\smallskip}
F  &  6.4$^{+0.6}_{-0.5}$ & 1.3$\pm$0.1 & 3.1$\pm$0.2 & 1.07$\pm$0.07 & 1.09$\pm$0.07  & 1.17/188	& 2.1 \\
\noalign{\smallskip}
G  &  7.7$^{+3.0}_{-2.6}$ & 1.1$\pm$0.4 & 1.6$^{+0.5}_{-0.3}$ & 1.0$\pm$0.2 & 1.0$\pm$0.2  & 0.52/28	& 0.7 \\
\noalign{\smallskip}
H  &  5.8$^{+1.8}_{-1.4}$ & 1.1$\pm$0.3 & 2.0$^{+0.4}_{-0.3}$ & 1.2$\pm$0.2 & 1.1$\pm$0.2  & 1.23/26	& 0.4 \\
\noalign{\smallskip}
L  &  9.0$^{+1.3}_{-1.1}$ & 1.5$\pm$0.2 & 1.8$^{+0.3}_{-0.2}$ & 1.0$\pm$0.1 & 1.1$\pm$0.1  & 0.88/107	& 2.0 \\
\noalign{\smallskip}
M  &  6.0$^{+1.3}_{-1.0}$ & 1.0$\pm$0.3 & 2.3$^{+0.3}_{-0.2}$ & 1.0$\pm$0.1 & 0.9$\pm$0.1  & 0.70/55	& 0.7 \\
\noalign{\smallskip}
  \hline  
        \multicolumn{8}{c}{Observation ID.~0701230701} \\
   \hline 
 \noalign{\smallskip} 
A	&	4.9$\pm$0.4 & 1.4$\pm$0.1 & 2.4$\pm$0.1 & 1.07$\pm$0.07 & 1.05$\pm$0.07  & 0.95/175	& 2.3 \\
\noalign{\smallskip}
B  &  6.4$\pm$0.9 & 1.4$\pm$0.2 & 1.9$\pm$0.2 & 1.1$\pm$0.1 &  1.1$\pm$0.1 & 1.01/56 & 0.9 \\
\noalign{\smallskip}
C  &  5.0$\pm$0.6 & 1.2$\pm$0.2 & 2.0$\pm$0.2 & 1.2$\pm$0.1 & 1.1$\pm$0.1 & 1.26/103	& 1.2 \\
\noalign{\smallskip}
D  &  5.0$\pm$0.6 & 1.4$\pm$0.2 & 1.2$\pm$0.2 & 1.2$\pm$0.1 & 1.1$\pm$0.1  & 1.09/93	& 1.8 \\
\noalign{\smallskip}
E  &  4.9$^{+1.5}_{-1.3}$ & 0.9$\pm$0.3 & 0.9$\pm$0.1 & 1.0$\pm$0.2 & 1.1$\pm$0.2  & 0.77/39	& 1.7 \\
\noalign{\smallskip}
F  &  5.4$\pm$0.3 & 1.3$\pm$0.1 & 3.8$\pm$0.2 & 1.00$\pm$0.05 & 1.00$\pm$0.05 & 1.06/237	& 2.5 \\
\noalign{\smallskip}
G  &  6.1$^{+1.6}_{-1.4}$ & 1.1$\pm$0.3 & 3.7$^{+0.6}_{-0.4}$ & 1.0$\pm$0.1 & 0.9$\pm$0.1  & 1.01/49	& 0.4 \\
\noalign{\smallskip}
H  & 8.4$^{+1.8}_{-1.5}$  & 1.1$\pm$0.3 & 4.2$^{+0.7}_{-0.5}$ & 1.0$\pm$0.1 & 1.0$\pm$0.1 & 1.02/62	& 0.6 \\
\noalign{\smallskip}
L  &  5.3$\pm$0.5 & 0.8$\pm$0.1 & 5.7$\pm$0.3 & 1.02$\pm$0.07 & 1.04$\pm$0.07 & 1.02/187	& 1.1 \\
\noalign{\smallskip}
M  &  5.8$\pm$0.4 & 1.1$\pm$0.1 & 5.0$\pm$0.2 & 1.05$\pm$0.05 & 1.03$\pm$0.05 & 1.24/236	& 1.8 \\
\noalign{\smallskip}
N  &  6.4$^{+0.6}_{-0.5}$ & 1.3$\pm$0.1 & 2.8$\pm$0.2 & 1.00$\pm$0.06 & 1.04$\pm$0.06  & 0.98/186	& 2.5 \\
\noalign{\smallskip}
O  &  5.2$\pm$0.4 & 1.4$\pm$0.1 & 4.5$\pm$0.3 & 1.05$\pm$0.07 & 1.01$\pm$0.07 & 1.03/167	& 1.2 \\
\noalign{\smallskip}
    \hline 
     \hline 
  \end{tabular}
  }
    \tablefoot{$a$: For these spectra 
  no EPIC-pn data were available and the two MOS spectra were merged together to improve the S/N.}
  \end{center}
  \end{table}
 
 \begin{table}[h!]
   \scriptsize	
 \begin{center} 	
 \caption{Same as Table~\ref{tab:j1845_fit2}, but in the case of the IGR\,J16328-4732 observation ID.~0679810201
 (see Fig.~\ref{fig:J16328_lc}) and ID.~0679810301 (see Fig.~\ref{fig:J16328_lc2}).}	
 \label{tab:J16328_fit} 	
  \resizebox{\columnwidth}{!}{%
 \begin{tabular}{@{}lllllllll@{}} 
 \hline 
 \hline 
 \noalign{\smallskip}  
 S.   & $N_{\rm H}$ & $\Gamma$ & Flux & C$_{\rm MOS1}$ & C$_{\rm MOS2}$ & $\chi^2_{\rm red}$/ & Exp  \\ 
            &  &  &  & & & d.o.f.& (ks)      \\
 \noalign{\smallskip} 
 \hline 
        \multicolumn{8}{c}{Observation ID.~0679810201} \\ 
  \hline 
 \noalign{\smallskip} 
A	&	18.0$\pm$0.7 & 1.3$\pm$0.1 & 15.6$_{-0.8}^{+0.9}$ & 0.90$\pm$0.03 & 1.1$\pm$0.04 & 0.91/279	& 2.62 \\
\noalign{\smallskip}
B  &  16.2$_{-3.8}^{+4.4}$ & 1.7$\pm$0.5 & 2.0$_{-0.5}^{+1.1}$ & 1.0$\pm$0.2 & 1.2$\pm$0.2 & 0.82/38 & 1.10 \\
\noalign{\smallskip}
C  &  21.8$\pm$1.5 & 1.8$\pm$0.2 & 3.1$_{-0.4}^{+0.5}$ & 1.03$\pm$0.06 & 1.03$\pm$0.06 & 0.89/168	& 5.17 \\
\noalign{\smallskip}
D  &  18.2$\pm$1.7 & 1.8$\pm$0.2 & 3.7$_{-0.5}^{+0.7}$ & 1.06$\pm$0.08 & 1.06$\pm$0.08 & 1.02/135 & 2.57 \\
\noalign{\smallskip}
E  &  11.5$_{-4.8}^{+5.4}$ & 1.2$\pm$0.5 & 7.0$_{-1.7}^{+3.4}$ & 1.1$\pm$0.2 & 1.2$\pm$0.2 & 0.70/25	& 0.18 \\
\noalign{\smallskip}
F  &  16.6$\pm$2.0 & 1.3$\pm$0.2 & 9.2$_{-1.4}^{+2.0}$ & 0.96$\pm$0.09 & 1.09$\pm$0.09 & 0.80/106	& 0.68 \\
\noalign{\smallskip}
G  &  18.8$\pm$3.5 & 1.7$\pm$0.4 & 12.7$_{-3.1}^{+5.3}$ & 1.1$\pm$0.1 & 1.3$\pm$0.2 & 0.94/51	& 0.27 \\
\noalign{\smallskip}
H  &  13.6$\pm$1.2 & 1.5$\pm$0.2 & 4.3$_{-0.4}^{+0.6}$ & 1.00$\pm$0.07 & 1.00$\pm$0.07 & 0.99/145	& 1.92 \\
\noalign{\smallskip}
L  &  11.8$_{-3.0}^{+3.8}$ & 1.1$\pm$0.4 & 0.8$_{-0.1}^{+0.2}$ & 1.1$\pm$0.2 & 1.1$\pm$0.2 & 0.79/41	& 2.18 \\
\noalign{\smallskip}
M  &  14.1$_{-3.7}^{+4.0}$ & 1.1$\pm$0.4 & 5.8$_{-1.1}^{+1.8}$ & 1.0$\pm$0.1 & 1.3$\pm$0.2 & 0.90/43	& 0.33 \\
\noalign{\smallskip}
N  &  11.9$\pm$0.9 & 1.1$\pm$0.1 & 8.4$\pm$0.5 & 0.94$\pm$0.5 & 1.02$\pm$0.05 & 0.90/217	& 1.60 \\
\noalign{\smallskip}
O  &  15.1$_{-4.1}^{+4.5}$ & 1.2$\pm$0.4 & 4.9$_{-1.1}^{+1.8}$ & 1.0$\pm$0.1 & 1.0$\pm$0.1 & 0.82/43	& 0.44 \\
\noalign{\smallskip}
P  &  14.8$_{-2.9}^{+3.4}$ & 1.5$\pm$0.4 & 1.1$_{-0.2}^{+0.4}$ & 1.1$\pm$0.1 & 1.0$\pm$0.1 & 0.97/47	& 2.25 \\
\noalign{\smallskip}
Q  &  14.7$_{-3.6}^{+4.2}$ & 1.5$\pm$0.5 & 1.3$_{-0.3}^{+0.6}$ & 1.1$\pm$0.2 & 1.2$\pm$0.2 & 0.68/34	& 1.44 \\
\noalign{\smallskip}
R  &  16.5$_{-3.7}^{+4.1}$ & 1.8$\pm$0.4 & 2.2$_{-0.6}^{+1.1}$ & 0.9$\pm$0.1 & 0.9$\pm$0.1 & 1.16/48	& 1.39 \\
\noalign{\smallskip}
S  &  14.6$\pm$1.1 & 1.4$\pm$0.2 & 1.9$\pm$0.2 & 1.02$\pm$0.07 & 1.03$\pm$0.07 & 0.96/177	& 5.49 \\
\noalign{\smallskip}
  \hline
        \multicolumn{8}{c}{Observation ID.~0679810301} \\ 
  \hline
  \noalign{\smallskip}
A	&	14.4$_{-3.6}^{+3.9}$ & 1.2$\pm$0.5 & 1.8$^{+0.7}_{-0.4}$ & 1.0$\pm$0.2 & 1.0$\pm$0.2 & 1.18/32	& 1.04 \\
\noalign{\smallskip}
B  &  12.3$_{-3.2}^{+4.1}$ & 1.6$\pm$0.5 & 0.5$^{+0.3}_{-0.1}$ & 1.0$\pm$0.2 & 1.0$\pm$0.2 & 0.97/32 & 3.27 \\
\noalign{\smallskip}
C  &  16.0$_{-3.8}^{+4.3}$ & 1.7$\pm$0.5 & 2.2$^{+1.2}_{-0.6}$ & 1.1$\pm$0.2 & 1.2$\pm$0.2 & 1.08/36	& 1.18 \\
\noalign{\smallskip}
D  &  $<$7.0 & 0.3$^{+0.6}_{-0.4}$ & 1.4$^{+0.3}_{-0.2}$ & 1.2$\pm$0.3 & 1.4$\pm$0.3 & 1.00/19 & 0.46 \\
\noalign{\smallskip}
E  &  10.5$_{-1.4}^{+1.6}$ & 1.1$\pm$0.2 & 2.3$^{+0.3}_{-0.2}$ & 1.00$\pm$0.09 & 1.00$\pm$0.09 & 1.33/113	& 2.15 \\
\noalign{\smallskip}
F  &  9.0$_{-3.0}^{+3.5}$ & 0.8$\pm$0.4 & 2.2$^{+0.5}_{-0.3}$ & 1.0$\pm$0.2 & 1.0$\pm$0.2 & 1.00/34	& 0.78 \\
\noalign{\smallskip}
G  &  13.6$_{-3.1}^{+3.4}$ & 1.2$\pm$0.4 & 4.8$^{+1.3}_{-0.9}$ & 0.8$\pm$0.1 & 1.1$\pm$0.1 & 0.80/48	& 0.50 \\
\noalign{\smallskip}
H  & 14.5$_{-2.9}^{+3.4}$  & 1.6$\pm$0.4 & 14.6$^{+5.4}_{-3.0}$ & 0.8$\pm$0.1 & 1.1$\pm$0.1 & 0.93/48	& 0.21 \\
\noalign{\smallskip}
L  &  12.1$_{-2.2}^{+2.4}$ & 0.9$\pm$0.2 & 4.8$^{+0.8}_{-0.6}$ & 0.9$\pm$0.1 & 1.1$\pm$0.1 & 1.00/92	& 0.93 \\
\noalign{\smallskip}
M  &  14.1$\pm$4.0 & 1.1$\pm$0.4 & 1.2$^{+0.4}_{-0.2}$ & 0.9$\pm$0.1 & 1.0$\pm$0.1 & 0.97/48	& 2.22 \\
\noalign{\smallskip}
N  &  13.1$_{-0.8}^{+0.9}$ & 1.3$\pm$0.1 & 12.3$^{+1.0}_{-0.9}$ & 0.87$\pm$0.05 & 1.09$\pm$0.06 & 0.97/204	& 1.03 \\
\noalign{\smallskip}
O  &  13.5$_{-3.0}^{+3.3}$ & 1.6$\pm$0.4 & 6.8$^{+2.5}_{-1.4}$ & 1.0$\pm$0.1 & 1.1$\pm$0.1 & 0.93/51	& 0.41 \\
\noalign{\smallskip}
P  &  14.0$\pm$1.4 & 1.5$\pm$0.2 & 3.3$^{+0.5}_{-0.4}$ & 1.00$\pm$0.08 & 1.00$\pm$0.08 & 0.96/129	& 2.24 \\
\noalign{\smallskip}
Q  &  13.0$_{-4.5}^{+4.9}$ & 1.3$^{+0.6}_{-0.5}$ & 1.0$^{+0.3}_{-0.5}$ & 1.1$\pm$0.2 & 1.0$\pm$0.2 & 1.15/30	& 1.64 \\
\noalign{\smallskip}
\noalign{\smallskip}
M+N & 13.1$\pm$0.8 & 1.2$\pm$0.1 & 4.6$\pm$0.3 & 0.90$\pm$0.05 & 1.08$\pm$0.05 & 0.96/217	& 3.25 \\
  \hline
    \hline
    \noalign{\smallskip}
  \end{tabular}
  }
  \end{center}
  \end{table}    

\end{document}